\documentclass[preprint]{cimento}
\usepackage{graphicx}
\usepackage{epsfig}
\title{Precision Kaon and Hadron Physics with KLOE}
\author{F. Bossi, E. De Lucia,  J. Lee-Franzini, S. Miscetti, M. Palutan\\\ETC
\atque
KLOE Collaboration\footnote{Appendix A}}
\instlist{\inst{}
Laboratori Nazionali di Frascati dell'INFN, Via E. Fermi 40, I-00044 Frascati, Italia}

\PACSes{\PACSit{11.30.Er}{Charge conjugation, parity, time reversal, and other discrete symmetries}
\PACSit{12.15.Hh}{Determination of Kobayashi-Maskawa matrix elements}
\PACSit{13.66.Bc}{Hadron production in $e^+e^-$ collisions}
\PACSit{14.40.Aq}{$\pi$, K, $\eta$ mesons}}

\begin{document}
\def\etc{{\it etc.}} \def\eg{{\it e. g.}}
\def\ie{{\it\kern-2pt i.\kern-.5pt e.\kern-2pt}}  
\def\BR{\hbox{BR}}  \def\prl{Phys. Rev. Lett}
\def\up#1{$^{#1}$}  \def\dn#1{$_{#1}$}
\def\ifm#1{\relax\ifmmode#1\else$#1$\fi}
\def\Bbar{\ifm{\rlap{\kern.22em\raise1.9ex\hbox to.58em{\hrulefill}} B}}
\def\bs{\ifm{B_s}} \def\bsbar{\ifm{\Bbar_s}}  \def\deg{\ifm{^\circ}}
\def\bo{\ifm{B^0}} \def\bra#1;{\ifm{\langle\,#1\,|}} 
\def\bob{\ifm{\overline B\vphantom B^0}}
\def\to{\ifm{\rightarrow}} \def\sig{\ifm{\sigma}}   \def\plm{\ifm{\pm}}
\def\K{\ifm{K}} \def\LK{\ifm{L_K}}
\def\ff{$\phi$--factory}  \def\DAF{DA\char8NE}
\def\f{\ifm{\phi}}  \def\klo{KLOE}
\def\pb{{\bf p}} \def\pic{\ifm{\pi^+\pi^-}} \def\pio{\ifm{\pi^0\pi^0}}
\def\pim{\ifm{\pi^-}}  \def\pip{\ifm{\pi^+}} \def\po{\ifm{\pi^0}}
\def\pipm{\ifm{\pi^\pm}}  \def\pimp{\ifm{\pi^\mp}}
\def\lpm{\ifm{\ell^\pm}}  \def\lmp{\ifm{\ell^\mp}}
\def\mupm{\ifm{\mu^\pm}}  \def\mump{\ifm{\mu^\mp}}
\def\epm{\ifm{e^\pm}}  \def\emp{\ifm{e^\mp}} \def\epem{\ifm{e^+e^-}}
\def\km{\ifm{K^-}}  \def\kp{\ifm{K^+}} \def\po{\ifm{\pi^0}}
\def\ks{\ifm{K_S}} \def\kl{\ifm{K_L}}
\def\kls{\ifm{K_{L,\,S}}}  \def\ksl{\ifm{K_{S,\,L}}}
\def\eps{\ifm{\epsilon}}
\def\rep{\ifm{\Re(\eps'/\eps)}}  \def\imp{\ifm{\Im(\eps'/\eps)}}
\def\Kb{\ifm{\rlap{\kern.2em\raise1.9ex\hbox to.6em{\hrulefill}} K}}
\def\kpm{\ifm{K^\pm}}  \def\kmp{\ifm{K^\mp}}
\def\ken{$K^\pm \rightarrow e^\pm \nu_e$}
\def\kmn{$K^\pm \rightarrow \mu^\pm \nu_\mu$}
\def\rk{$\Gamma(K^\pm \rightarrow e^\pm \nu_e) / \Gamma(K^\pm \rightarrow \mu^\pm \nu_\mu)$}
\def\C{\ifm{C}}  \def\P{\ifm{P}}  \def\T{\ifm{T}} \def\CP{\ifm{CP}}
\def\noc{\relax\hglue0pt{\rlap{$C$}\raise.15ex\hbox{$\kern
.18em\backslash$}}}
\def\nop{\relax\hglue0pt{\rlap{$P$}\raise.15ex\hbox{$\kern
.18em\backslash$}}}
\def\noT{\relax\hglue0pt{\rlap{$T$}\raise.15ex\hbox{$\kern
.18em\backslash$}}}
\def\slash#1{\rlap{$#1$}/\:} \def\nocp{\noc\nop} \def\nocpt{\noc\nop\noT}
\def\ko{\ifm{K^0}}  \def\kob{\ifm{\Kb\vphantom{K}^0}}
\def\gam{\ifm{\gamma}} \def\kkb{\ifm{\ko\kob}}
\def\dt{ \ifm{{\rm d}t} } \def\ab{\ifm{\sim}}  \def\x{\ifm{\times}}
\def\sta#1;{\ifm{|\,#1\,\rangle}} \def\ket#1;{\ifm{|\,#1\,\rangle}}
\def\L{\ifm{{\mathcal L}}}  \def\R{\ifm{{\cal R}}}
\def\pt#1,#2,{\ifm{#1\x10^{#2}}}
\def\kon{\ifm{K_1}} \def\ktw{\ifm{K_2}}
\def\minus{$-$}  \def\dif{\hbox{d}}   \def\Gam{\ifm{\Gamma}}
\def\Rh{\ifm{R_\eta'}}
\def\ord#1;{\ifm{{\mathcal O}(#1)}}
\def\cl{\centerline}
\def\Ref#1{reference \citen{#1}} \def\Sec#1{section #1}
\let\text=\rm
\def\black{\color{black}}
\def\red{\color{red}}
\def\blue{\color{blue}}  \def\L{\ifm{{\cal L}}}  \def\Lpb{\hbox{pb\up{-1}}}
\def\bbb{\blue\bf\mathversion{bold}}
\def\figb#1;#2;{\parbox{#2cm}{\epsfig{file=#1.eps,width=#2cm}}}
\def\BRo#1;{\ifm{\hbox{BR}(#1)}}  \def\BRu#1;{\ifm{\hbox{BR\up{(0)}}(#1)}}

\def\bye{\end{document}}
\newcommand{\abs}[1]{\ensuremath{\left|#1\right|}}
\newcommand{\Vus}{\ensuremath{V_{us}}}
\newcommand{\Vud}{\ensuremath{V_{ud}}}
\newcommand{\be}{\begin{equation}}
\newcommand{\ee}{\end{equation}}

\def\dd{{\rm d}}  \def\GeV{{\ \rm GeV\ }}
\def\MeV{{\ \rm MeV\ }} \def\keV{{\ \rm keV\ }}
\catcode`@=11 
\newdimen\z@ \z@=0pt 
\newskip\z@skip \z@skip=0pt plus0pt minus0pt
\def\m@th{\mathsurround=\z@}
\def\ialign{\everycr{}\tabskip\z@skip\halign} 
\def\eqalign#1{\null\,\vcenter{\openup\jot\m@th
  \ialign{\strut\hfil$\displaystyle{##}$&$\displaystyle{{}##}$\hfil
      \crcr#1\crcr}}\,}
\def\gev{\hbox{ GeV}}  \def\mev{\hbox{ MeV}}
\def\kev{\hbox{ keV}}  \def\ev{\hbox{ eV}}
\def\baba{$~e^+e^-\to e^+e^-$}
\def\mumu{$~e^+e^-\to \mu^+\mu^-$}
\def\tot{$~e^+e^-\to e^+e^-,~\mu^+\mu^-$}
\font\euler=eufm10 at 10pt    \def\Ma{\hbox{\euler M}}
\newcommand{\SN}[2]{\ensuremath{#1\times10^{#2}}}
\newcommand{\VSS}[3]{\ensuremath{#1\pm#2_{\rm stat}\pm#3_{\rm syst}}}
\def\muiii{\ifm{\pi^\pm \mu^\mp\nu}}
\newcommand{\pvec}{\ensuremath{\mathbf{p}}}
\newcommand{\PK}{\ensuremath{K}}
\newcommand{\PKbar}{\ensuremath{\bar{K}}}
\newcommand{\fzhat}{\mbox{$\tilde{f}_0(t)$}}
\newcommand{\lamp}{\mbox{$\lambda'_+$}}
\newcommand{\lampp}{\mbox{$\lambda''_+$}}
\newcommand{\lamo}{\mbox{$\lambda_0$}}
\newcommand{\lam}{\mbox{$\lambda_+$}}
\newcommand{\keg}{\mbox{$K^0_{e3\gamma}$}}
\newcommand{\kegf}{\mbox{$K^0_{e3(\gamma)}$}}
\newcommand{\Estar}{\ensuremath{E^*_{\gamma}}}
\newcommand{\qstar}{\ensuremath{\theta^*_{\gamma}}}
\newcommand{\X}{\ensuremath{\langle X \rangle}}
\newcommand{\kmpipigall} {K^{-}\rightarrow \pi^{-}\pi^0~(\gamma)}
\newcommand{\Dktau}[1] {\ensuremath{K^{{#1}}\to\pi^{#1}\pi^{+}\pi^{-}}}
\newcommand{\ktaup}[1] {\ensuremath{K^{{#1}}_{\tau '}}}
\newcommand{\mtrenta} {\ensuremath{p^{\ast}_{\pi\mu}}}
\newcommand{\fphat}{\mbox{$\tilde{f}_+(t)$}}
\newcommand{\lamop}{\mbox{$\lambda'_0$}}
\newcommand{\lamopp}{\mbox{$\lambda''_0$}}
\newcommand{\fVus}{\ensuremath{f_+(0)\,V_{us}}}
\newcommand{\fVusVud}{\ensuremath{V_{us}/V_{ud}\times f_K/f_\pi}}
\newcommand{\fo}{\ensuremath{f_+(0)}}
\newcommand{\fkfp}{\ensuremath{f_K/f_\pi}}
\newcommand{\etappp}{\ensuremath{\eta\rightarrow\pi^+\pi^-\pi^0}}
\newcommand{\etatrepi}{\ensuremath{\eta\rightarrow3\pi}}

\newcommand{\cA}{\mathcal{A}}
\def\GAML{\ifm{\Gamma_L}}
\def\GAMS{\ifm{\Gamma_S}}
\def\vv{\vphantom{H^I_I}}
\def\Re{{\rm Re}}
\def\Im{{\rm Im}}
\newcommand{\kzero}       {\ensuremath{K^{0}}}
\newcommand{\kzerob}      {\ensuremath{\bar{K}^{0}}}
\newcommand{\Mno}{m_{K_0}^{\phantom{p}}}
\newcommand{\Mnb}{m_{\overline K_0}}
\newcommand{\Gno}{\Gamma_{K_0}^{\phantom{p}}}
\newcommand{\Gnb}{\Gamma_{\overline K_0}}
\newcommand{\dis}{\displaystyle}
\newcommand{\no}{\nonumber}

\newcommand{\Eq}[1]{Eq.~\ref{#1}}
\newcommand{\eq}[1]{eq.~\ref{#1}}
\newcommand{\fig}[1]{fig.~\ref{#1}}
\newcommand{\Fig}[1]{Fig.~\ref{#1}}

\catcode`@=12 

\maketitle

\begin{abstract}
We describe the KLOE detector at \DAF, the Frascati \ff, and its physics program. We begin with a brief description of the detector design and operation. Kaon physics is a major topic of investigation with KLOE thanks in part to the unique availability of pure \ks,\kl, \kpm\ beams at a \ff. We have measured all significant branching ratios of all kaon species, the \kl\ and \kpm\ lifetimes and the $K\to\pi$ form factor's $t$ dependence. From the measurements we verify the validity of Cabibbo unitarity and lepton universality. We have studied properties of light scalar and pseudoscalar mesons with unprecedented accuracy. We have measured the $e^+e^-$\to\pic\ cross section necessary for computing the major part of the hadronic contribution to the muon anomaly. The methods employed in all the above measurements as well as the \f\ leptonic width, precision mass measurements and searches for forbidden or extremely rare decays of kaons and $\eta$-mesons are described. The impact of our results on flavor and hadron physics to date, as well as an outlook for further improvement in the near future, are discussed.
\end{abstract}
\tableofcontents{}
\section{Introduction}
\label{sec:intro}
Late in 1989 the Istituto Italiano di Fisica Nucleare, INFN, decided to construct an \epem\ collider meant to operate around 1020 MeV, the mass of the \f-meson. The \f-meson decays mostly to kaons, neutral and charged, in pairs. Its production cross section peaks at about 3 $\mu$barns. Even with a modest luminosity, \L=100 $\mu$b\up{-1}/s, hundreds of kaon pairs are produced per second. This collider, called a \ff\ and christened \DAF, is located at the Laboratori Nazionali di Frascati, LNF, INFN's high energy physics laboratory near Rome \cite{DAFNE+90:prop}.
\subsection{\DAF}
\label{sec:daf}
\begin{figure}[htb]
\begin{center}
\includegraphics[width=0.8\textwidth]{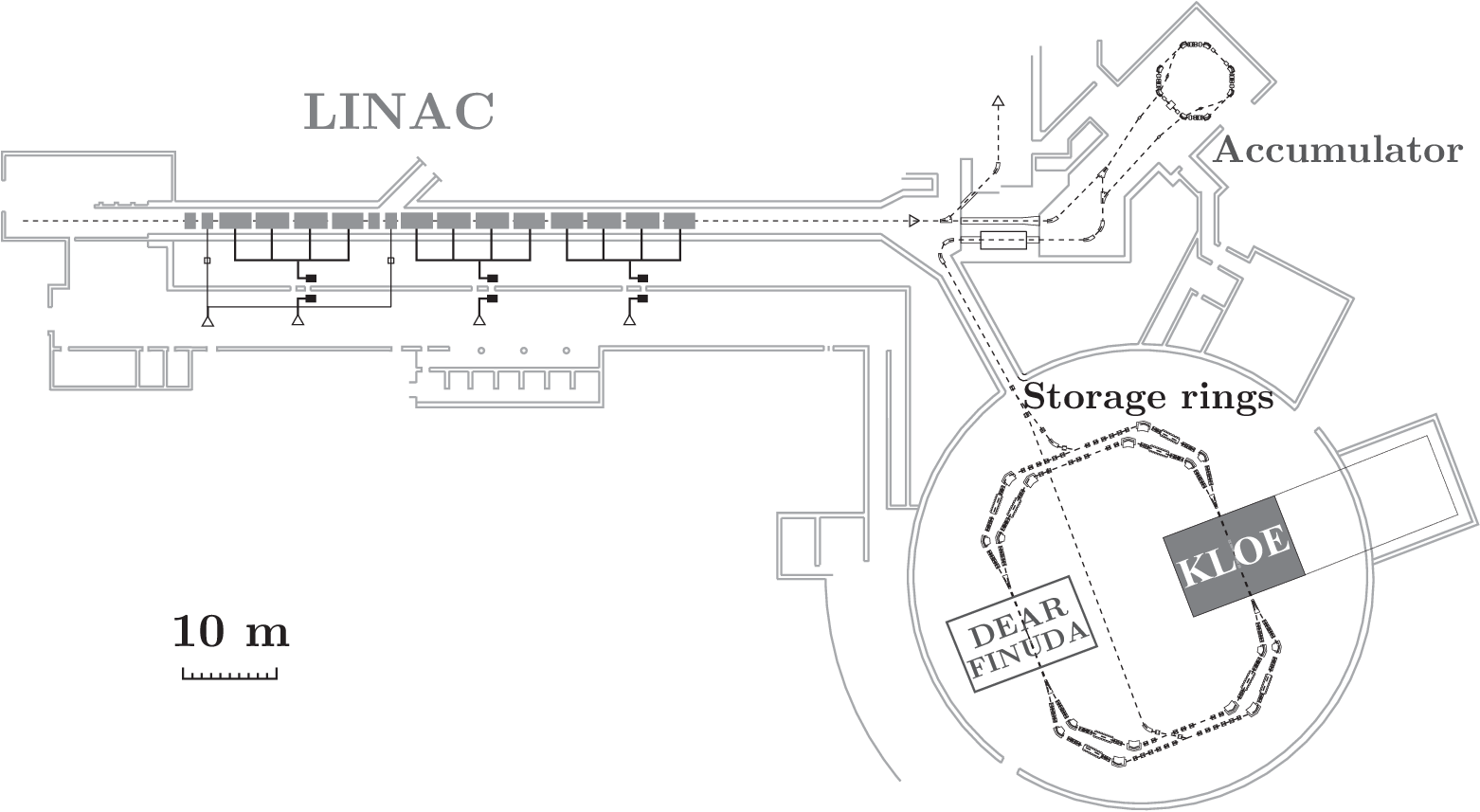}
\caption{The \DAF\ complex.}
\label{dafne}
\end{center}
\end{figure}

The \DAF\ layout is shown in \fig{dafne}.
The heart of the collider are two storage ``rings'' in which 120 bunches of electrons and positrons are stored.
Each bunch collides with its counterpart once per turn, minimizing the perturbation of each beam on the other.
Electrons and positrons are injected in the rings at final energy, \ab 510 MeV. Electrons are
accelerated to final energy in the Linac, accumulated and cooled in the accumulator and transferred to a single
 bunch in the ring.
Positrons require first accelerating electrons to about 250 MeV to an intermediate station in
the Linac,
where positrons are created. The positrons then follow the same processing as electrons.

\begin{figure}[htb]
\begin{center}
\includegraphics[width=0.8\textwidth]{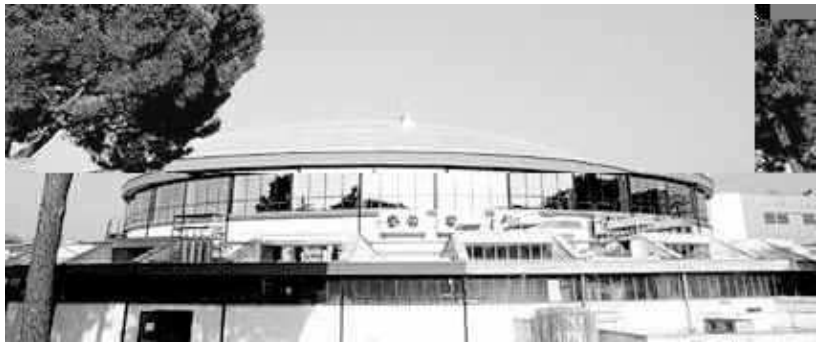}
\caption{The Adone building.}
\label{adone}
\end{center}
\end{figure}

Since the stored beam intensities decay rapidly in time, the cycle is repeated several times per hour. A collider is characterized by its luminosity \L\ defined by: event rate=\L\x cross section, beam lifetime and beam induced radiation background.
The latter is large for short lifetimes, which also result in the average luminosity being smaller than the peak value, in a word, lower event yield and large, variable background. \DAF\ is housed in the ADONE building at LNF, see \fig{adone}.

\begin{figure}[htb]
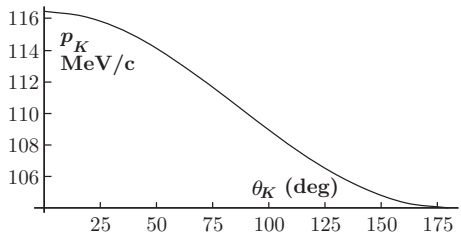

\cl{\figb p-vs-angle;6;}
\caption{Laboratory momentum vs angle to the $x$-axis for \ko-mesons.}
\label{fig:p-th}
\end{figure}

\DAF\ runs mostly at center of mass energy $W = m_\f \approx 1019.45~{\rm MeV}$.
The \f\ meson decays dominantly to charged kaon pairs (49\%), neutral kaon
pairs (34\%), $\rho\pi$ (15\%), and $\eta\gam$ (1.3\%).
A \f\ factory is thus a copious source of tagged and monochromatic
kaons, both neutral and charged.
In most of the following we will assume that all analysis are carried out in the \f-meson
center of mass.
While this is in fact what we do, it is appropriate to recall that the electron
and positron collide at an angle of $\pi$-0.025 radians.
The \f-mesons produced in the collisions therefore move in the laboratory system toward
the center of the storage rings with a momentum of about 13 MeV corresponding to
$\beta_\phi$\ab0.015,
$\gamma_\phi$\ab1.0001.
$K$ mesons from \f-decay are therefore not monochromatic in the laboratory,
\fig{fig:p-th}.

All discussion to follow refers to a system of coordinates with the $x$-axis in the horizontal
plane,
toward the center of \DAF, the $y$ axis vertical, pointing upwards and the $z$-axis
bisecting the angle of the two beam lines. The momentum of neutral kaons varies between 104 and 116 MeV and is a single valued function of the angle between the kaon momentum in the laboratory and the \f-momentum, \ie\ the $x$-axis. Knowledge of the kaon direction to a few degrees allows to return to the \f-meson center of mass, \fig{fig:p-th}. The mean charged kaon momentum is 127 MeV.

\subsection{KLOE}
\label{sec:kloe}
A state-of-the-art detector is needed to collect data at \DAF, from which physics results can be extracted. In summer 1991, the KLOE Collaboration was initiated proposing a detector and a program. The KLOE detector was not formally approved nor funded for another couple of years, because the authorities had not envisioned fully the magnitude of \klo. By mid Summer 1998 however, \klo\ was designed, constructed, completely tested and, complete with all of its electronics for signal processing, event gathering and transmission, was practicing with Cosmic Rays. Between Christmas and 1999 New Year, \klo\ was moved from its own assembly hall onto the  \DAF's South Interaction Region.

The detailed KLOE saga, with its concomitant joys and travails, could be an entertaining tale that should be written up somewhere. Our purpose in this paper, is instead, chapter \ref{sec:detector}, to extract and describe those properties of the detector, chapter \ref{sec:recoperf}, its performance,  and chapter \ref{sec:kb}, the special kaon ``beams'' available at \DAF, that made the physics measurements possible.

In the subsequent seven years KLOE, after beginning with a trickle of beam, by 2006 had produced data of such precision that the kaon, the first fundamental particle to introduce the concept of ``flavor" into our current way of thinking, got its definitive 21st century portrait re-mapped with high precision, 60 years after its discovery. These are the subjects covered in the next chapters: chapter \ref{sec:KPPD} details the new measurements of kaon physical parameters and decays; chapter \ref{kloevus}, discusses the KLOE determination of $V_{us}$ of the CKM mixing matrix , unitarity and lepton universality tests, as well as search for new physics; chapter \ref{sec:qinterf} is on quantum interferometry , and chapter \ref{sec:cpt} is on tests of CPT. Concurrently, capitalizing on the profuse production of light mesons resulting from electromagnetic transitions from the $\phi$-mesons KLOE has improved scores of light meson properties often by orders of magnitude and these are discussed in chapter \ref{sec:lightms}. Besides neutral kaon decays, symmetry tests have also been performed using light meson and charged kaon decays, see sections \ref{sec:eta3pi} \ref{sec:pipipi0}, and \ref{sec:kpmlife}. Finally, utilizing initial state radiation KLOE has measured the dipion cross section, $\sigma_{\pi\pi}$, over the whole range from the $\pi\pi$ threshold to the $\phi$ mass energy, see chapter \ref{sec:sighad}.

Most KLOE data were taken during two running periods. The 2001--2002 run
yielded 450 \Lpb\ of data, corresponding to approximately 140 million
tagged \ks\ decays, 230 million tagged \kl\ decays, and 340 million
tagged \kpm\ decays. Analysis of this data set is essentially complete
and most of the results are described in this report.
KLOE also took data in 2004--2005, after completion of a number of
\DAF\ modifications intended to increase the luminosity. About 2000~\Lpb\
were collected, giving a fivefold increase in statistics.
In total, \klo\ has collected during the years 1999 - 2005 \ab2.5 fb$^{-1}$ luminosity.
Analysis of the 2004--2005 data is very advanced and prognosis of the improvements in accuracy in the measurements that are statistically limited are included either in the specific sections or in the final chapter \ref{sec:concl}.

\section{Detector}
\label{sec:detector}
\begin{figure}[htb]
\begin{center}
\includegraphics[width=0.7\textwidth]{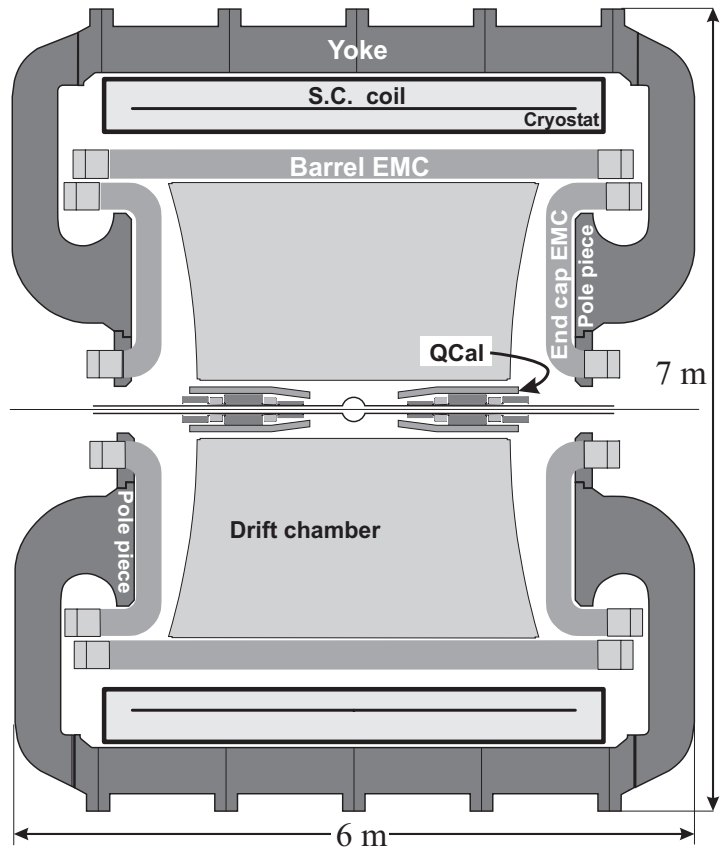}
\caption{Vertical cross section of the KLOE detector, showing the interaction region, the drift chamber (DC), the electromagnetic calorimeter (EMC), the superconducting coil, and the return yoke of the magnet.}
\label{kloe}
\end{center}
\end{figure}

\begin{figure}[htb]
\begin{center}
\includegraphics[width=0.8\textwidth]{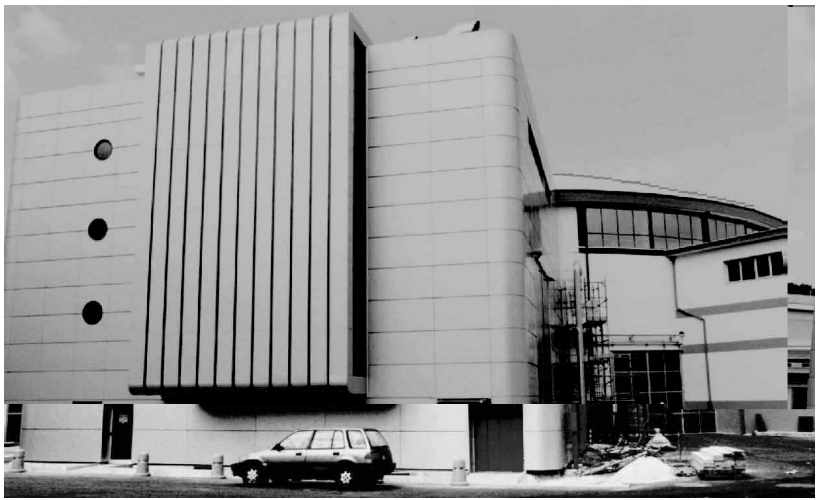}
\caption{KLOE Assembly Hall. The KLOE control room is on the top floor. The Adone cupola is just visible behind.}
\label{kloehall}
\end{center}
\end{figure}

The KLOE detector has a giant inner core, approximately 125 cubic meters, see \fig{kloe}, and was built in Frascati in the KLOE assembly building, see \fig{kloehall}.
It is on the same scale and complexity as the ``general purpose detectors'' of its time that were operating at the world's foremost laboratories such as LEP at CERN. If we consider that LEP operated at about one hundred times \DAF's energy, producing much more energetic particles and with correspondingly greater multiplicities, \klo's large size would seem unwarranted. We shall see however that while KLOE's complexity was necessitated  by its intent of being a definitive high precision experiment, its size, instead, was dictated by the quirk of one of the two neutral kaons, called the \kl, of living an extraordinarily long time (for a short lived particles) , about 51 nanoseconds. Kaons from \f-mesons decaying at rest, travel at approximately one fifth of the speed of light. This low speed is a great bonus for many reasons. However the mean path travelled by a \kl-meson, $\lambda_L=\gam\beta c\tau$ is 3.4 m. Thus if we want a detector which can catch $1-1/e$=63\% of all decaying neutral long-lived kaons, it must have a radius of some three and a half meters. We compromise on two meters and catch some 40\% of those decays. Less than that in fact, since some volume is inevitably lost around the beam interaction point, IP, and some more is lost at the outer edge, in order to be able to recognize the nature of the decays close to it.

The major components of \klo\ that enclose the decay volume around the IP are, going outwards radially, a large tracking device to reconstruct the trajectory of charged particles, the Drift Chamber DC, surrounded by a hermetic calorimeter to measure the energy and the entry point of photons, the Barrel EMC and two End-cap EMC.
The tracking chamber and the calorimeter are immersed in a mostly axial magnetic field of 0.52 T, provided by a superconducting coil enclosed in its iron yoke.

\subsection{Drift Chamber}
\label{sec:dc}
%
%
In the design of the KLOE tracking chamber one is constrained by the desire to catch all the charged secondary products from a decay and measure their properties with great precision, without jeopardizing the need to achieve the same for the neutral secondary products.
Recalling that when a charged particle travels in a medium, a gas, it
undergoes repeated collisions with electrons, releasing part of their
energy, resulting in a string of ion pairs created marking its
passage. Thus when the particle enters a drift chamber, DC (a gas filled
chamber traversed by a large number of wires, some kept at +2000 volt, anode wires, some at ground),
 electrons of the ion pairs created along its trajectory drift to the
 positive voltage wires and from an avalanche multiplication mechanism a detectable signal appears at the wire's end.
KLOE custom designed sophisticated electronics that are mounted at the wire's end which senses the tiny signals and provide us with the drift charge and time measurement. Furthermore, the total integrated charge collected at the wire's end gives information on the energy released from the initial particle.

Furthermore, particles of different mass
travelling in the same medium and with the same momenta have different energy releases: for the same $p$ they have different $\beta$ and
approximately the specific ionization $\dif E/\dif x\propto 1/\beta^2$.
Therefore this information is suitable to perform particle identification (PID). Finally, measuring the drift time locates the distance of the track from the anode wires which is then used to reconstruct the trajectory of the particles.

A caveat, the actual path of the particle is altered by multiple scattering, the multiple scattering angle being $\propto 1/\sqrt X_0$, where $X_0$ is the radiation length which is roughly proportional to $1/Z^2$, in gr/cm$^2$ (and one must not forget the density). Finally, while moving towards the anode, the drifting ionization undergoes diffusion, the amount of which is a function of the density of the gas mixture of the DC.

After judicious balancing, in order to avoid the \kl\to\ks\ regenerations
(\ks\ generated from the strong interaction of a \kl\ with the traversed medium that might simulate CP-violating decays), a helium based gas mixture was chosen to minimize multiple scattering and density, despite the fact diffusion of drifting ionization is larger in He than certain other gases. The average value of the radiation length in the DC volume is $X_0\ab900$ m, including the wire's contribution.

Again, for the mechanical structure low-$Z$ and low density materials were used
to minimize \kl\ regeneration, tracks' multiple coulomb
scattering and the absorption of photons before they can reach the
calorimeter. To achieve this the chamber is constructed out of a carbon fiber composite. The chamber volume is delimited by an outer cylinder of 2 m radius, an inner cylinder of 25 cm radius and closed by two 9 mm thick annular endplates 9.76 m radius of curvature and a 25 cm radius hole in the center. Due to their large size, the endplates were fabricated in two halves then glued together. About 52,000 holes were drilled on each endplate for the insertion of the feedthroughs holding the DC wires.
The total axial load on the end plates, about 3500 kg, would deform the end plates by several mm at the end of stringing. The deformation problem was solved by using rings of rectangular cross-section which surround the L-shaped plate rims, made of the same material. The ring is used to apply a tangential pull to the spherical plate via 48 tensioning screws made of titanium alloy pulling on tapped aluminum inserts in the L-shaped flange. The net result of the pull is to approximate a complete sphere beyond the outer edge of the plate which greatly reduces the deformation. Each screw was instrumented with two strain gauges, calibrated in order to compensate for bending effects on the screws, so that the pulling force was always known. At the end of stringing the deviation from the spherical shape
of the endplates was measured to be less than 1 mm and the hole positions to be accurate at the level of 60 $\mu$m. The KLOE DC is indeed a tour de force achievement thereafter copied in other experiments ({\it albeit} never to the same large dimensions).

\begin{figure}[htb]
\begin{center}
\includegraphics[width=0.7\textwidth]{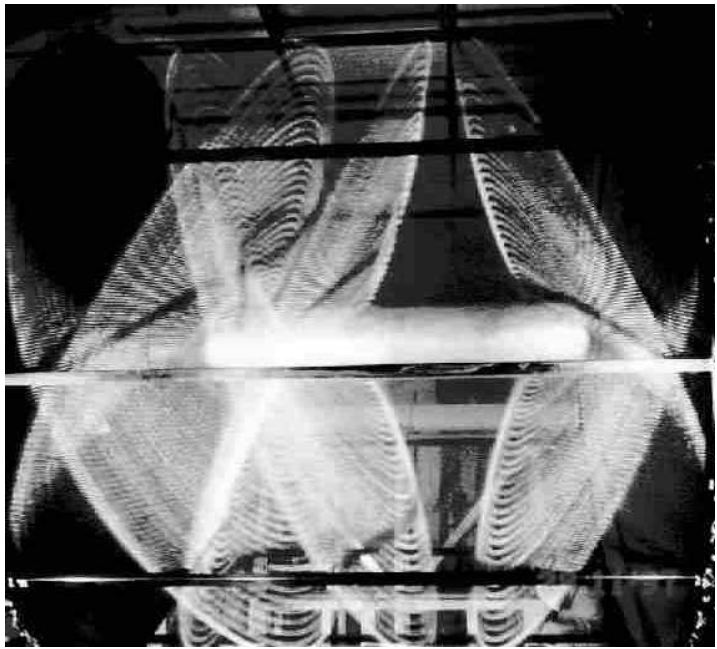}
\caption{DC all stereo geometry.}
\label{fig:dcbw}
\end{center}
\end{figure}

\begin{figure}[htb]
   \begin{center}
   \mbox{ \epsfig{file=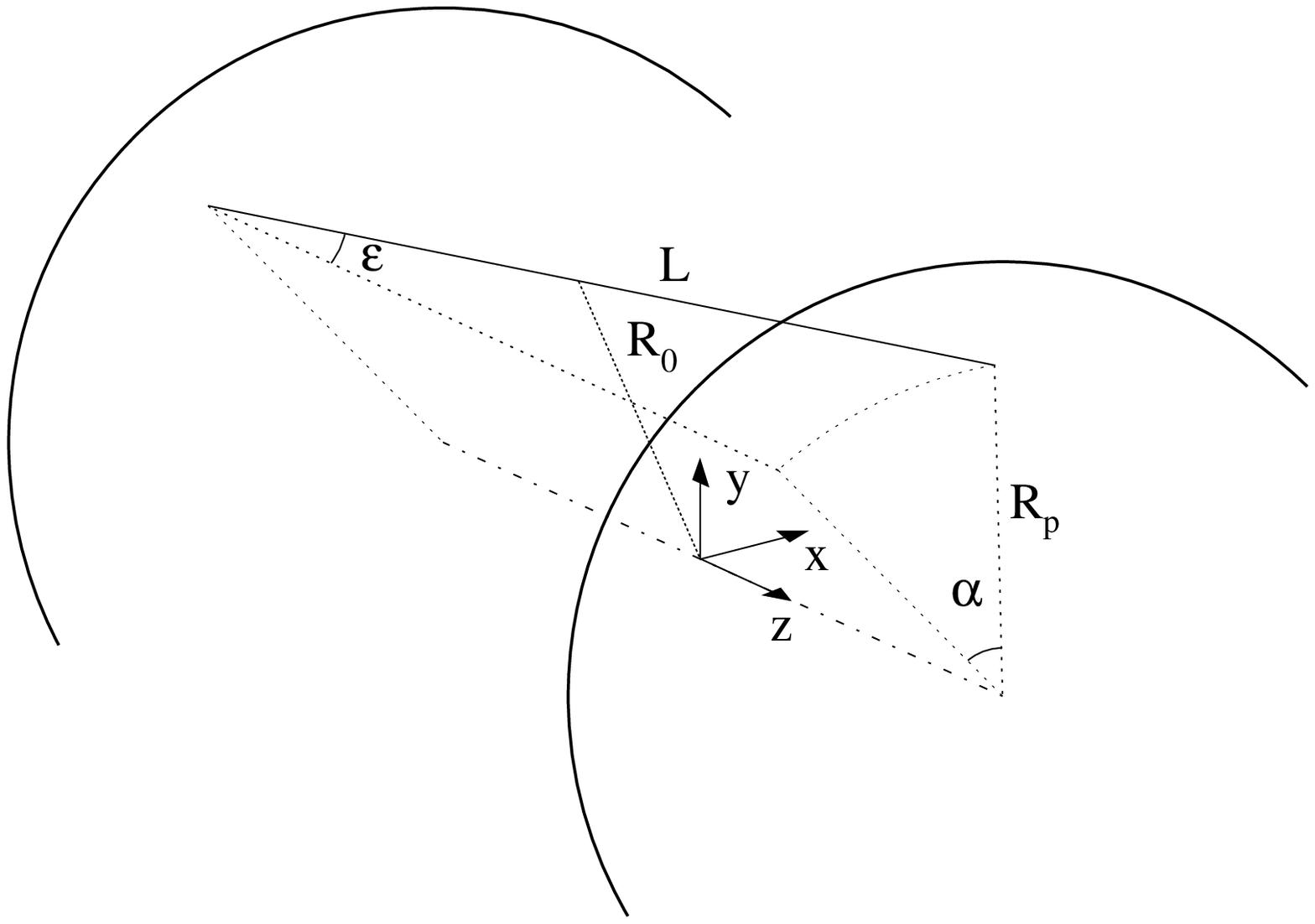,width=0.55\textwidth}}
   \mbox{ \epsfig{file=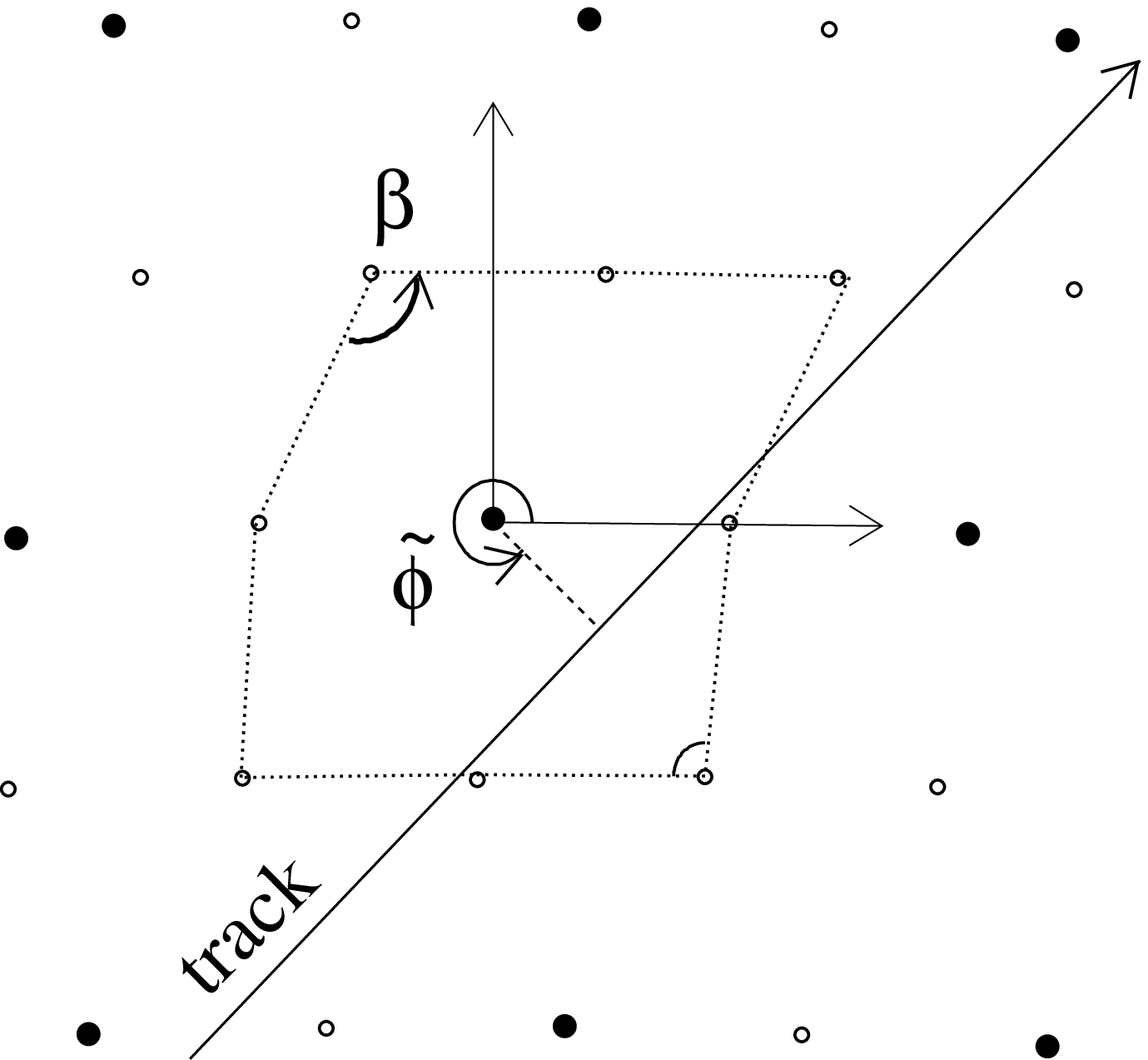,width=0.4\textwidth}}
   \end{center}
   \caption{{\it Left:} Wire geometry with the definition of stereo angle
   $\epsilon$ between the wire of length $L$ and the $z$-axis.
            {\it Right:} Definition of $\beta$ and $\tilde{\phi}$
            characterizing the shape of the cell and the angle of the
            incident track.}
   \label{fig:cells}
\end{figure}

The \klo\ DC~\cite{DCNIM} has \ab52,000 wires, about 12,500 of them are sense wires,
the remaining ones shape the electric field in 12,500 almost square cells
located on cylindrical layers around the $z$-axis (defined by the bisector of the two crossing beams).
Particles from the $\phi$ decays are produced with small momenta and therefore track density is much higher at small radii.
This motivated the choice of having the 12 innermost layers made up of $2\times2$ cm\up2 cells and the remaining 48 layers with $3\times3$ cm\up2 cells.
For a particle crossing the entire chamber we get dozens of space points from which we reconstruct a track:
the track of a  pion from $K_{S} \rightarrow \pi^+ \pi^-$ decay has $\sim$60 space points in average.
An all stereo geometry allows the measurement of the $z$ coordinate. \Fig{fig:dcbw} shows the wire geometry during the DC construction as
illuminated by light.
In this particular arrangement, shown in \fig{fig:cells} left,
the wires form an angle with the $z$-axis, the stereo angle $\epsilon$.
Consecutive layers have alternating signs of this angle
and its absolute value increases from 60 to 150 mrad with the layer radius
while keeping fixed for all the layers the difference between $R_{p}$ and
$R_{0}$ distances (1.5 cm), defined in \fig{fig:cells} left.

This design results in a uniform filling of the
sensitive volume with almost-square drift cells, with shape slowly changing along $z$,
which increases the isotropy of the tracking efficiency.

To extract the space position
from the measured drift time of the incident particle, we need to know the {\it space-time relation}
describing the cell response. The whole DC can be described using
232 {\it space-time relations} only, parametrized in terms of the angles $\beta$ and $\tilde{\phi}$
defined in \fig{fig:cells} right. The $\beta$ angle characterizes the geometry of the cell,
directly related to the electric field responsible of the avalanche multiplication mechanism.
The $\tilde{\phi}$ angle gives the orientation of the particle trajectory in the
cell's reference frame, defined in the transverse plane and with origin in the sense wire of the
cell.
The presence of a signal on a wire and its drift time and distance is called hit.
Once we have the hits, we can proceed with the event reconstruction using the wire geometry, {\it space-time relations}
and the map of the magnetic field. This procedure follows three steps: (i) pattern recognition,
(ii) track fit, and (iii) vertex fit.
The pattern recognition associates hits close in space to form track candidates and gives a first estimate of the track parameters.
Then track fit provides the final values of these parameters minimizing a $\chi^2$ function based on the
difference between the fit and the expected drift distances, residuals, as evaluated from
a realistic knowledge of the cells' response (measured drift times and {\it space-time relation}).
This is an iterative procedure because the cells' response depend on the
track parameters.
Finally the vertex fit searches for possible primary and secondary vertices, on the basis of the distance of closest approach
between tracks.

%
%
Besides the drift time and position information, hits in the DC are
also associated to a charge measurement directly related to the energy released
in the cell from the incident particle.
The amount of collected charge depends on the
particle's track length in the crossed cell. The energy deposited in a
sample of finite thickness fluctuates and exhibits a Landau distribution,
with a long tail at high energies. Therefore to improve the energy
resolution, the energy deposited is sampled many times for each track and
a truncated mean technique is used. To use the charge information for
particle identification purposes it is necessary to normalize the charge
to the track length and define the specific ionization dE/dx. If the energy
is measured in counts, we have on average 15 counts/cm for
pions and muons from $K^{\pm}_{\mu2}$ and $K^{\pm}_{\pi2}$ decays,
with momentum within 180 and 300 MeV. Instead $K^{\pm}$ tracks
exhibit a larger number,
$\sim$120 counts/cm.

The drift chamber provides tracking in three dimensions with resolution in the bending plane better than 200 $\mu$m, resolution on the $z$ measurement of \ab2 mm and of \ab1 mm on the decay vertex position.The particle's momentum is determined from the curvature of its trajectory in the magnetic field with a fractional accuracy better than \ab0.4\% for polar angles larger than 45$^{\circ}$.
The calibration of the absolute momentum scale has a fractional accuracy of
few $10^{-4}$ using several two- and three-body decays ($e^+e^-\to
e^+e^-$,$e^+e^-\to\mu^+\mu^-$, $K_{L}\to \pi l \nu$, $K_{L}\to \pi^+ \pi^-
\pi^0$, $K^{\pm}\to\mu^{\pm}\nu$,$K^{\pm}\to\pi^{\pm}\pi^0$) covering a wide momentum
range. Deviations from nominal values of invariant mass, missing mass or
momentum of the charged decay particle in the rest frame of the decaying
particle are used as benchmarks for the calibration procedure. The resolution on the specific ionization measurement obtained using
$K^{\pm}$ two- and three-body decays and Bhabha scattering events,
is $\sim 6\%$ for pions and $\sim 7\%$ for electrons~\cite{DCADCNIM},
using 60 truncated samples.

The DC system's complexity, as well as the demands placed upon it requires
the response of each cell to an incoming particle to be known accurately.
The cell response is a function of its geometry and of the voltage
applied to the wires. Thus the position of the DC,
mechanical structure as well as wires, had to be surveyed, in situ,
together with the other \klo\ components.

The Slow Control monitoring system (sect.~\ref{sec:slow}) fully embedded in
the Data Acquisition process (sect.~\ref{sec:daq}), sets and registers the
voltages applied to the wires, sends the information to the read-out
electronics and controls its proper functioning \cite{DAQNIM}.
Moreover the gas mixture parameters, namely composition, temperature and pressure,
enter directly in the {\it space-time relations} used in reconstructing the
particle trajectory, to extract the space position from the measured drift
time of the incident particle.
The Slow Control System also monitors the gas parameters.

To ensure the stability in time of the DC performance,
the system has to be calibrated periodically by acquiring
samples of cosmic-ray muons suitable for the
measurement of the \ab200 different {\it space-time relations}.
The calibration program, incorporated into the KLOE online system,
automatically starts at the beginning of each run and selects about 80,000
cosmic-ray events. These events are tracked using the existing
{\it space-time relations} and the average value of the residuals for hits
in the central part of the cells is monitored. If the residuals exceed 40
$\mu$m then about 300,000 cosmic-ray events are collected, with about 30 Hz
rate, and a new set of calibration constants is obtained.
At the beginning of each data-taking period a complete calibration of the
{\it space-time relations} is needed together with the measurement of
the DC time offsets, the latter from some 10$^7$ cosmic-ray events.
Finally, during data taking the DC performances are
monitored using selected samples of Bhabha scattering events.

%
%
\subsection{Electromagnetic Calorimeter}
\label{sec:emc}
The neutral decay products of the \kl's, its short-lived partner \ks\
(lifetime \ab\ 89 picoseconds) and of those of their charged brethren,
the \kpm's, which we need to detect, are photons, either directly
produced or from the decay of neutral pions into two photons.
It is important to recall that photons for energies larger than a few MeV, as in most of our cases, interact with matter by electron positron pair production that in turn radiate photons when traversing matter, a process quite similar to pair-production. Thus, a photon, or an electron for that matter, travelling through dense, high $Z$  matter, repeatedly undergoes radiation and pair production in cascade, until all its energy is expended in a so called shower of \epem\ and photons. From detection of these particles and their energy one measures the original photon's (or electron's) energy. This is done with a ``calorimeter'',
which also usually pinpoints the position of the electromagnetic
cascade (or shower) as well.

The \klo\ calorimeter, EMC, has been designed and built to satisfy
a set of stringent requirements such as providing a hermetic
detection of low energy photons  (from 20 to 500 MeV) with
high efficiency, a reasonable energy resolution  and an excellent
time resolution to reconstruct the vertex for the \kl\ neutral decays.
The calorimeter response has also to be fast since its signals are
used to provide the main trigger of the events.

The chosen solution for the EMC is that of a sampling calorimeter,
composed of lead passive layers, accelerating the showering
process, and of  scintillating fiber sensing layers.
It is made of cladded 1 mm scintillating
fibers sandwiched between 0.5-mm-thick lead foils.
The foils are imprinted with grooves wide enough to accommodate
the fibers and some epoxy, without compressing the fibers.
This precaution prevents damage to the fiber-cladd\-ing interface.
The epoxy around the fibers also provides structural strength
and removes light travelling in the cladding.
In more detail, the basic structure of 0.5 mm, grooved Pb
layers between which are embedded 1 mm diameter scintillating fibers are
positioned at the corners of almost equilateral triangles.
\ab 200 such layers are stacked, glued, and pressed, resulting
in a bulk material. The resulting composite has a fibers:lead:glue volume ratio of
48:42:10 which corresponds to a density of 5 g/cm$^3$, an equivalent
radiation length $X_0$ of 1.5~cm
and an electromagnetic sampling fraction of \ab13\%.
The large ratio of active material to
radiator, the frequent signal sampling and the special fiber arrangement
result in a factor $\sqrt{2}$ improvement in energy resolution
with respect to calorimeter with slabs of equivalent scintillator:lead
ratio. The chosen fibers, Kuraray SCSF-81 and Pol.Hi.Tech. 00046,
are blue emitting scintillating fibers with  a fast emission time
whose larger component is described by an exponential distribution
with a $\tau$ of 2.5 ns. In addition, the special care in design and
assembly of the Pb-fiber composite ensures that the light propagates
along the fiber in a single mode, resulting in a greatly reduced spread
of the light arrival time at the fiber ends.  Propagation speed in the
fiber is $\sim$ 17 cm/ns.
These fibers produce more than 5$\times 10^3$ photons for one MeV of
deposited energy by ionizing particles. 3\% of the light reaches the fiber ends. Fibers transmit 50\% of the light over 2 m ($\lambda \sim 4 $m).

\begin{figure}[htb]
\begin{center}
\includegraphics[width=0.8\textwidth]{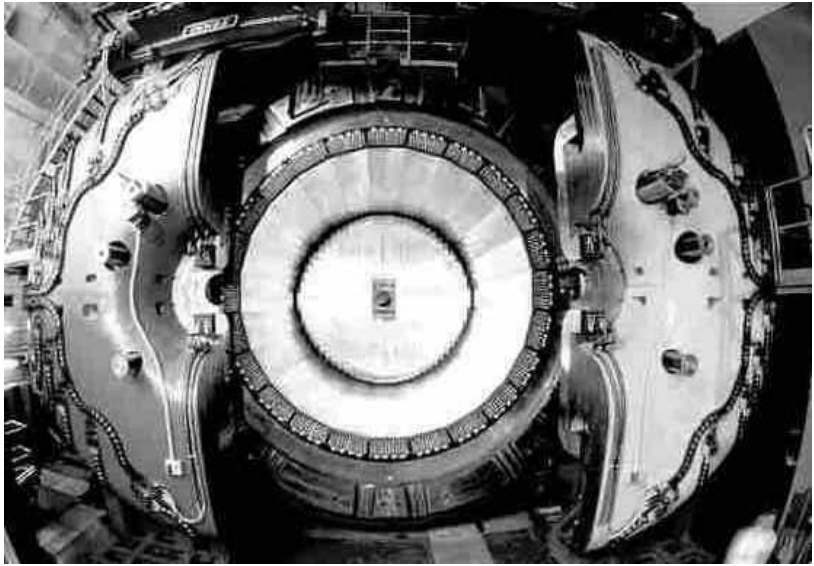}
\caption{EMC viewed before DC insertion. Note end-caps attached to pole
  pieces seen in profile.}
\label{fig:emcbw}
\end{center}
\end{figure}

This material is shaped into modules 23~cm thick (\ab$15X_0$).
24 modules of trapezoidal cross section are arranged in azimuth to form
the calorimeter barrel, aligned with the
beams and surrounding the DC. An additional 32 modules, square or rectangular
in cross section, are wrapped around each of the
pole pieces of the magnet yoke to form the endcaps,
which hermetically close the calorimeter up to \ab98\% of
4$\pi$, see \fig{fig:emcbw}. The unobstructed solid-angle coverage of the calorimeter
as viewed from the origin is \ab94\%~\cite{EmCNIM}.

The fibers run parallel to the axis of the detector in the barrel,
vertically in the endcaps, and are read out at both ends viewed by lucite light guides of area of
4.4$\times$4.4 cm$^2$. The other end of the guides are coupled to the photocathodes of
fine mesh photomultipliers of 1.5 inches diameter with quantum
efficiency of \ab20\%.  The fiber length of 4 m corresponds
to an average photoelectron yield of 35 p.e./layer for
a minimum ionizing particle crossing the calorimeter at the
fiber center. The photomultiplier tubes at each fiber's end transform the
light into electric pulses. Their amplitudes, $A_{i}$, is proportional to
the amount of energy  deposited while the recorded times, $t_{i}$,
are related to the time of flight of the particle.
For each cell the position of
the readout elements  and the  difference of the arrival times
at the two fiber ends  determine the shower position with a
${\cal O}$(1) cm accuracy.
The full stack of 23 cm (15 $X_0$) fully contains a shower of
500 MeV. 4880 photomultiplier tubes view the 24+32+32 modules
into which the calorimeter is subdivided for a grand total of
15,000 km used optical fibers.

As a first step, the reconstruction program makes the average of time
and energy of the recorded $t_{i}$ and $A_{i}$ for the two
sides of each cell and compute the hit position.
Corrections for attenuation length, energy scale, time offsets
and light propagation speed are taken into account at
this stage. A clustering procedure then groups together nearby clumps
of energy  deposition and calculates the average quantities over all
the participating cells. The calibration constants to transform $t_{i}$ and $A_{i}$ from raw quantities to time in nanoseconds and energy in MeV  are evaluated with
dedicated on-line and off-line algorithms.

The energy calibration starts by a first equalization in cell response
to minimum ionizing particles (m.i.p)  at calorimeter center and by
determining the attenuation length of each single cell with a dedicated
cosmic ray trigger. This is done before the start of each long data taking period.
The determination of the absolute energy scale in MeV relies instead
on a monochromatic source of 510 MeV photons: the $e^+e^- \to \gamma \gamma$
sample. This last calibration is routinely carried out each 200-
400 nb$^{-1}$ of collected luminosity.

For the timing, the relative time offsets of each channel,
$T^0_{i}$, related to cable lengths and electronic delays
and the light velocity in the fibers are evaluated every
few days with high momentum cosmic rays selected
with DC information. An iterative procedure
uses the extrapolation of the tracks in the calorimeter to
minimize the residuals between the expected and measured times
in  each cell. A precision of few tens of picoseconds is obtained
for these offsets.

To run at the design luminosity, DA$\Phi$NE operated
with a bunch-crossing period equal to the machine radio frequency (RF)
period, $T_{RF}=2.715$~ns. Due to the spread of the
particle's arrival times, the trigger is not able to
identify the bunch crossing related to each event,
which has to be determined offline. The start signal for
the electronics devoted to time measurement (TDC)
is obtained by phase-locking the trigger level-1 signal (sect.~\ref{sec:trig}) to an RF replica, with a
clock that has a period of $4\times T_{RF}$. The calorimeter
times are then related to  the time of flight
of the particle from IP to the calorimeter, $T_{tof}$, by
$T_{cl} = T_{tof}+\delta_{C}-N_{bc} T_{RF}$,
where $\delta_C$ is a single number accounting for
the overall electronic offset and cable delay,
and  $N_{bc}$ is the number of bunch-crossings needed
to generate the TDC start.
The values of $\delta_C$ and $T_{RF}$
are determined for each data taking run with $e^+e^- \to \gamma\gamma$
 events
by looking at the distribution of $T_{cl}-R_{cl}/c$
where $R_{cl}/c$ is the expected time of flight:
well separated peaks correspond
to different values of $N_{bc}$. We arbitrarily define
$\delta_C$ as the position of the peak with the largest
statistics, and determine $T_{RF}$ from the peaks' distances.
Both quantities are evaluated with a precision better than 4 ps for
200 nb$^{-1}$ of integrated luminosity. This
measurement of  $T_{RF}$ allows us to set the absolute
calorimeter time scale to better than 0.1 \%.
During offline processing, to allow the cluster times to be related
to the particle time of flight,
we subtract $\delta_C$ and determine, on an event
by event basis, the ``global'' event start time $T_0$,
i.e. the quantity $N_{bc} T_{RF}$.
A starting value for all analysis is evaluated by
 assuming that the earliest cluster
in the event is due to a photon coming from the
interaction point. Further corrections
are analysis dependent.

The high photon yield and the frequent sampling enable
cluster energies to be measured with a resolution of
$\sigma_E/E = 5.7\%/\sqrt{E\ {\rm(GeV)}}$, as determined
with the help of the DC using radiative Bhabha events.
The absolute time resolution
$\sigma_t = 57~{\rm ps}/\sqrt{E\ {\rm (GeV)}}$
is dominated by photoelectron statistics, which is
well parametrized by the energy scaling law.
A constant term of 140 ps has to be added in quadrature, as
determined from $e^+e^- \to \gamma\gamma$, radiative \f\ decays
and $\phi \to \pi^+\pi^-\pi^0$ data control samples.
This constant term is shared between a channel by channel
uncorrelated term and a  common term to all channels.
The uncorrelated term is mostly due to the  calorimeter
calibration while the common term is related to the
uncertainty of the event $T_0$, arising from the \DAF\ bunch-length
and from the jitter in the trigger phase-locking to the machine RF.
By measuring the average and the difference of
$T_{cl}-R_{cl}/c$ for the two photons in $\phi \to \pi^+\pi^-\pi^0$
events, we estimate a similar contribution of \ab~100 ps for the
two terms.
Cluster positions are measured with resolutions of 1.3~cm in the
coordinate transverse to the fibers, and, by timing, of
$1.2~{\rm cm}/\sqrt{E\ {\rm(GeV)}}$ in the longitudinal coordinate.
These performances enable the $2\gamma$ vertex in
$\kl\to\pic\po$ decays to be localized with $\sigma \approx 2$~cm
along the \kl\ line of flight, as reconstructed from the tagging \ks\
decay. Incidentally, the thin lead layers used and the high photon yield
allows high reconstruction efficiency for low energy photons.

\subsection{Beam Pipe and Quadrupole Calorimeter}
\begin{figure}[htb]
\begin{center}
\includegraphics[width=0.8\textwidth]{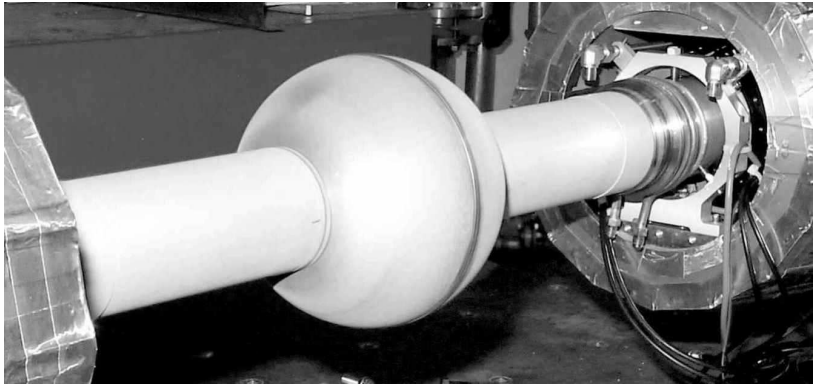}
\caption{Beam pipe at the KLOE interaction region. Just visible on either side are the final focus quadrupole ends.}
\label{beamp}
\end{center}
\end{figure}

At \DAF\ the mean \ks, \kl\ and $K^{\pm}$ decay path lengths are $\lambda_S = 0.59$~cm, $\lambda_L = 3.4$~m and $\lambda_{\pm} = 95$~cm.
To observe rare \ks\ decays and \kl\ks\ interference with no background from
$\ks \rightarrow \kl$ regeneration, a decay volume about the interaction
point with $r > 15 \lambda_S$ must remain in vacuum. The beam pipe, see
\fig{beamp}, surrounds the interaction point, with a sphere of 20 cm inner
diameter, with walls 0.5 mm thick made of a Be-Al sintered compound. This
sphere provides a vacuum path $\sim 10\times$ the $K^0$ amplitude decay length,
effectively avoiding all $\ks \rightarrow \kl$ regeneration.
Permanent quadrupoles for beam focusing are inside the apparatus
at a distance of 46 cm from IP, and surrounding the beam pipe.
The quadrupoles of this low-$\beta$ insertion are equipped with two
lead/scintillating-tile calorimeters, QCAL, of $\sim 5 X^0$ thickness, with the
purpose of detecting photons that would be otherwise absorbed
on the quadrupoles~\cite{QNIM}.
Specifically, the QCAL main task is to
identify and reject  photons from $K_L \to 3 \pi^0$ decays
when selecting the CP violating $K_L \to 2 \pi^0$ events.
Each calorimeter consists of a sampling structure of lead and
scintillator tiles arranged in 16 azimuthal sectors. The readout
is performed by plastic optical fibers, Kuraray Y11, which
shift light from blue to green (wavelength
shifter, WLS), coupled to mesh photomultipliers.
The special arrangement of WLS fibers allows also the measurement of the
longitudinal $z$ coordinate by time differences.
Although the tiles are assembled in a way which maximizes
efficiency for $K_L$ photons, a high efficiency is in fact also obtained
for photons coming from the IP. This allows us to extend
the EMC coverage  of the solid angle down to $\cos\theta$=0.99
for prompt decays. The time resolution is related
to the light yield and emission time/photoelectron of the
Kuraray Y11 fibers.
For a m.i.p. crossing the whole stack of 16 tiles, we had obtained
during tests a light yield of $\sim$ 50 photoelectrons for entry points
close to the PM's side. Considering that a m.i.p. deposits
an energy equivalent to that of a 75 MeV photon, this
corresponds to $\sim 0.5$ p.e./MeV and in turn implies
a time resolution for electromagnetic showers
of 205 ps$/\sqrt{\rm E(GeV)}$. However during running,
the presence of the B-field reduces the effective light
yield by a factor 1.4, thus worsening the timing to
240 ps$/\sqrt{\rm E(GeV)}$, which however is
high enough to yield an efficiency
above 90\% for photon energies  greater than 20 MeV.
\subsection{Trigger}
\label{sec:trig}
Event rates at \DAF\ are high; at the maximum luminosity of $10^{32}$ cm$^{-2}$s$^{-1}$ up to $300 \phi$ s$^{-1}$
and $30,000$ Bhabha s$^{-1}$ are produced within the KLOE acceptance window. The trigger design~\cite{trigNIM}
has been optimized to retain all $\phi$ decays. Moreover, all Bhabha and $\gamma\gamma$ events produced at large polar
 angles are accepted for detector monitoring and calibration, as well as a downscaled fraction of cosmic-ray particles,
which cross the detector at a rate of $\sim 3000$ Hz. Finally, the trigger provides efficient rejection
on the two main sources of background: small angle Bhabha events, and particle lost
from the \DAF\ beams, resulting in very high photon and electron fluxes in the interaction region.

The trigger electronic logic system rapidly recognizes topologies and energy releases of interest,
processing the signals coming both from EMC and DC.
Since $\phi$ decay events have a relatively high multiplicity, they can be efficiently selected by the EMC trigger
by requiring two isolated energy deposits (trigger sectors) in the calorimeter above a threshold of 50 MeV in the
barrel and 150 MeV in the endcaps. Events with only two fired sectors in the same endcap are rejected, because this
topology is dominated by machine background.  Moreover, events with charged particles in the final state give a
larger number of hit wires in the DC than do background events. The DC trigger employs this information, requiring
the presence of \ab 15 hits in the DC within a time window of 250~ns from beam crossing.
An event satisfying at least one of the two above conditions generates a first level trigger, T1, which is produced
with minimal delay and is synchronized with the \DAF\ master clock. The T1 signal initiates conversion in the
front-end electronics modules, which are subsequently
read out following a fixed time interval of $\ab 2 \mu$s, which is ultimately driven by the typical drift distances
travelled by electrons in the DC cells.
In case of DC trigger, a validation of the first level decision is required
by asking for \ab 120 hits within a 1.2-$\mu$s time window.

The KLOE trigger also implements logic to flag cosmic-ray events,
which are recognized by the presence of two energy deposits
above 30~MeV in the outermost calorimeter layers.
For most data taking, such events were rejected after partial
reconstruction by an online software filter.

\subsection{Data acquisition and processing}
\label{sec:daq}
\begin{figure}[htb]
\begin{center}
\includegraphics[width=0.4\textwidth]{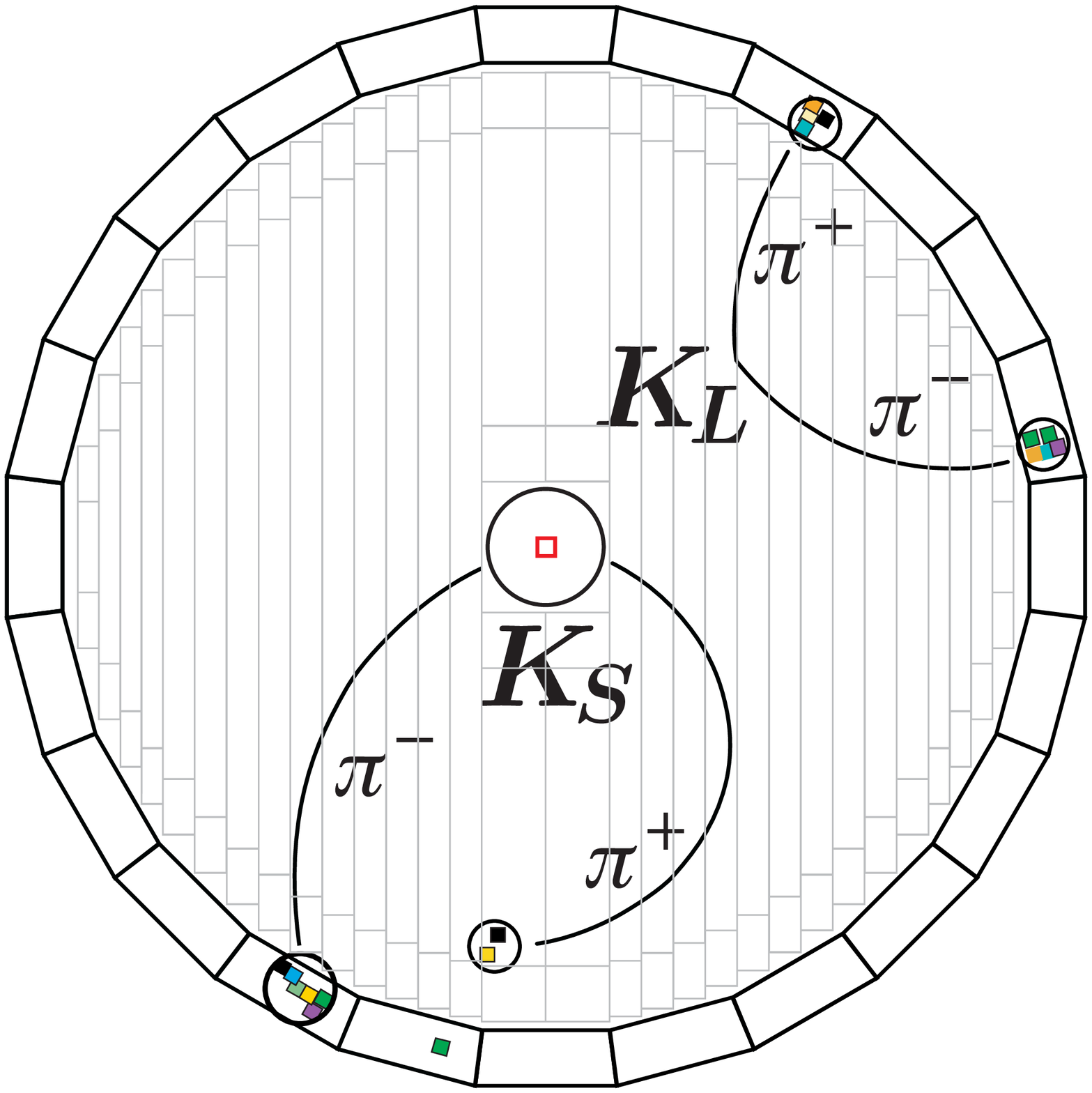}
\includegraphics[width=0.4\textwidth]{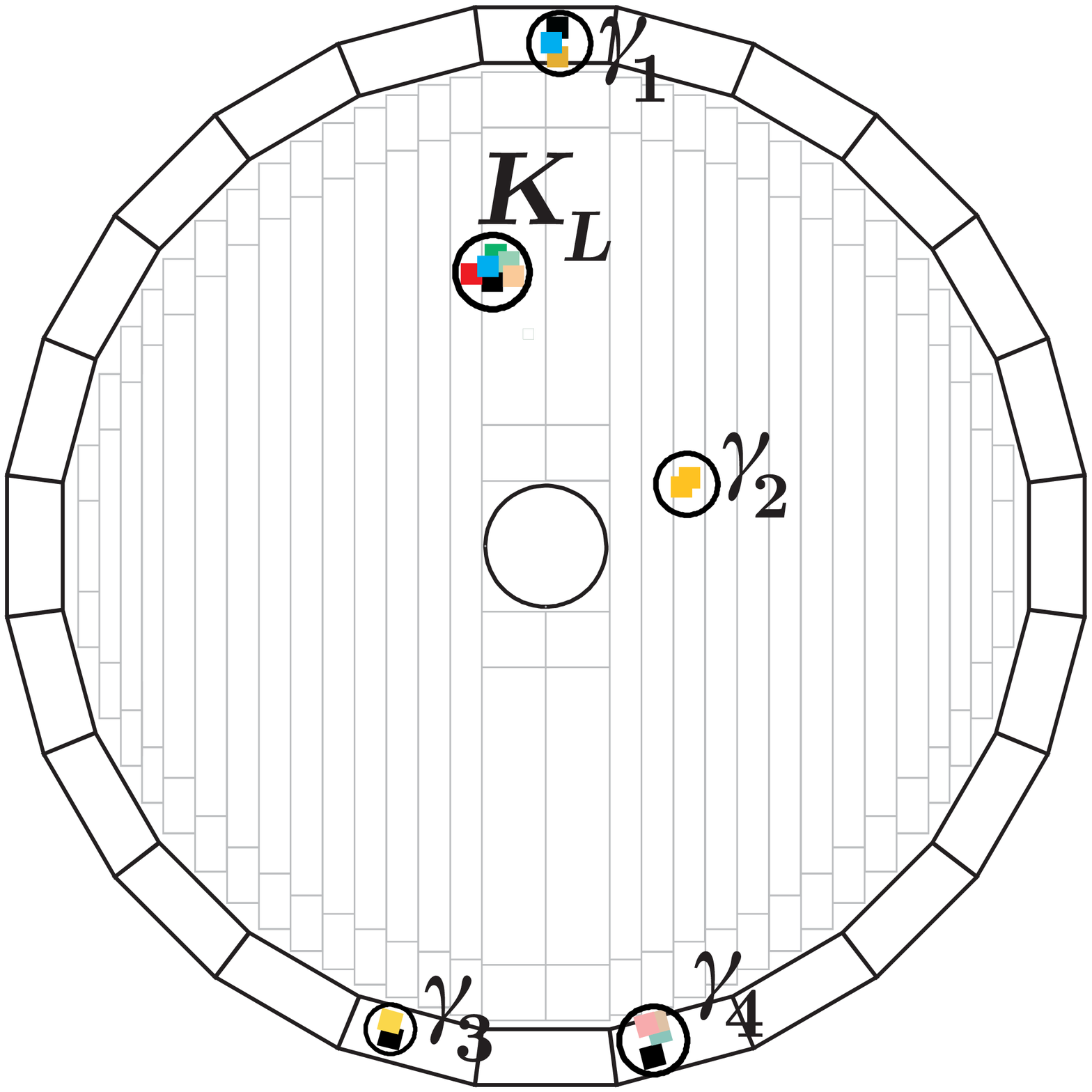}
\caption{\ks\kl\ events in \klo. Left: both \kl\ and \ks\ decay to charged pions \pip\pim. Color patches indicate energy in the calorimeter. Right: 5 neutral particles. \gam\dn1 trough \gam\dn4 are showers from 4 photons from \ks\to\pio\to4\gam. The \kl-mesons reaches the calorimeter before decaying. It literally crashes in the EMC, producing a huge energy release.}
\label{kkevent}
\end{center}
\end{figure}

At a luminosity of \ab100 $\mu$b\up{-1}/s, the trigger rate is \ab2000 Hz.
Of these \ab500~Hz are from \f\ decays or Bhabha scattering.
All digitized signals from the calorimeter, drift
chamber,
calibration and monitoring systems are fed via fiber optics
from the electronics, sitting on platforms mounted
on the \klo\ iron yoke,
to the computers in the next building.
The Data Acquisition architecture has been designed to sustain a throughput
of 50 Mbytes/s all along this path.
In the near-by control room one can see real time displays
of  reconstructed charged tracks traversing the DC and photons
depositing energy clusters in the EMC.
The display of two events is shown in \fig{kkevent}.

The events are reconstructed, meaning that charged tracks and energy deposits that occurred
at the same instant in time on the various subcomponents of \klo\  are associated together,
properly labelled and packed into a unique file. To optimize the event reconstruction we use continuously updated calibration
constants for the time and energy scales. Furthermore, background hits are acquired using a random trigger,
which are recorded in order to be used
in the simulation of \klo\ events (MC). The MC simulation program reproduces accurately
the detector geometry, the response of the apparatus and the trigger requirements , but of course
cannot foretell the moment to moment variation of the background.

All raw data, background, reconstructed and MC events are recorded on tape. \klo\ has a huge
tape library of \ab800~TB capacity \cite{DAQNIM}. At reconstruction time, events are classified
into broad categories, or ``streams''. Data summary tapes are produced with a separate procedure.

Any of the \klo\ analyses, even after event reconstruction is quite complex. It includes making
data summary tapes and extensive modelling of the detector response. This requires
generation
of large number of Monte Carlo (MC) events for each process under study.
``Offline'' processes run in a farm of computers of \ab\ 150 kSPECint2K CPU power, all residing in the dedicated computer center.
Both data and MC summaries
(\ab80~TB) are cached on disk for analysis \cite{offlineNIM}.
\subsection{Slow control system}
\label{sec:slow}
\klo\ has to be operated 24 hours a day, 7 days a week. Therefore it has to be kept under
continuous control, both to guarantee efficient data taking, and for safety reasons. Parameters such as high and
low voltage settings, the temperature of the electronics components, or the status of the drift chamber gas system are set and monitored
by a dedicated system fully embedded in the Data Acquisition
process: the Slow Control.

This system also records some machine parameters, and provides to the \DAF\ team
information about \klo\, including our on line measurements of the luminosity and of the
background levels.

\section{Reconstruction Performances}
\label{sec:recoperf}
KLOE  provides a continuous monitoring of the machine working point, providing feedback to \DAF\ continuously as well. The most
important parameters are the beam energies and crossing angle, which are
obtained from the analysis of Bhabha scattering events with $e^{\pm}$ polar
angles above 45 degrees. The average value of the center-of-mass energy is
evaluated online during data taking with a precision of $\sim 50$~keV for
each 200 nb$^{-1}$ of integrated luminosity. This determination is further
refined with offline analysis to achieve a precision of $\sim 20$~keV, as discussed later.
The position of the $e^+e^-$ primary vertex,
with coordinates $X_{PV}$, $Y_{PV}$, and $Z_{PV}$, is reconstructed
run-by-run from the same sample
of Bhabha events. $X_{PV}$ and $Y_{PV}$ are determined with typical accuracy
of about $10$~$\mu$m, and have widths $L(X)$ and $L(Y)$ which are about $1$~mm
and few tens of microns, respectively. $Z_{PV}$ is also reconstructed online
with $100-200$~$\mu$m accuracy, but it has a natural width of $12-14$~mm,
determined by the bunch length itself.
\subsection{Overview of EMC and DC performances}
\label{sec:det_sum}
We note that while KLOE is a general
purpose apparatus with  good performances
of the two main detectors, the DC and the EMC,
its real strength manifests when joining the reconstruction capability of both
detectors. To give an overall view at a glance,
we first summarize in table \ref{TAB:recon}
the parametrized resolution  for neutral particles
(mainly photons) and for charged tracks
(mainly electrons/pions) for prompt decays and $K_L$,$K^{\pm}$.
Resolution on invariant masses are also reported
in the same table as well as the reconstruction
capability for neutral and charged vertices.
\begin{table}
\begin{center}
\begin{tabular}{cc|cc}
\hline
 &  EMC &  &  DC \\
 & parametrization &  & parametrization \\
\hline
$\sigma_{\rm E}/{\rm E}$ & 5.7\%$/\sqrt{\rm E(GeV)}$ & $\sigma_{\rm p}/{\rm
 p}(\theta > 45^0)$ & $ \ab0.4$ \% \\
$\sigma (M)$ (MeV) & 15 ($\pi^0$), 40 ($\eta$,K) & $\sigma (M)$ (MeV) & 1
($\eta$,K) \\
$\sigma_{x,y}$ (cm)  &  \ab1.3  & $\sigma_{x,y}$ ($\mu$m) & 150 \\
$\sigma_{z}$ (cm)  & 1.2 $/\sqrt{\rm E(GeV)}$ & $\sigma_{z}$ (mm) & 2  \\
$\sigma_{t}$ (ps)  & 57 $/\sqrt{\rm E(GeV)}$ + 140  & $<\frac{ \rm dE}{\rm dx}>$
$\frac{\rm counts}{cm}$ & \ab120 (K$^{\pm}$), \ab15 ($\mu,\pi$
 from $K^{\pm}_{\mu2,\pi2}$) \\
$\sigma_{R_v}$ (cm)  & 2-3  & $\sigma_{Xv,Yv}$ (mm)
 & \ab2 (@IP), \ab3 (@L \ab1m) \\
\hline
\end{tabular}
\caption{Summary table of the EMC and DC performances.}
\label{TAB:recon}
\end{center}
\end{table}

In KLOE, the single particle reconstruction efficiency can be determined from
data control samples, as well as from Monte Carlo simulation. As an example,
the photon reconstruction efficiency as a function of the photon energy
can be estimated from
 $\phi \to \pi^+ \pi^- \pi^0$ events, where a redundant determination of the decay photon kinematics
is possible using the charged pion momenta as measured from the DC.
For charged particles produced at the interaction point, the track efficiency can be measured
from $\ks \to \pi^+ \pi^-$ events selected with the requirement of having at least one track
fulfilling the two-body decay kinematics.
A high reconstruction efficiency is observed
for both detectors on data and Monte Carlo.
For each specific analysis, the event selection efficiency is evaluated from Monte Carlo. To
take into account data-MC difference in the cluster/track efficiencies, the calculation
is usually performed by weighting each particle in the final state with the ratio of
data/MC reconstruction efficiencies.

For particle identification, PID,  we start extrapolating tracks on
the calorimeter surface and associating tracks to clusters.
The identification of photons is based on clusters
not associated to any track and on the
difference between expected and reconstructed time of flight, ToF,
for straight trajectories.  The PID of charged tracks,
which at the considered energy are electrons, pions and muons, is based on a combined usage of EMC and DC,
by looking at ToF, $E/p$ and cluster shape.
At low momenta, $P<200$ MeV, the ToF is usually the winning
handle while at higher momenta variables based on cluster shape
are used. Often these information  are combined together in a likelihood
identification variable \cite{sighad} or using a
neural network \cite{testamoriond08}. To
identify charged kaons we can use instead dE/dx,
given the excellent separation provided by the specific ionization
measurement between kaons and the other charged particles produced in the apparatus.

%

\subsection{Absolute energy scale}
\label{sec:escale}
 An improved determination of the center-of-mass energy, $W$, is obtained for
each run by fitting  the $e^+ e^-$ invariant-mass distribution
for Bhabha events to a Monte Carlo generated function, which includes radiative effects.
Initial state radiation (ISR),
where one or both initial colliding
particles radiate a photon before interacting, affects the $e^+ e^-$
center-of-mass energy and therefore the final state invariant mass.
The ISR is mostly collinear to the beam, and in general is not detected.
MC Bhabha events were generated using the
 \texttt{Babayaga}~\cite{Carloni Calame:2000pz} event generator,
which accounts for both  final and  initial state radiation.

The absolute energy scale is calibrated by measuring the visible cross section for the $\phi \to \ks \kl$ process.
The cross section peak is fitted to a theoretical function~\cite{CMD2par},
which depends on the $\phi$ parameters, takes into account the
effect of ISR, and includes the interference with the $\rho(770)$ and
the $\omega(782)$ mesons.
The $\phi$ mass, total width, and peak cross section are the
only free parameters of the fit, the $\rho(770)$ and  the $\omega(782)$
parameters being fixed.
The results of the fit to the data are shown in \fig{lineshape}. The
$\phi$ mass value obtained from the fit is $m_{\phi} = 1019.329 \pm 0.011$ MeV,
which has to be compared with
$ m_\phi= 1019.483 \pm 0.011 \pm 0.025 $ MeV, measured by CMD-2
at VEPP-2M ~\cite{CMD2}.  For CMD-2 the absolute energy scale is accurately determined
from beam induced resonant depolarization, so that we use
the ratio $m_{\phi}^{CMD}$/$m_{\phi}^{KLOE} = 1.00015 $ to correct our determination of
the center of mass energy. This corresponds to a shift in the value of $W$ of \ab150 keV.

\def \phidec {{\phi \to   K_S K_L }}

\begin{figure}[htb]
      \cl{\figb lineshape;6;}
      \cl{\figb residual;6;}
      \caption {Top: Cross section for $e^+ e^- \to \phidec$ as a function
              of the center-of-mass energy (line, fit). Bottom:  Fit residuals.}
      \label{lineshape}
\end{figure}

\subsection{Luminosity}
\label{sec:Lumi}
\begin{figure}[htb]
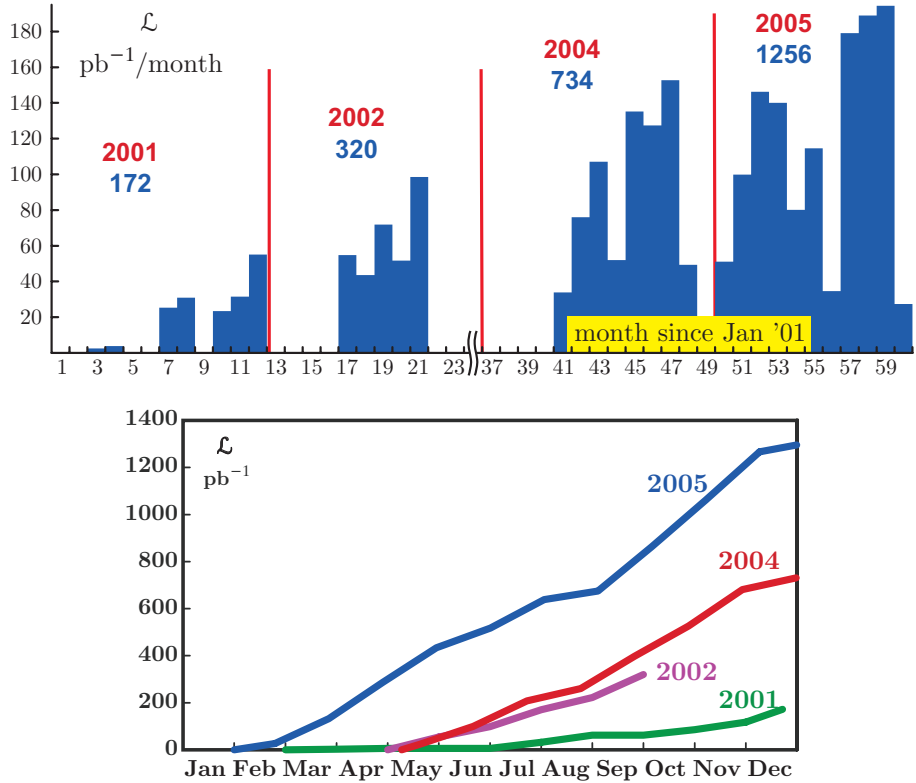

\centering
\centering\figb lumibymonth;12;\\[.4cm]
\figb lumibyyeara;9;
\caption{Integrated luminosity since January 2001: by month (top) and by year (bottom).}
\label{fig:lumi}
\end{figure}

The tagging capability of the calorimeter trigger allowed us to perform fast
luminosity measurements, providing a robust estimate of the \DAF\ luminosity with few per cent
accuracy. This was achieved by counting Bhabha events in the acceptance of the
calorimeter barrel. For this purpose, we selected events with two trigger sectors fired on the barrel, imposing a higher discrimination threshold than
the one used in the standard acquisition chain. The tight angular selection, which reduces the available statistics by a factor of $\sim 6$, was motivated
by the need to keep under control the overwhelming machine background, which is
concentrated at small angle.
Additional cuts on the position of the fired sectors and on the time difference
between the two signals were added to improve rejection on the residual background,
from cosmic-ray events. When operating at \L=100 $\mu$b\up{-1}/s, the luminosity
measurement was updated every 15~s, with a statistical error of $\sim$ 3\%.
\Fig{fig:lumi} top shows the monthly
integrated luminosity by KLOE since January 2001, while in the bottom the year
integrated luminosity is shown. These data have been collected
almost exclusively at the $\phi$ peak. In early 2006, KLOE also
collected \ab200 pb\up{-1} of data at $\sqrt s = 1000$~MeV, at
which $\sigma(e^+ e^-\to\f\to\pic\po)$ drops to 5\% of its peak value.
%
Finally, it is important to note that already in year 2000 KLOE collected 20 pb$^{-1}$,
producing quite a few publications which improved the knowledge in both
hadronic and kaon physics \cite{ks2piold,ksleptold,f0,a0,etapr}.

A more accurate measurement of the integrated luminosity is performed
offline~\cite{Ambrosino:2006te}, using
Bhabha events in the polar angle range $55^\circ<\theta<125^\circ$, the so-called VLAB
events. The effective cross section for these events, $\sim 430$~nb, is large enough
to reduce the statistical error at a negligible level.
The luminosity is obtained by counting the number of VLAB candidates, $N_{\rm VLAB}$, and
normalizing it to the effective Bhabha cross section, $\sigma_{\rm VLAB}^{\rm MC}$,
obtained from MC simulation, after subtraction of the background, $\delta_{\rm bkg}$:
\begin{equation}
\int \L dt = \frac{N_{\rm VLAB}}{\sigma_{\rm VLAB}^{\rm MC}}(1 - \delta_{\rm bkg}).
\label{eq:lumi}
\end{equation}
The precision of the measurement depends on the correct inclusion of higher-order
terms in computing the Bhabha cross section. For this purpose we use  \texttt{Babayaga},
which include
the QED radiative corrections in the framework of the parton-shower method.
The quoted precision is 0.5\%, although
 an updated version of this generator,
\texttt{Babayaga@NLO}~\cite{Balossini:2006wc}, has been released recently.
The new predicted cross section
decreased by 0.7\% and the theoretical systematic uncertainty
improved from 0.5\% to 0.1\%.

The VLAB events are selected with requirements on variables that are well reproduced by
the KLOE MC. The acceptance cut on the electron and positron polar angle,
$55^{\circ} < \theta_{+,-} < 125^{\circ}$ is based on the calorimeter clusters, while
the momentum selection, $p_{+,-} > 400$~MeV, is based on the DC determination. The
background from $\mu^+\mu^-(\gamma)$, $\pi^+\pi^-(\gamma)$, and $\pi^+\pi^-\pi^0$ events
is well below $1\%$ and is subtracted. All selection efficiencies (trigger, cluster and
track reconstruction) are greater than $99\%$, as obtained from MC and confirmed with data
control samples. Finally, corrections are applied on a run-by-run basis to follow the
small variations in the center-of-mass energy and in the detector calibration.
In the end, we quote a 0.3\% uncertainty in the determination of the acceptance due
to experimental effects.

\section{Kaon beams}
\label{sec:kb}
The neutral kaon pair from $\f\to\ko\kob$ is in a pure $J^{PC}=1^{--}$ state. Therefore the initial two-kaon state can be written, in the \f-rest frame, as
\begin{equation}\eqalign{
\sta K\Kb, t=0; &= \left( \sta\ko,\pb;\sta \kob,-\pb;-\sta\kob,\pb;\sta\ko,-\pb;\right) /\sqrt2  \cr &=
\left(\sta\ks,\pb;\sta \kl,-\pb;-\sta\kl,\pb;\sta\ks,-\pb;\right) / \sqrt2 ,\cr }
\label{eq:state}
\end{equation}
where the identity holds even without assuming $CPT$ invariance. Detection of a \ks\ thus signals the presence of,  ``tags'', a \kl\
and vice versa. Thus at \DAF\ we have pure \ks\ and \kl\ beams of precisely known momenta (event by event) and flux, which can be used
to measure absolute $K_S$ and $K_L$ branching ratios. In particular \DAF\ produces the only true pure \ks\ beam and the only \kl\
beam of known momentum. A \ks\ beam permits studies of suppressed \ks\ decays without overwhelming background from the \kl\ component.
 A \kl\ beam allows lifetime measurements. Similar arguments hold for $K^+$ and $K^-$ as well, although it is not hard to produce pure,
monochromatic charged kaon beams.
In the following sections we briefly discuss the criteria adopted to tag the \ks, \kl, and charged kaon beams.
\subsection{\ks\ beam}
\label{sec:ksbeam}
\begin{figure}[htb]
\begin{center}
\includegraphics[width=0.5\textwidth]{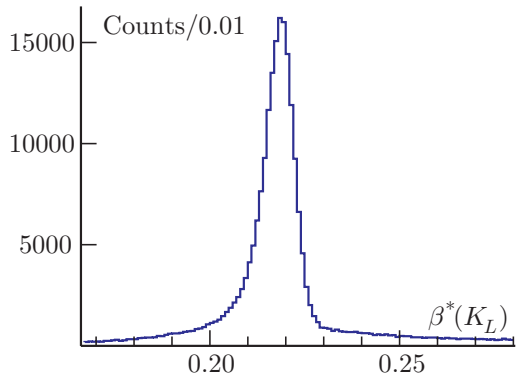}
\caption{\kl\ velocity in \f-frame from time of flight.}
\label{bkl}
\end{center}
\end{figure}

Because of the exceptional timing capabilities of the EMC and the slowness of the kaons
from \f-decays, we use a ToF technique to tag \ks-mesons in a unique way.
As already mentioned, neutral kaons have a velocity $\beta=0.2162$ in the \f\ frame and $\beta$=0.193 to 0.239
in the lab, and 60\% of them reach the calorimeter in a time $\geq$31 ns.
\kl-mesons interact in the calorimeter with an energy release up to their mass, 497 MeV.
For an energy release of 100 MeV we have a time resolution of \ab0.3 ns,
\ie\ about a \ab1\% accuracy on \kl\ velocity, shown in \fig{bkl}.

In \klo, we call a \kl\ interaction in the EMC a ``K-crash'', and we use it as a tag for the \ks.
Note that the above error on $\beta$(\kl) corresponds to an error on the kaon
energy of \ab0.25 MeV or \ab1 MeV for its momentum, with just one event.
Adding the information about the position of the energy release, the direction of the \kl\ line of flight is determined with
$\ab 1^\circ$ angular accuracy.
Using the previous value of \kl\ momentum, and the run-average value of the $\phi$ momentum as evaluated from Bhabha events,
we determine the \ks\ momentum from $p_S = p_{\phi} - p_L$, with the same accuracy as for \kl.
\subsection{\kl\ beam}
\label{sec:klbeam}
All of the produced \ks\ mesons decay within few centimeters from the interaction point.
Therefore, in principle, all of them can be detected in the apparatus. On the
contrary, \kl 's can travel several meters before decaying, thus escaping detection.
From the above, it follows that a pure sample of nearly monochromatic \kl's can be selected by
identification of \ks\ decays. The best performances in terms of efficiency and momentum resolution
are obtained by selecting $\ks \rightarrow \pi^+\pi^-$ decays. These are identified as
two-track vertices close to the $e^+ e^-$ interaction point, with
invariant mass in a 5 MeV acceptance window around the nominal kaon mass.
The $\ks \rightarrow \pi^+\pi^-$ decay provides also a good measurement of \kl\ momentum,
$p_L = p_{\phi} - p_S$, where $p_{\phi}$ is again the $\phi$ momentum as determined from Bhabha
scattering events. The resolution is \ab 0.8 MeV on the momentum, while the \kl\ flight direction
is determined within $\ab 2^\circ$.

More interestingly, events with both a $\ks \to \pi^+\pi^-$ decay and a
K-crash identified allow us
to measure the \DAF\ beam energy spread. This is extracted by comparing for each event
the two independent determinations of center-of-mass energy that are evaluated using the \ks\ and \kl\
momenta, respectively. Using this technique we find a beam energy spread of \ab 220 keV, in agreement
with the machine model expectation.

\begin{figure}[htb]
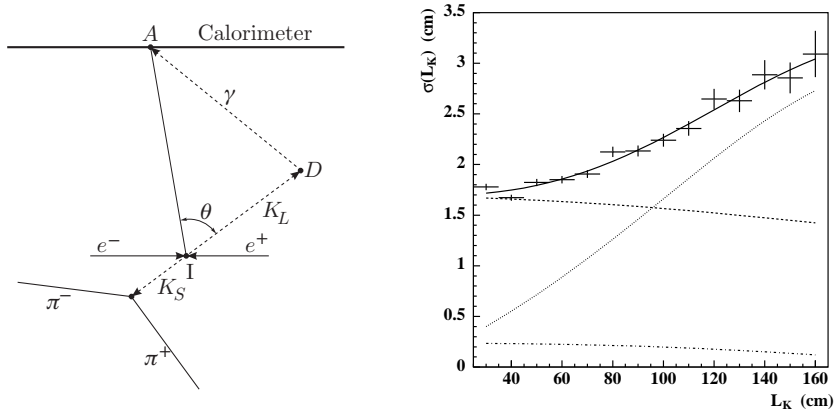

\begin{center}
\centering\figb kl-path;4.5;\kern1cm\figb sigma_kl;5.5;
\caption{Left: $D$ is the \kl\to\pio\to4\gam\ decay point. The \kl\ path is obtained from the time for
each photon to arrive at the calorimeter, $t$(I \to A), and the \ks\
direction.
Right: Resolution $\sigma(L_{\rm K})$  on the determination of the \kl\ decay length using photon
 vertices, as a function of $L_K$, in $\kl \to \pi^+\pi^-\pi^0$ events. The contributions from the
uncertainties on the point of photon incidence on the calorimeter, the cluster time, and the \kl\
flight direction are shown as the dot-dashed, dashed, and dotted lines, respectively.}
\label{klp}
\end{center}
\end{figure}

For many \klo\ measurements, it is of fundamental importance to accurately
determine the exact amount of lost/retained decays, or, in other words,
determining the \kl\ decay fiducial volume.
For decays involving two charged particles this is obtained relatively easily
by the observation of the vertex of the two related tracks in the drift chamber,
measured with an accuracy of \ab\ 3 mm.
The task is more difficult for totally neutral decay channels, but can still be achieved
using the excellent timing accuracy of the EMC.
As an example, we can locate the \kl\to\pio\ decay point, the ``neutral vertex'', with  accuracies of  \ord2; cm,
as illustrated in \fig{klp}, left.
Each photon from the \kl\ decay point (D) defines a time-of-flight triangle: the first side is
the segment from the interaction point to the \kl\ decay vertex (ID); the second is the segment
from the \kl\ decay
vertex to the centroid of the calorimeter cluster (DA); and the third is the segment from the
interaction point to
the cluster centroid (IA). The \kl\ direction is initially known, because the \kl\ decay is tagged.
The photon vertex position is specified by the distance ID, which is determined from
\begin{eqnarray}
& & IA^2 + ID^2 -2 {\bf IA}\cdot {\bf ID} = DA^2 \nonumber \\
& & ID/\beta_K  + DA = ct_{\gamma},
  \label{eq:neuvtx}
\end{eqnarray}
where $t_{\gamma}$ is the cluster time and $\beta_K$ is the \kl\ velocity. The position of the
interaction point is obtained by backward extrapolation along the \ks\ flight path.
The \kl\ decay vertex position is evaluated for each neutral cluster separately, and
the energy-weighted average of the values for each cluster is then taken as its final determination.
The accuracy in the location of the photon vertex has been studied using $\kl \to \pi^+\pi^-\pi^0$
decays, in which the decay position can be independently determined using clusters and tracks,
with much greater precision in the latter case. The dependence of the position resolution on
decay distance ($L_K$) is illustrated in \fig{klp}, right.
\subsection{$K^{\pm}$ beam}
\label{sec:kpmbeam}
\begin{figure}[htb]
\centering
\includegraphics[width=0.5\textwidth]{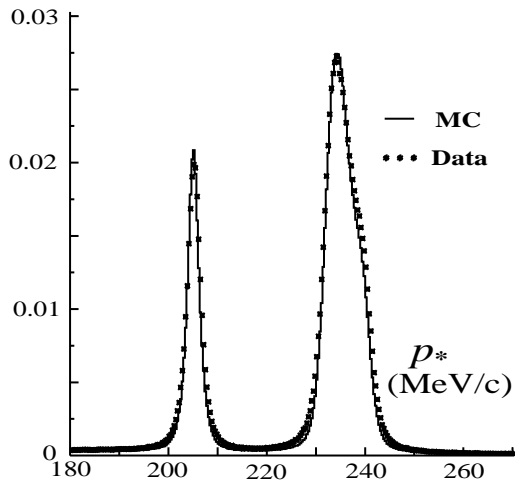}
\caption{ Momentum spectrum in the kaon rest frame of the negative charged
decay particle, assuming the particle has the pion mass for data
(dots) and MC (lines). The distribution is normalized to
unity. The two peaks correspond to pions and muons
from $K^- \rightarrow \pi^- \pi^0$ (205 MeV/c) and $K^-\to\mu^-\nu_\mu$ (236 MeV/c). The muon peak is broadened by the use of the incorrect mass.}
\label{fig:tag_spectrum}
\end{figure}

%
%
The $\phi$-meson decays 50$\%$ of the times into $K^{+}K^{-}$ pairs;
these are quasi anti-collinear in the laboratory, due to the small crossing
angle of the e$^+$e$^-$ beams, with momentum ranging between  120 and
133 MeV/c and decay length $\lambda_{\pm}\ab 95$ cm.
While moving towards the DC volume,
the average $K^{\pm}$ momentum decreases to \ab100 MeV/c due to the energy loss
in the beam pipe and DC inner wall materials.
As for neutral kaons, the identification of a $K^{\mp}$ decay tags a $K^{\pm}$
beam with known flux, thus giving the possibility to measure kaon
absolute branching ratios.
Charged kaons are tagged using their two-body decays,
$K^{\pm}\rightarrow \mu^\pm\rlap{\raise1.2ex\hbox{\scriptsize($-$)}}
\kern.3em\nu_{\,\mu}$ and $K^{\pm}\rightarrow \pi^{\pm} \pi^0$, accounting for
$\sim$85\% of the total decay channels.
These decays are observed as ``kinks'' in a track originating at the IP.
The charged kaon track must satisfy $70 <p_{\rm K^{\pm}}< 130$~MeV/c.
The momentum of the decay particle in the kaon rest frame evaluated using
the pion mass, $p^{\ast}$, is 205 and 236~MeV, for $K^{\pm}$ decays to $\pi^{\pm}\pi^0$
and $\mu^{\pm}\nu$, respectively. \Fig{fig:tag_spectrum} shows the $p^{\ast}$ distribution obtained assuming the decay particle
is a pion.
Cutting on the spectrum of the decay particle, we select about  $1.5 \times 10^6 K^+K^-$ events/pb$^{-1}$ of integrated luminosity.

\section{Kaon physical parameters and decays}
\label{sec:KPPD}
Since the kaon's discovery about sixty years ago, knowledge of its properties, from masses, lifetimes, and how
often each species decays into to which particles (the so called branching ratios, BR),
 has been accumulated piecemeal literally over hundreds of
individual experiments, performed at various laboratories all over the world. Each experiment has its own
peculiarity, contingent upon the particle beam available and its own equipment's acceptance range, which can result
in highly accurate measurements from a statistical point of view (namely based on thousands of data events) but with
 hidden corrections peculiar to it perhaps unknown even to its authors. The particle data group~\cite{pdg06}
 in charge of compiling the results, often adopt a ``democratic'' way of weighing the
various results according to their nominal errors and literally applying a scale factor to these errors when
faced with two inconsistent measurements. This procedure skirts the problems of judging the validity of  results
 obtained over time and sometimes ends up with a hodge-podge of inconsistent results forced into a mold of dubious
 scientific validity.

 The mission of \klo\ is to measure most, if not all, of the properties of the kaon system to high accuracy,
with a single detector simultaneously.
One problem that consistently plagues the interpretation of older measurements is the lack of clarity
 about accounting for radiative contributions.
All of our measurements of kaon decays with charged particles in the final state are fully inclusive of
radiation. Radiation is automatically accounted for in the acceptance correction. All our MC generators
 incorporate radiation as described in \cite{Gat06:rad}.

Because of the availability of tagged and pure kaon beams of known momenta, \klo\ is the only experiment
 that can at once measure the complete set of experimental inputs, branching ratios, lifetimes and form factor
parameters for both charged kaons and long lived neutral kaons. In addition \klo\ is the only
experiment that can measure \ks\ branching ratios at the sub-percent level.

\subsection{\ko\ Mass}
\label{sec:k0mass}
 Kaon masses are about 500 MeV, very close to one half the \f-meson mass, known to an accuracy of
 0.019/1019.460 \ab\pt2,-5, by the use of the $g$-2 depolarizing resonances, \cite{pdg06}.
The events $\phi \to \ks \kl$ offer a unique possibility to obtain a
precise value of the neutral kaon mass. Indeed we observe that if the
$\phi$-meson is at rest the kaon mass can be extracted from the kaon momentum
using the relation:
\begin{equation}
m_{K} = \sqrt{\frac{m_{\phi}^2}{4} - p_{K}^2};  \qquad \frac{\Delta m_{K}}{m_{K}} \simeq \frac{p_K^2}{m_{K}^2}
\frac{\Delta p_K}{p_K} \sim \beta^2 \frac{\Delta p_K}{p_K}
 \label{eq:kmasssimple}
\end{equation}
Since $p_K \simeq 110$ MeV, measuring it at  1\% level, well within the KLOE capability, results in a measurement of
the $K^0$ mass better than $0.1$\%. 50,000 events are enough to reach a statistical accuracy of about 1 keV.

$\phi$ mesons are produced with a momentum along the $x$ axis,
$p_{\phi} = 12.5$ MeV at \DAF.
From the measured momenta of the two pions from $K_S \to \pi^+ \pi^-$, we  measure the $K_S$ momentum. The $K_L$ momentum is given by
$\vec{p}_{K_L} = \vec{p}_{\phi} - \vec{p}_{K_S}$, where $\vec{p}_{\phi}$ is the average $\phi$ momentum
 measured with Bhabha events collected in the same runs.
The center of mass energy of the $K_S K_L$   pair ($W_{KK}$) is related to the kaon mass $m_K$, according to:
\[
W_{KK}(m_K) =\sqrt{2M^2_K+2E_{K_S}E_{K_L}-2\vec{p}_{K_S}\cdot\vec{p}_{K_L}}
\]
with
\[
E_{K_S} = \sqrt{p_{K_S}^2+m^2_K} \qquad \qquad E_{K_L} = \sqrt{p^2_{K_L}+m^2_K}.
\]
On the other hand the  collision center of mass energy $W$ is computed from Bhabha events, as described in
sect.~\ref{sec:recoperf}.

 Corrections due to ISR have to be taken into account
 when relating  $W$ to   $W_{KK}$.
 The correction function $f_K(W)$
 has been evaluated using a full detector simulation where
 the radiation from both beams has been implemented, and $W_{KK}$
 is reconstructed as in the data.
 The expression of the radiator function
 has been taken from \cite{nicrosini}, including
 $\mathcal{O}(\alpha^2)$ corrections.
 The correction $|1-f_K(W)|$ is very small below the resonance, corresponding to a shift
in $W_{KK}$ of 40~keV.  For $W$ above the $\phi$ mass, $W_{KK}$ increases up to 100~keV.
The neutral kaon mass is then obtained solving the equation:
 \begin{equation}
  W = f_K(W) \times W_{KK}(m_K)
\label{eq:kaonmass}
 \end{equation}
 The single event mass resolution is  about 430~keV. Contributions to the mass resolution are: about 370~keV from experimental resolution,
about 220~keV from beam energy spread, as measured by KLOE
 in agreement with machine theory, and about 100~keV from ISR.

The systematic error due to the momentum miscalibration has been evaluated by
changing the momentum scale in computing the pion momenta.
A momentum miscalibration, $\delta p/p$ , translates to a miscalibration on
the mass $\delta m_{K} /m_{K} = 0.06 \delta p/p$,
in agreement with the above qualitative calculation.
The momentum scale is obtained by using several processes covering a wide momentum
 range, from 50 to 500 MeV ($K_L \to \pi^+ \pi^- \pi^0, ~K_L \to \pi \ell \nu, ~\phi \to \pi^+ \pi^- \pi^0$).
We obtain a fractional
accuracy below $2 \times 10^{-4}$, in agreement  with the estimate obtained
 using Bhabha's \cite{offlineNIM}, which results in a systematic error
$\delta m_K$ of 6~keV.

The systematic error coming from theoretical uncertainty on the radiator function
 has been evaluated considering the contribution  from higher order terms in $\alpha$.
The correction function $f_K (W )$ has been evaluated by excluding the constant
term in the $\mathcal{O}(\alpha^2)$.
The corresponding change in $f_K(W)$ is $1.3 \times 10^{-5}$, corresponding to a
variation on $m_K$ of 7~keV.
Further checks have been made by using the  function given in \cite{Kuraev}:
no significant differences were observed.
Additional systematics come from the dependence of the measured mass from
the value of $W$. They have been evaluated by
comparing  the mass values obtained with data collected at $W<1020$ MeV
and $W>1021$ MeV,
 where the value of $f_K(W)$ is more than a factor two larger.
 The difference between the two mass values is
$m_K(W<1020)-m_K(W>1021)=9\pm10$ keV, consistent with zero.

Other sources of systematics are due to the uncertainties on the W calibration, i.e.,
the statistic and systematic error on
$m_{\phi}^{CMD-2}$ and on $m_{\phi}$ obtained from our fit, discussed in
 sect.~\ref{sec:escale}. The
total contribution from these sources amounts to a mass uncertainty of 15~keV.
Systematic uncertainties are treated as uncorrelated. The result is \cite{ketamass}:
\begin{equation} \label{eq:EQK0}
m_K = 497.583 \pm 0.005_{\mathrm{stat}} \pm 0.020_{\mathrm{syst}} \quad \mathrm{MeV}.
\end{equation}

\subsection{\kl\ lifetime}
\label{sec:KLlife}
Precision measurements of the lifetime of neutral kaons, especially in presence
of many multibody decay modes, is particularly difficult since it is in
general not possible to prepare monochromatic beams of neutral particles
nor to stop them. \klo\ enjoys the availability of such monochromatic
beams. While the \ks\ lifetime is very well known, \pt(0.8958 \pm0.0005),-10,
s~\cite{pdg06}, this is not the case for the \kl\ lifetime, $\tau_L$.

A total of approximately 13 million tagged \kl\
decays are used for the measurement of the four major \kl\ BRs,
as discussed in sect.~\ref{sec:brl}.
Since the geometrical efficiency for the detection of \kl\ decays depends on
$\tau_L$, so do the values of the four BRs:
\begin{equation}
{\rm BR}(\kl\to f)/{\rm BR}^{(0)}(\kl\to f) = 1 + 0.0128~\mbox{ns}^{-1}\:\left(\tau_L -
\tau_L^{(0)}\right),
\label{eq:fvdep}
\end{equation}
where ${\rm BR^{(0)}}$ is the value of the branching ratio evaluated
for $\tau_L^{(0)}$, an assumed value for $\tau_L$.
The four relations defined by \eq{eq:fvdep}, together with the condition
that the sum of all \kl\ BRs must equal unity, allow
determination of the \kl\ lifetime and the four BR values
independent of $\tau_L$. This approach, followed in \cite{klbr}, gives
 $\tau_L = 50.72 \pm 0.11_{\rm stat} \pm 0.13_{\rm syst-stat} \pm
0.33_{\rm syst}$ ns.

\begin{figure}
\centering
\parbox{8cm}{\centerline{\epsfig{file=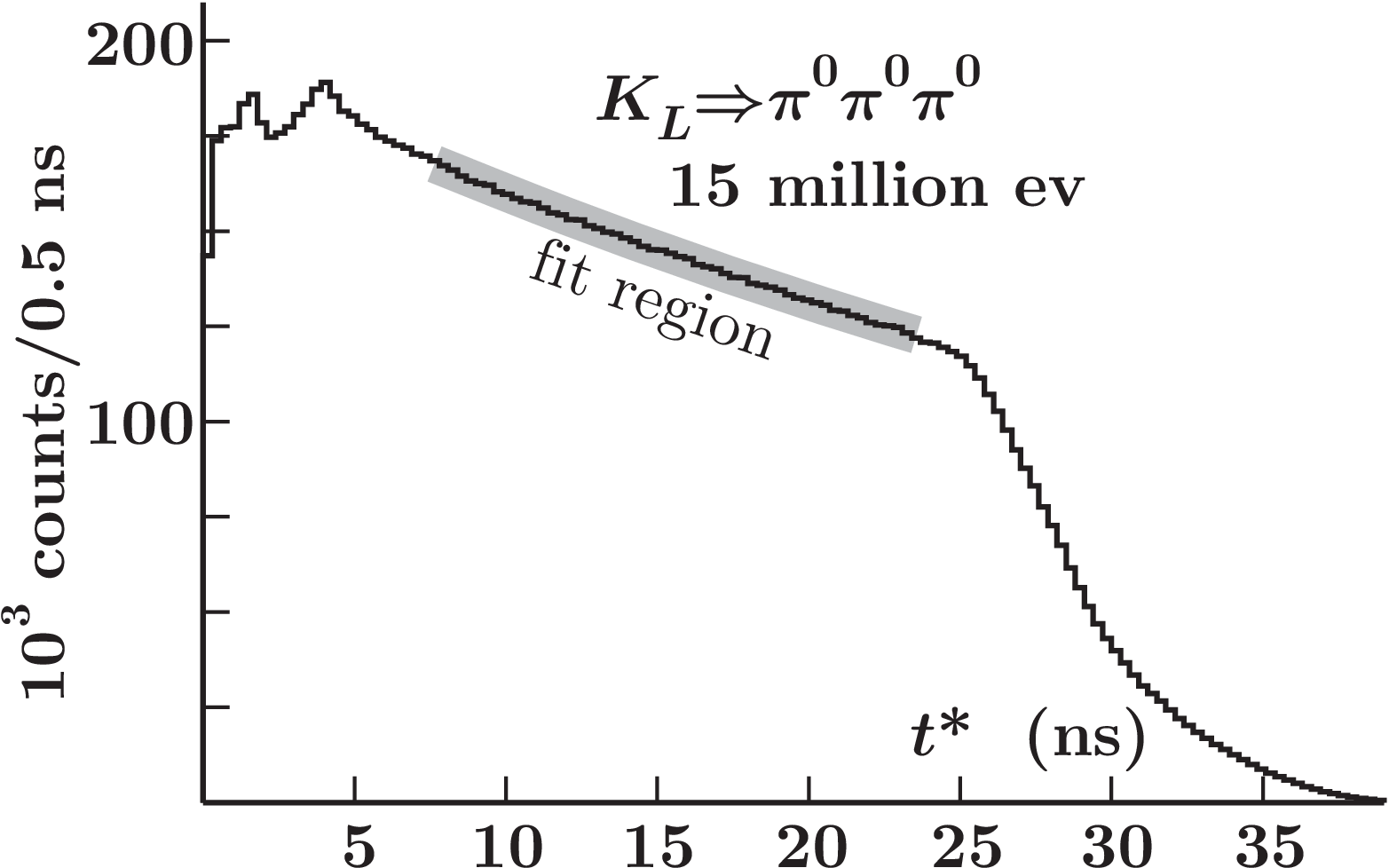,width=6cm}}}
\caption{Proper-time distribution for $\kl\to 3\po$ decays.}
\label{fig:tkl}
\end{figure}

In addition, the lifetime can be measured by observing the time dependence
of the decay frequency to a single mode. Because of tagging, the
lifetime can also be obtained by simply counting the total number of
decays in a given time interval from a known \kl sample.
This second, independent input comes from our separate analysis of the
proper decay-time distribution for
$\kl\to 3\pi^0$ events, for which the reconstruction efficiency
is high and uniform over a fiducial volume of
$\sim$$0.4\,\lambda_L$. The position of the \kl\ decay position is determined
from the photon arrival times, with the technique described in
sect.~\ref{sec:klbeam}.
The \kl\ proper decay-time, $t^{\ast}$, is obtained
from the decay length, $l_K$, divided by the Lorentz factor $\beta \gamma c = p_{\kl}/m_K$.
The \kl\ momentum is evaluated event-by-event using the information from the
$\ks \to \pic$ tagging decay.
About 8.5 million decays are observed within the proper-time interval
from 6.0 to 24.8~ns. We fit the distribution in this interval
 (\fig{fig:tkl}), and obtain
$\tau_L=50.92\pm0.17_{\rm stat}\pm0.25_{\rm syst}$~ns ~\cite{kllife}.
The systematic error of 0.49\% is presently dominated by the uncertainty on the
dependence of the tagging efficiency on $l_K$, and by background subtraction.

This latter measurement is included together with the results
for the four main \kl\ BRs and the \kl\ lifetime in a fit to determine
the \kl\ BRs and lifetime.
We also use our measurements of $\BR(\kl \to \pic)/\BR(K_L\mu3)$
and $\BR(\kl \to \gam\gam)/\BR(\kl \to 3\po)$, reported in sections
\ref{sec:klpp} and  \ref{sec:klggover3p0}, requiring that the seven
largest \kl\ BRs add to unity. The only non-\klo\ input to the fit is the
2006 PDG ETAFIT result $\BR(\kl \to \po\po)/\BR(\kl \to \pic) =
0.4391 \pm 0.0013$, based on relative amplitude measurements for
$K \to \pi\pi$.
The results of the fit are presented in table \ref{tab:klbr2}; the fit gives
$\chi^2/{\rm ndf} = 0.19/1$ ($CL=66\%$).
The BRs for the $K_{Le3}$ and $K_{L\mu3}$ decays are determined to within
0.4\% and 0.5\%, respectively.
\begin{table}[hbt]
\begin{footnotesize}
\begin{center}
\begin{tabular}{lcccccccc}
\hline
Parameter & Value & \multicolumn{5}{c}{Correlation coefficients} \\
\hline
\BRo K_{e3};          & 0.4008(15)                &         &         &         &         &         &         &          \\
\BRo K_{\mu3};        & 0.2699(14)                & $-0.31$ &         &         &         &         &         &          \\
\BRo 3\pi^0;          & 0.1996(20)                & $-0.55$ & $-0.41$ &         &         &         &         &          \\
\BRo \pi^+\pi^-\pi^0; & 0.1261(11)                & $-0.01$ & $-0.14$ & $-0.47$ &         &         &         &          \\
\BRo \pi^+\pi^-;      & $1.964(21)\times 10^{-3}$ & $-0.15$ & $+0.50$ & $-0.21$ & $-0.07$ &         &         &          \\
\BRo \pi^0\pi^0;      & $8.49(9)\times 10^{-4}$   & $-0.15$ & $+0.48$ & $-0.20$ & $-0.07$ & $+0.97$ &         &          \\
\BRo \gam\gam;        & $5.57(8)\times 10^{-4}$   & $-0.37$ & $-0.28$ & $+0.68$ & $-0.32$ & $-0.14$ & $-0.13$ &          \\
$\tau_L$              & 50.84(23)~ns              & $+0.16$ & $+0.22$ & $-0.14$ & $-0.26$ & $+0.11$ & $+0.11$ & $-0.09$  \\
\hline
\end{tabular}
\end{center}
\end{footnotesize}
\caption{Final KLOE measurements of main \kl\ BRs and $\tau_L$.}
\label{tab:klbr2}
\end{table}
%
\subsection{\kpm\ lifetime}
\label{sec:kpmlife}
The measurements of the \kpm\ lifetime listed in the PDG compilation, \cite{pdg06}, exhibit poor consistency.
The PDG fit has a confidence level of \pt1.5,-3,
and the error on the recommended value is enlarged by a scale factor of 2.1.

\noindent
KLOE has measured the decay time for charged kaons in two ways both using events
tagged by $K^{\pm}_{\mu 2}$ decays \cite{kpmlife}.
The first method was to obtain the proper time from the kaon path length in the DC.
The second method was based on the precise measurement of the kaon decay time from the arrival times of the photons
from kaon decays with a $\pi^0$ in the final state.
These two methods reach comparable accuracy and allow us to cross-check systematics.
%
%
The method relying on the measurement of the charged kaon decay length
requires the reconstruction of the kaon decay vertex
using DC information only. The signal is given by a
$K^\pm$, moving outwards from the IP in the DC with momentum
$70 < p_{K^\pm} < 130$ MeV/c. The kaon decay vertex (V) has to be  in the DC  volume
defined by $40 < \sqrt{x^2_V +y^2_V} < 150$ cm, $|z_V| < 150$ cm.
Once the decay vertex has been identified, the kaon track is extrapolated
backwards to the IP into 2 mm steps, taking into account the
ionization energy loss $dE/dx$ to evaluate its velocity $\beta c$.
Then the kaon proper time ($t^\ast$) is obtained from:
$t^\ast = \sum_i \Delta t_i =
\sum_i \frac{\sqrt{1-\beta^2_i}}{\beta_i} \Delta l_i $.
The reconstruction and selection efficiency of the decay vertex
has been evaluated directly from data. The control sample is
selected using calorimetric information only: neutral vertices are identified
looking for two clusters in time fired by the photons coming from the $\pi^0\to\gamma\gamma$ decay.
The proper time distribution is fit between 16 and 30 ns
 correcting for the measured efficiency.
Resolution effects are taken into account in the convolution with the exponential decay function
used to fit the $t^\ast$ distribution.
The result we have obtained, which is the weighted mean between
 the $K^+$ and the $K^-$ lifetimes, is $\tau^\pm = (12.364 \pm 0.031_{\rm
   stat} \pm 0.031_{\rm syst})$ ns.
The fit window of this method covers 1.1 $\tau^\pm$ .
\begin{figure}[htb]
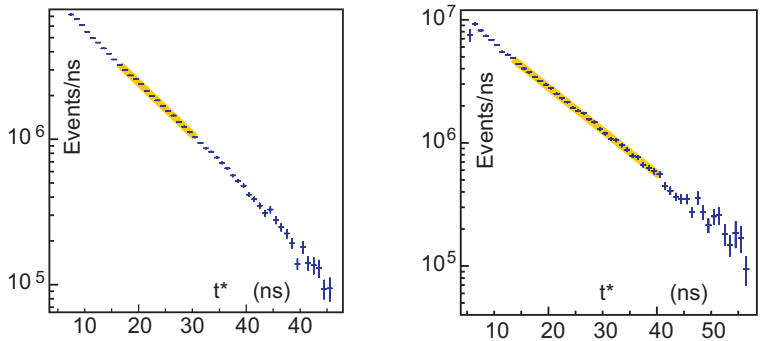

\centering\figb fit_length_new;4.5;\kern1cm\figb fit_time_new;4.5;
\caption{Charged kaon proper time distribution obtained with the two methods: the kaon decay length
         (Left) and the kaon decay time (Right).
         The yellow line shows the fit performed with the convolution of an exponential decay function and
         the resolution function.}
\label{fig:kpm_tau}
\end{figure}

\noindent
The second method relies on the measurement of the kaon decay time
and uses charged kaon decays with a $\pi^0$ in the final state,
$K^\pm \to X \pi^0$. In these decays the kaon time of flight is obtained
from the time of the EMC clusters of the photons from the $\pi^0$ decay.
We require the backward extrapolation of the tagging
kaon track to the IP. Then exploiting the kinematic closure of the $\phi\to K^+ K^-$ decay,
knowing the momentum of $\phi$ and of the $K^+$($K^-$)
we can build the expected helicoidal trajectory of the $K^-$($K^+$)
on the signal side (virtual helix). Stepping along this helix we look for the
$\pi^0\to\gamma\gamma$ decay.
For each photon it is then possible to measure the kaon proper time $t^\ast$ using:
$t^* = (t_\gamma - \frac{r_\gamma}{c} - T_0) \cdot \sqrt{1-\beta^2_K}$
with $t_\gamma$ the time of the cluster associated to the photon, $r_\gamma$ the distance between
$V_0$ and the position in EMC of the photon cluster, $T_0$ the production
time of the $\phi$  and $\beta^2_K$  the beta of the signal kaon. The efficiency has been evaluated directly from data.
The control sample has been selected using DC information only,
selecting the kaon decay vertex in the volume defined above.
The proper time is fit between 13 and 42 ns, about 2.3 $\tau^\pm$,
correcting for the
efficiency and using the convolution of an exponential decay function and the resolution function.
The weighted mean between the $K^+$ and the $K^-$ lifetimes gives the
result $\tau^\pm = (12.337 \pm 0.030_{\rm stat} \pm 0.020_{\rm syst})$ ns.
Given a statistical correlation of $\rho=30.7\%$,
the weighted mean between the two charges and methods is:
\begin{center}
\begin{equation}
\tau^\pm = (12.347 \pm 0.030)\ ns
\label{eq:taukpm}
\end{equation}
\end{center}
Both fits are shown in \fig{fig:kpm_tau}. The comparison of K$^+$ and K$^-$ lifetimes is a test of CPT invariance
which guarantees the equality of the decay lifetimes for
particle and antiparticle. The average of the two methods are
 $\tau^+ = (12.325 \pm 0.038 )~\rm{ns}$ and $\tau^- = (12.374 \pm 0.040 )~\rm{ns}$.
From these measurement we obtain:
$\tau^- / \tau^+ = 1.004 \pm 0.004$.
The result agrees well with CPT invariance at the 4 per mil level.

\subsection{\ks\ decays}
As already mentioned, central to \ks\ studies is the use of
K-crash tag (sect.~\ref{sec:ksbeam}), thanks to which
it has been possible to measure the \ks\ branching ratios,
 whose values span six orders of magnitude.\vglue4mm
%
%
\subsubsection{\ks\to\pic($\gamma$), \pio}
\label{sec:kspp_ratio}
The \ks\ decays overwhelmingly (99.9\%) into two pions: \pio\ and \pic.
The ratio of the charged decay mode to that of the neutral,
$\R^{\pi}_S \equiv \Gamma(\ks\to\pic(\gam))/\Gamma(\ks\to\pio)$, is a fundamental
parameter of the \ks\ meson. First of all, it provides the BRs for the $\ks\to\pio$
and $\ks\to\pic$ with only small corrections. The latter BR is a convenient normalization
for the BRs of all other \ks\ decays to charged particles.
From $R^{\pi}_S$ one can also derive phenomenological parameters of the
kaon system, such as the differences in magnitude and phase of the
isospin   $I=0,\,2$ $\pi\pi$ scattering amplitudes.
Finally, this ratio together with the corresponding one from the \kl\ determine the amount
of direct CP violation in $K\to\pi\pi$ transitions.
Prior to KLOE, the fractional error on $\R^{\pi}_S$ was 1.2$\%$, as a result of
an average on various measurements, each of  $\sim$5$\%$ accuracy.
This
averaging was somehow questionable, since the various experiments did not
clearly describe their procedure for handling radiative events.
During the last six years we performed two precise measurements of
$\R^{\pi}_S$ with increasing systematical accuracy, and with proper
treatment of radiative corrections.
  The final combined value has a precision of three parts per mil~\cite{ks2pi}.

Given the tag, the $\ks \to \pic(\gam)$ events are selected by requiring
the presence of two tracks of opposite charge with their point of closest approach
to the origin
inside a small fiducial volume around the IP, and with momentum in the range
$120 < p < 300$~MeV. $\ks \to \pio$ events are identified by the prompt photon
clusters from \po\ decays. A calorimeter cluster is considered as a prompt photon if
its velocity, evaluated by ToF, is compatible with $\beta = 1$ within the
expected resolution. Moreover, it must not be associated to any track.
To accept a $\ks \to \pio$ event, three or more prompt photons are required, with a
minimum energy of 20~MeV, to reduce contamination from machine background.

To reach a high precision, all of the possible systematic
effects  have been studied
in great detail, making use of
several independent data control samples.
Systematic errors can arise from imperfections in the detector simulation,
limitations in the methods used to evaluate the detector efficiencies, and
uncertainties in the cross sections and branching ratios used to estimate
the fraction of background events. Moreover, a small difference in the
tagging efficiency between charged and neutral decays, due to the different
time response of the two categories of events, has been taken
into account. Actually, the $e^{+}e^{-}$ interaction time,
the so called $T_{0}$, is obtained from the fastest particle reaching
the calorimeter, assuming a velocity $\beta = 1$ and a straight flight path
starting from the interaction point (sect.~\ref{sec:emc}). This assumption is correct
 for photons
coming from \ks\to\pio events, while most of the pions from their
charged counterpart arrive $\sim$3 ns later: therefore their
$T_{0}$ is delayed by one RF period, $\sim$2.7 ns. As a consequence of this,
the velocity of the $K_L$ crash clusters is overestimated by $\sim$10$\%$
for \ks\to\pic events.

Another subtle effect that has to be taken into consideration, is
the different energy response of the EmC to pions and photons.
This, in turn, causes a
small difference in the trigger efficiencies for the two categories of events
($\sim$98.5$\%$ for \pic, $\sim$100$\%$ for \pio), which has to be
properly taken into account.

We obtain $\R_{\pi}$ = $2.2549 \pm 0.0054$. Using this result,
and the measurement of  $\Gamma(K_S\rightarrow \pi e\nu)/\Gamma(
K_S\rightarrow \pi^{+}\pi^{-}(\gamma))$ discussed in 5.4.2,
we extract the dominant \ks\ branching ratios (see table~\ref{tab:ksdata}).
To this end, we exploit unitarity: the sum of the BR's for the $\pi\pi$ and
$\pi l \nu$ modes has been assumed to be equal to one, the remaining decays
accounting for less than 10$^{-4}$.

To extract the value of the phase shift difference $\delta_0 -\delta_2$
between $I=0$ and $I=2$ amplitudes we use the method described in
reference~\cite{cirigliano:delta02}.

In the isospin limit, we find:
\begin{equation}
\delta_0 - \delta_2 = (44.5 \pm 1.0)^\circ.
\end{equation}
Effects of isospin breaking increase this value by $\sim 10^\circ$, although the correction
is determined with a large error (see Ref.~\cite{cirigliano:delta02}).

\subsubsection{$K_S\rightarrow \pi e\nu(\gamma)$}
\label{sec:ksl3}
The \ks\ decays semileptonically less than one per cent of the time.
To pick out such decays where the event contains an unseen neutrino is non-trivial.
Yet, KLOE has isolated a very pure sample of \ab13,000 semileptonic \ks\
decays and accurately measured the BRs for $\ks\to\pi^+e^-\bar\nu(\gam)$
and $\ks\to\pi^-e^+\nu(\gam)$ \cite{kslept}.

The basic steps in the analysis are: tag \ks\ decays by the \kl\ crash,
apply a cut on the $\pi\pi$ invariant mass (this removes 95\% of the
$\ks\to\pic$ decays and reduces the background-to-signal ratio to about 80:1),
impose several geometrical cuts to further improve the purity of the sample
and, in particular, remove contamination by events with early $\pi\to\mu\nu$ decays.
Finally, stringent requirements are imposed on the particle's ToF,
which very effectively separates electrons from pions and muons and allows
the charge of the final state to be assigned.

\begin{figure}[htb]
\begin{center}
\includegraphics[width=0.45\textwidth]{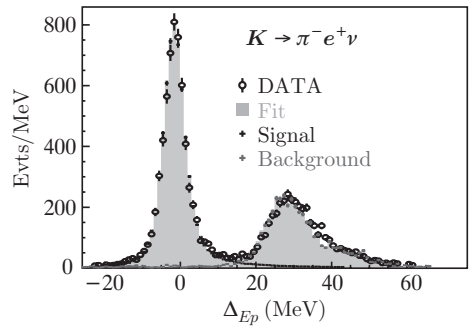}
\caption{$\Delta_{Ep}$ distribution for $\ks\to\pi^-e^+\nu$ candidates,
showing signal and background.}
\label{fig:kslept}
\end{center}
\end{figure}

\Fig{fig:kslept} shows the signal peak and the residual background in the
distribution of $\Delta_{Ep} = E_{\rm miss}-|{\bf p}_{\rm miss}|$
for the $\pi^-e^+\nu$ channel, where $E_{\rm miss}$ and ${\bf p}_{\rm miss}$
are respectively the missing energy and momentum at the vertex, evaluated
in the signal hypothesis. For signal events, the missing particle is a
neutrino and $\Delta_{Ep}=0$.

The numbers of $\pi e\nu$ decays for each charge state are normalized to
the number of observed \pic\ events, resulting in the ratios in the first
column of table \ref{tab:ksdata}.
Using the result for $\R_\pi$\ of the previous section
(also in table \ref{tab:ksdata}),
the absolute BRs for $\ks\to \pi\pi$ and $\ks\to\pi e\nu$
reported in the second column of the table are obtained.
\begin{table}[htb]
\begin{center}
\begin{tabular}{@{}lcc@{}}
\hline
 Mode & BR(mode)/BR(\pic) & BR(mode) \\
\hline
 $\pic$ & --- & ($69.196\pm0.024\pm0.045$)\% \\
$\pio$ & 1/($2.2549\pm0.0054$) & ($30.687\pm0.024\pm0.045$)\% \\
 $\pi^-e^+\nu$ & \pt(5.099\pm0.082\pm0.039),-4, & \pt(3.528\pm0.057\pm0.027),-4,\\
$\pi^+e^-\bar{\nu}$ & \pt(5.083\pm0.073\pm0.042),-4, & \pt(3.517\pm0.050\pm0.029),-4, \\
$\pi e \nu$ & \pt(10.19\pm0.11\pm0.07),-4, & \pt(7.046\pm0.076\pm0.051),-4,\\
\hline
\end{tabular}
\caption{KLOE measurements of $K_S$ branching ratios.}
\label{tab:ksdata}
\end{center}
\end{table}
These ratios give also the first measurement of the
semileptonic charge asymmetry for the \ks:
\begin{equation}
A_S = \frac{\Gamma (\ks \to \pi^-e^+\nu_e) - \Gamma(\ks \to \pi^+e^-\overline{\nu}_e)}{{\rm SUM }} = (1.5\pm9.6\pm2.9)\times10^{-3}.
\label{eq:aske3}
\end{equation}

The comparison of $A_{S}$ with the asymmetry $A_{L}$ for $K_L$ decays
allows test of the $CP$ and $CPT$ symmetries to be performed. Assuming $CPT$ invariance,
 $A_{S}\!=\!A_{L}\!\sim\!2\,\mathrm{Re}\,\epsilon\!\simeq\!3\!\times\!10^{-3}$,
 where
$\epsilon$ gives the $CP$ impurity of the $K_{S}$, $K_{L}$ mass eigenstates
due to $CP$ violation in $\Delta$S=2 transitions (for an exact definition of $\epsilon$ see sect.~\ref{sec:cptsm}).
To evaluate $A_S$ and $A_L$ without making any assumptions on $CPT$ symmetry, we consider the
standard decomposition of the semileptonic amplitudes~\cite{maiani_semilep}:
\begin{eqnarray}
  \mathcal{A}(\kzero  \to \pi^- l^+ \nu ) &=&  \mathcal{A}_0(1-y)~, \nonumber \\
  \mathcal{A}(\kzero  \to \pi^+ l^- \nu ) &=&  \mathcal{A}_0^* (1+ y^*) (x_+ - x_-)^*~,  \nonumber \\
  \mathcal{A}(\kzerob \to \pi^+ l^- \nu ) &=&  \mathcal{A}_0^* (1+ y^*)~, \nonumber \\
  \mathcal{A}(\kzerob \to \pi^- l^+ \nu ) &=&  \mathcal{A}_0(1-y) (x_+ + x_-)~,
\label{ampkl}
\end{eqnarray}
 where $x_+$ ($x_-$) describes the violation of the $\Delta S = \Delta Q$ rule in $CPT$ conserving
(violating) decay amplitudes, and $y$ parametrizes $CPT$ violation for $\Delta S = \Delta Q$
transitions.

The difference between the charge asymmetries,
\begin{equation}
A_{S}-A_{L}=4\,\left(\mathrm{Re}\,\delta+\mathrm{Re}\,x_{-}\right)
\label{eq:rexm}
\end{equation}
signals $CPT$ violation either in the mass matrix ($\delta$ term, see sect.~\ref{sec:cptsm}) or in
the decay amplitudes with $\Delta S \neq \Delta Q$ (Re$x_{-}$ term).
The sum of the asymmetries,
\begin{equation}
A_{S}+A_{L}=4\left(\mathrm{Re}\,\epsilon-\mathrm{Re}\,y\right),
\label{eq:rey}
\end{equation}
is related to $CP$ violation in the mass matrix ($\epsilon$ term) and to
$CPT$ violation in the
$\Delta S=\Delta Q$ decay amplitude ($y$ term).

Using $A_L = (3.34\pm0.07)\times10^{-3}$~\cite{pdg06} and our $A_S$
measurement (\eq{eq:aske3}), we obtain
from \eq{eq:rexm}
\begin{equation}
\mathrm{Re}\,x_{-}+\mathrm{Re}\,\delta=\left(-0.5\pm2.5\right) \times 10^{-3}.
\end{equation}
Current knowledge of these two parameters is dominated by results from CPLEAR
\cite{CPLEAR_redelta:98}:
the error on $\mathrm{Re}\,\delta$ is
$3\times10^{-4}$ and that on $\mathrm{Re}\,x_{-}$ is $10^{-2}$. Using
$\mathrm{Re}\,\delta=(3.0\pm3.3_{\rm stat}\pm0.6_{\rm syst})\times10^{-4}$
from CPLEAR, we obtain:
\begin{equation}
\mathrm{Re}\,x_{-}=\left(-0.8\pm2.5\right) \times 10^{-3},
\end{equation}
thus improving on the error of $\mathrm{Re}\,x_{-}$ by a factor of four.

From \eq{eq:rey} we obtain
\begin{equation}
\label{eq:reymwnoree}
  \mathrm{Re}\,\epsilon-\mathrm{Re}\,y=\left(1.2\pm2.5\right) \times 10^{-3}.
\end{equation}
We calculate $\mathrm{Re}\,\epsilon$ using values from~\cite{pdg06}
that are obtained without assuming $CPT$ conservation:
$\mathrm{Re}\,\epsilon=|\epsilon|\times\cos{\phi_{\epsilon}}=(1.62\pm0.04)\times10^{-3}$.
Subtracting this value from \eq{eq:reymwnoree}, we find
\begin{equation}
  \mathrm{Re}\,y=\left(0.4\pm2.5\right) \times 10^{-3},
\end{equation}
which has precision comparable to that ($3\!\times\!10^{-3}$) obtained from the unitarity relation by CPLEAR~\cite{CPLEAR:bell}.

\subsubsection{\ks\to$\gamma\gamma$}
\label{sec:ksgg}
%
%
A precise measurement of the $K_S\to \gam\gam$ decay rate is an
important test of Chiral Perturbation Theory (ChPT) predictions.
The decay amplitude of $\ks \to \gam\gam$ has been evaluated
at leading order of ChPT~\cite{ambrosio},  $\mathcal{O}(p^4)$,
providing a precise estimate of
$\BR(\ks \to \gam\gam) = 2.1 \times 10^{-6} $, with 3\% uncertainty.
This estimate is $\sim $ 30\% lower with respect to the latest determination
from NA48~\cite{na48}, thus suggesting relevant contributions from higher order
corrections.

We measured the  $\ks \to \gam\gam$ rate using 1.9~fb$^{-1}$ of integrated luminosity.
A MC background sample of comparable statistics and a large MC signal
sample (equivalent  to $\sim$ 50 fb$^{-1}$
of collisions) have been used in the analysis.
After K-crash tag, we select $\sim$~700 million events, out of which
$\sim 1900$ are $\ks \to \gam\gam$ decays.

\begin{figure}[htb]
\begin{center}
\includegraphics[width=0.5\textwidth]{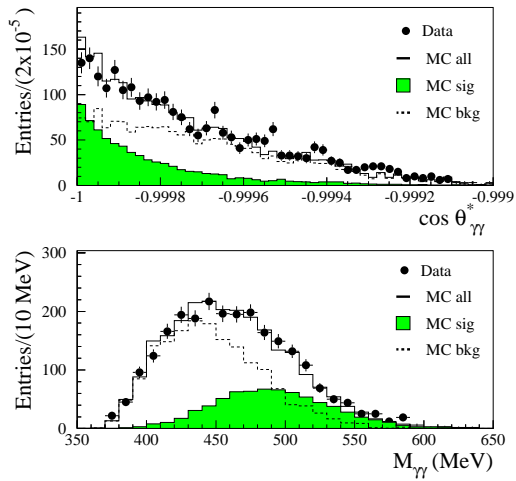}
\caption{Distributions of cos$\theta^{*}_{\gam \gam}$ (top) and
$M_{\gam\gam}$ (bottom) at the end of the analysis chain; black points are data,
solid line is the sum of MC signal and background shapes as weighted by the fit.}
\label{2dresult}
\end{center}
\end{figure}

The signal is selected by requiring exactly two
prompt photons, with an efficiency of $\sim 83\%$.
After photon counting,
the background composition is dominated by $\ks \to 2\pi^0$, with
two photons undetected by the EMC. This background is strongly
reduced, with negligible signal loss, by vetoing the events with photons
reaching the QCAL in a $5$~ns coincidence window with respect to
the event-$T_0$.
At the end of this selection, we are left with $157\times 10^3$ events
and a signal over background ratio S/B=1/80.
Further background reduction is obtained by performing
a kinematic fit to the event in the signal hypothesis, and using as constraints
the total 4-momentum conservation, the kaon mass and the photon ToF.
We retain events with $\chi^2<20$, reaching S/B$\sim 1/3$ with an efficiency on signal of $\sim 63 \%$.
After this cut we count the signal events by fitting
the 2-D distribution of the photon invariant mass,
$M_{\gam\gam}$, and of the photon opening angle
in the $\ks$ center of mass, $\theta^{*}_{\gam\gam}$, and using MC signal and
background shapes.
We count $N(\gam\gam)= 711 \pm 35$ signal events out of
2740 events, with $\chi^2/{\rm dof} = 854/826$, corresponding
 to a probability P$(\chi^2)\sim 24\%$.
In \fig{2dresult}, the observed distributions of $\cos \theta^{*}_{\gam\gam}$ and
$M_{\gam\gam}$ are compared with the simulated signal and background shapes as weighted
by the fit procedure.
To get the BR($\ks \to \gam\gam$), signal events are normalized to the number of
$K_S \to 2 \pi^0$ decays observed in the same data sample.

These are selected, after K-crash tag, by requiring three to five prompt photons.
An extensive study on systematics effects, which affect both the
efficiency correction and the signal counting, has been carried out.
As an example, the MC efficiency of the prompt photon selection
has been corrected for possible differences in the photon detection efficiency between
data and MC, and a systematic error has been evaluated which accounts for residual imperfections.
Moreover, a control sample of $K_L \to \gam \gam$ events decaying close to the IP has been selected,
which has been used to check systematics from the kinematic fit selection and from data-MC differences
in the EMC energy scale.
Systematic errors on signal counting have been evaluated by:
(i) performing a set
of fits to the $M_{\gam\gam} - \theta^{*}_{\gam \gam}$ distribution in regions with
an enhanced signal content, to evaluate the error
induced by the uncertainty in the background shape,
(ii) repeating the fit
with different bin-size and applying an
energy scale correction a factor of two greater with respect to that
measured with the $\kl \to \gam\gam$ control sample.

Finally, we obtain~\cite{ksgg}
\begin{equation}
\BR(\ks \to \gam \gam) = (2.26 \pm 0.12_{\rm stat} \pm 0.06_{\rm syst} ) \times 10^{-6},
\label{eq:ksgg}
\end{equation}
which differs by 3$\sigma$ from the previous best
determination. Our result is also consistent with $\mathcal{O}(p^4)$ ChPT prediction.

\subsubsection{\ks\to 3\po}
\label{sec:ks3p0}
%
%
The decay $\ks\to3\po$ violates $CP$ invariance. The parameter $\eta_{000}$,
defined as the ratio of \ks\  to \kl\ decay amplitudes, can be written
as $\eta_{000} = \mathcal{A}(\ks \to 3\pi^0) / \mathcal{A}(\kl \to 3\pi^0) = \epsilon +
\epsilon^{\prime}_{000}$, where $\epsilon$ quantifies the \ks\ $CP$ impurity and
$\epsilon^{\prime}_{000}$ is due to a direct $CP$-violating term.
Since we expect
$\epsilon^{\prime}_{000} << \epsilon $~\cite{ambrosio000}, it follows that
$\eta_{000} \sim \epsilon$. In the Standard Model (SM), therefore,
$\BR(\ks\to3\po) \sim 1.9 \times 10^{-9}$ to an accuracy of a few \%, making
the direct observation of this decay quite a challenge.

We performed a search for the $\ks \to 3\po$ decay using 450~pb$^{-1}$ of integrated luminosity.
A MC background sample $\sim 2.5$ times larger than data and a high statistics MC signal
sample  have been used in this analysis as well.
After K-crash tag, we select the signal by requiring six prompt photon clusters, and no tracks from
the IP, which is useful
to reject background from $\ks \to \pic$ with an early $\kl \to 3\po$ decay.
 At this stage of the analysis, the residual background is
dominated by $\ks\to\pio$ events with two spurious clusters from
shower fragments ({\em splitting}) or accidental coincidence
with clusters produced by machine activity.
Further background rejection is obtained by performing a kinematic fit to the event in the signal
 hypothesis, and cutting at $\chi^2/{\rm d.o.f}<3$.

After the previous rejection cuts, the signal search is performed by defining
 two $\chi^2$-like discriminating variables, $\zeta_3 - \zeta_2$, optimized
 for $\ks \to 3\po$ and $\ks \to 2\po$ selection, respectively. $\zeta_2$ is defined by
selecting the four out of six photons which provide the best agreement in terms of kinematic
variables with the $\ks \to 2 \po$ hypothesis. $\zeta_3$ is defined as
$$ \zeta_3 = \frac{\Delta m_1^2}{\sigma^2_m} + \frac{\Delta m_2^2}{\sigma^2_m} +
\frac{\Delta m_3^2}{\sigma^2_m}, $$
where $\Delta m_i = m_i - m_{\po}$ is the difference between the nominal
$\pi^0$ mass and the invariant masses of the $i$th photon pair, chosen among the six clusters
of the event. $\zeta_3$ is close to zero for a $\ks \to 3 \po$ event, and is expected to be large
for a six-photon background event. A signal box is defined in the $\zeta_3 - \zeta_2$ plane, while side-bands are used to check the MC prediction for the background.
At the end of the analysis chain we observe two candidates in the signal box, with an
expected background of  $3.1\pm0.8\pm0.4$.
The signal selection efficiency, after tagging, is $\sim 24\%$.
To obtain an upper limit for  $\BR(\ks \to 3\po)$, we normalize the number of signal
candidates to $\ks \to \pio$ decays observed in the same data sample.
For this purpose, we select events with three to five prompt photon clusters.
Finally, we obtain~\cite{ks3pi0}
\begin{equation}
\BR (\ks \to 3\po) \leq \pt 1.2,-7, \quad {\rm at \, 90\% \, CL},
\label{eq:ks3pi0}
\end{equation}
which is a factor of $\sim 6$ improvement with respect
to the best previous limit~\cite{na48ks3pi0}.
This limit on the BR can be directly translated into a limit on $|\eta_{000}|$:
\begin{equation}
|\eta_{000}| = \left| \frac{\mathcal{A}(\ks \to 3\po)}{\mathcal{A}(\kl \to 3\po)} \right|
 = \sqrt{\frac{\tau_L}{\tau_S} \frac{\BR(\ks\to3\po)}{\BR(\kl\to3\po)}}<0.018 \, {\rm at\,90\%\,CL}
\end{equation}
This result can be visualized in the complex plane as a circle of radius 0.018 centered at zero in the $\Re(\eta_{000})$, $\Im(\eta_{000})$ plane,
and is a 2.5 times better improvement with respect to the result of \cite{na48ks3pi0}.
%
%
\subsubsection{\ks\to\epem}
\label{sec:ksee}
This decay is a flavor-changing neutral current process, suppressed in the SM,
with an amplitude dominated by the two photon intermediate state.
Using ChPT to $\mathcal{O}(p^{4})$, one obtains the prediction
BR(\ks\to\epem) $\sim$2$\times$10$^{-14}$~\cite{kseeth}. A value significantly higher
could indicate new physics.
Prior to \klo\, the best experimental limit on this decay was set by CPLEAR,
 BR(\ks\to\epem) $\leq$ \pt{1.4},{-7}, at 90$\%$ CL~\cite{cplksee}.

We performed a search for the $\ks \to \epem$ decay using 1.9~fb$^{-1}$ of integrated luminosity.
A MC background sample of comparable statistics and a large MC signal sample have been also used in the
analysis. After K-crash tag, we search for the signal by requiring two tracks of opposite charge
originating near the IP. The two tracks are required to have
an invariant mass $M_{ee}$, evaluated in the electron hypothesis, in a $\sim20$~MeV window around the nominal
\ks\ mass.\footnote{With the invariant mass cut given above, we actually measure $\ks \to \epem (\gamma)$, with
$E_{\gamma}^{\ast}< 20$~MeV.} This cut is particularly effective on $\ks \to \pic$ events, which peak at
$M_{ee} \sim 409$~MeV. After it the background is dominated by $\ks \to \pic$ with at least
one pion wrongly reconstructed, and by $\phi \to \pic\po$ events. The \ks\ events are strongly reduced by
cutting on the track momentum in the \ks\ rest frame, which is expected to be $\sim 206$~MeV for
$\ks \to \pic$ decays.
To reject $\phi \to \pic \po$ events we use instead $|\vec{p}_{miss}| = |\vec{p}_{\phi} - \vec{p}_{\ks} -
\vec{p}_{\kl} |$, where \ks\ and \kl\ momenta  are evaluated from the charged tracks and from the K-crash tag,
respectively. The value of $|\vec{p}_{miss}|$ peaks at zero for the signal, within the
experimental resolution of few MeV. It spreads towards higher values for $\phi \to \pic\po$ events.
The residual background, both from $\ks\to\pic$ and $\phi \to \pic\po$ decays, is rejected identifying the two
electrons by ToF, and by using the properties of the associated calorimetric cluster. The reliability of the
MC background simulation is checked after each step of the selection on the invariant mass sidebands.
At the end of the analysis chain, we don't count any event in the signal box; the background estimate
is also compatible with zero.
The upper limit on the number of signal events is therefore $N_{ee} = 2.3$ at 90\% CL The signal selection efficiency, after tagging, is $\sim 47 \%$. Such performances in terms of
exceptional background rejection ($>10^8$) with an acceptable signal efficiency, have been achieved
 largely thanks to the very good momentum resolution of our DC.
To obtain an upper limit for the $\BR(\ks \to \epem)$, we normalize $N_{ee}$ to the $\ks \to \pic$ decays
observed in the same data sample.
For this purpose, we used the same selection criteria as for the measurement of $\BR(\ks \to \pic)$,
described in sect.~\ref{sec:kspp_ratio}.
Finally, we obtain as a preliminary result~\cite{kksee} 
\begin{equation}
\BR (\ks \to \epem) \leq \pt 9.3,-9, \quad {\rm at \, 90\% \, CL},
\label{eq:ksee}
\end{equation}
which represents a factor of $\sim 15$ improvement with respect
to the best previous limit.

\subsection{\kl\ Decays}
A measurement of the absolute \kl\ BRs is a unique possibility of the \ff, where a pure sample of
nearly monochromatic \kl's can be selected by identification of a simultaneous $\ks \to \pic$ decay.
The absolute BRs can be determined by counting the fraction of \kl's that decay into each channel,
and correcting for acceptances, reconstruction efficiencies, and background. More specifically, we
evaluate \kl\ BRs as:
\begin{equation}
\BR(\kl \to f) = \frac{N_f}{N_{\rm tag} \epsilon_{\rm FV} \epsilon_{\rm rec}}\frac{\langle \epsilon_{\rm tag}
\rangle}{\epsilon_{\rm tag}},
\label{eq:absolutebrl}
\end{equation}
where $N_f$ is the number of decays to $f$ identified in the FV after background subtraction,
$N_{\rm tag}$ is the number of tagged \kl's,
$\epsilon_{\rm FV}$ is the fraction of decays in the FV,
$\epsilon_{\rm rec}$ is the reconstruction efficiency for channel $f$,
and $\epsilon_{\rm tag}/\langle \epsilon_{\rm tag} \rangle$ is the  tag bias, $i.e.$ the ratio
of the $\ks \to \pic$ reconstruction efficiencies for $\kl \to f$ and independent of the \kl\ fate, respectively.
 We use a FV well within the DC, which gives an efficiency $\epsilon_{\rm FV} \sim 26\%$.
Losses of \kl's from interactions in the beam pipe and chamber walls are taken into
account in the evaluation of $\epsilon_{\rm FV}$, as well as its dependence on the \kl\
lifetime (see sect.~\ref{sec:KLlife}).
\subsubsection{\kl\ semileptonic and 3 pion decays}
\label{sec:brl}
The four major decay modes, $\pi^\pm e^\mp\nu$ ($K_{e3}$),
$\pi^\pm\mu^\mp\nu$ ($K_{\mu3}$), \pic\po, and 3\po\ account for more
than 99.5\% of all decays, and are measured simultaneously from a sample of
13 millions tagged \kl\ decays. For the first three modes, two
tracks are observed in the DC, whereas for the $3\po$ mode, only photons
appear in the final state. The analysis of two-track and all-neutral-particle
events is therefore different.

\begin{figure}[htb]
\begin{center}
\includegraphics[width=0.8\textwidth]{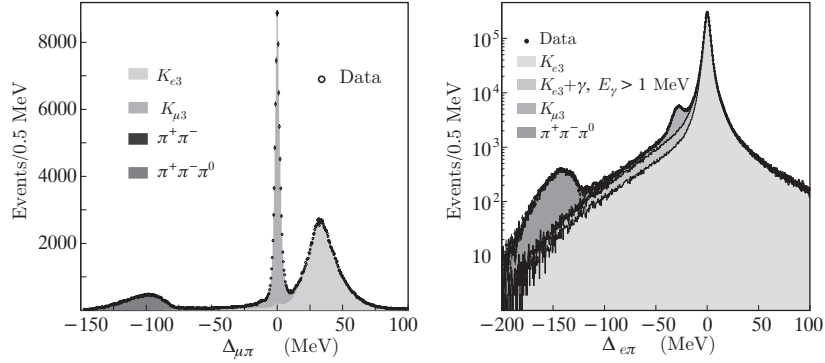}
\caption{Left: distribution of $\Delta_{\mu\pi}$ for a run set. Right: distribution of $\Delta_{e\pi}$ for events with an identified electron, for the entire data set.}
\label{depmiss}
\end{center}
\end{figure}

Two-track events are assigned to the three channels of interest by use of a single variable:
the smaller absolute value of the two possible values of
$\Delta_{\mu\pi}=|{\bf p}_{\,\rm
miss}|-E_{\,\rm miss}$, where
${\bf p}_{\,\rm miss}$ and $E_{\,\rm miss}$ are the missing momentum and
energy in the \kl\ decay, respectively, and are evaluated
assuming the decay particles are a
pion and a muon.
\Fig{depmiss}, left, shows an example of a $\Delta_{\mu\pi}$
distribution.
We obtain the numbers of $K_{e3}$, $K_{\mu3}$, and $\pic\po$ decays
by fitting the $\Delta_{\mu\pi}$ distribution with the corresponding
MC-predicted shapes.
The signal extraction procedure is tested using PID variables from the calorimeter.
This is illustrated in \fig{depmiss}, right, where the $\Delta_{e\pi}$ spectrum
is shown for events with
identified electrons, together with the results of a fit with MC shapes.
The $K_{e3(\gamma)}$ radiative tail is clearly evident.
The inclusion of radiative processes in the simulation
is necessary to obtain a good fit, as well
as to properly estimate the fully inclusive radiative rates.
The reconstruction efficiency is $\sim 54\%$ for $K_{e3}$, $\sim 52\%$ for $K_{\mu 3}$, and
$\sim 38\%$ for \pic\po, as evaluated from MC. These efficiencies are then corrected to account
for MC imperfections in reproducing the tracking efficiency. The correction factors range
between 0.99 and 1.03, depending on the decay channel and on the data taking period.

The decay $\kl\to3\po$ is easier to identify. Detection
of $\ge3$ photons originating at the same point along the \kl\ flight path is accomplished with
the technique described in sect.~\ref{sec:klbeam}. This selection guarantees very high
 reconstruction efficiency, $\sim99\%$, and small background contamination, $\sim1\%$.

The tag bias is mostly due to a dependence of the calorimeter trigger efficiency on
the \kl\ final state. Additional effects are introduced by the possible overlap between
\ks\ and \kl\ charged products, which undermines \ks\ reconstruction performances
for \kl\ decaying close to the IP. Both effects are reduced by additional requests to the
tag selection, which give an overall tagging efficiency of $\sim 9\%$.

Our measured BR values for the four modes~\cite{klbr},
corresponding to a reference value $\tau_L^{(0)} = 51.54$~ns, ${\rm BR}^{(0)}$,
are listed in table \ref{tab:klbr1}.
 \begin{table}[htb]
\centering
\begin{tabular}{lccccc}
\hline
Parameter & Value & \multicolumn{4}{c}{Correlation coefficients} \\
\hline
\BRu K_{e3};          & 0.4049(21) & 1    &         &      &   \\
\BRu K_{\mu3};        & 0.2726(16) & 0.09 & 1       &      &   \\
\BRu3\pi^0;          & 0.2018(24) & 0.07 & $-0.03$ & 1    &   \\
\BRu\pi^+\pi^-\pi^0; & 0.1276(15) & 0.49 & 0.27    & 0.07 & 1 \\
\hline
\end{tabular}
\caption{KLOE measurements of principal $K_L$ BRs assuming
$\tau_L = 51.54$~ns.}
\label{tab:klbr1}
\end{table}
The errors given for the absolute BRs do not include the contribution from the
\kl\ lifetime value, which enters into the calculation of the FV
efficiency. This dependence has been used in
sect.~\ref{sec:KLlife} to perform a combined fit to BRs and lifetime,
including also the $\tau_L$ measurement from the proper decay time
distribution of $\kl \to 3\po$ decays. The results of this fit have been
given in table \ref{tab:klbr2}.
\subsubsection{$K_L\rightarrow \pi l \nu(\gamma)$ form factors}
\label{sec:ff}
\begin{figure}[htb]
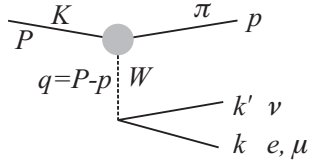

\centerline{\figb kel3ffgray; 4; }
\caption{Amplitude for kaon semileptonic decay}.
\label{fig:amp}%
\end{figure}

The semileptonic kaon decay, see \fig{fig:amp},
still provides the best means for the measurement of \abs{\Vus},
because only the vector part of the weak current contributes
to the matrix element $\bra\pi;J_\alpha\ket K;$. In general,
\begin{displaymath}
\bra\pi;J_\alpha\ket K; = f_+(t)(P+p)_\alpha + f_-(t)(P-p)_\alpha,
\end{displaymath}
where $P$ and $p$ are the kaon and pion four-momenta, respectively,
and $t=(P-p)^2$.
The form factors (FF) $f_+$ and $f_-$ appear because pions and kaons are
not point-like particles, and also reflect both SU(2) and SU(3)
breaking.
Introducing the scalar FF $f_0(t)$, the matrix element above is written as
\begin{equation}
\hspace{0.5cm}\langle\pi(p)|\bar u\gamma_\alpha s|K(P)\rangle=f(0)\left((P+p)_\alpha\,\tilde f_+(t)+(P-p)_\alpha\left(\tilde f_0(t){\Delta_{K\pi}\over t}-\tilde f_+(t){\Delta_{K\pi}\over t}\right)\right),
\label{eq:kl3}
\end{equation}
with $\Delta_{K\pi}=m_{K}^2-m_{\pi}^2$. The $f_+$ and $f_0$ FFs must have the same value at $t=0$.
We have therefore factored out a term $f(0)$. The functions $\tilde f_+(t)$ and $\tilde f_0(t)$ are both unity at $t=0$.
For vector transitions, the Ademollo-Gatto theorem \cite{AG64:f0} ensures that
SU(3) breaking appears only to second order in $m_s-m_{u,d}$.
In fact, $f_+(0)$ differs from unity by only \ab4\%.
The behavior of
 the reduced FFs \fphat\ and \fzhat\ as a function of $t$
can be measured from the decay spectra. At the level of
accuracy reached by the present experiments, the parameterization used to
extract the FF slopes is a relevant issue.

If the FFs are expanded in powers of $t$ up to $t^2$ as
\begin{equation}
\tilde{f}_{+,0}(t) = 1 + \lambda'_{+,0}~\frac{t}{m^2}+\frac{1}{2}\;\lambda''_{+,0}\,\left(\frac{t}{m^2}\right)^2
  \label{eq:ff2}
\end{equation}
 four parameters (\lamp, \lampp, \lamop, and \lamopp)
 need to be determined from the decay spectrum in order to be able to compute the phase space integral.
 However, this parametrization of the FFs is problematic, because the values
 for the $\lambda$s obtained from fits to the experimental decay spectrum
 are strongly correlated, as discussed in \cite{Fra07:Kaon}.
 In particular, the correlation between
\lamop\ and \lamopp\ is $-99.96\%$; that between
 \lamp\ and \lampp\ is $-97.6\%$. It is therefore impossible
 to obtain meaningful results using this parameterization.

Form factors can also by described by a pole form:
\begin{equation}
\tilde f_{+,0}(t) ={M_{V,S}^2\over M_{V,S}^2-t},
  \label{eq:pole}
\end{equation}
 which expands to $1+t/M_{V,S}^2+(t/M_{V,S}^2)^2$, neglecting powers of $t$
 greater than 2. It is not clear however what vector (V) and scalar (S) states
 should be used.

 Recent $K_{e3}$ measurements \cite{KTeV:ff,NA48:ff,KLOE+06:Ke3FF} show that the vector FF is
 dominated by the closest vector
 $(q\bar q)$ state with one strange and one light quark (or $K\pi$ resonance, in an older language).
 The pole-fit results are also consistent with predictions from a
 dispersive approach \cite{stern,pich1}.
 We will therefore use a parametrization for the vector FF based on a dispersion relation twice subtracted
 at $t=0$ \cite{stern}:
\begin{equation}
\fphat = \exp\left[\frac{t}{m_\pi^2}\left( \Lambda_+ + H(t)\right)\right],
  \label{eq:sternv}
\end{equation}
where $H(t)$ is obtained using $K-\pi$ scattering data.
An approximation to \eq{eq:sternv} is
\begin{equation}
\fphat =1+\lam{t\over m^2} + {\lam^2+p_2\over2}\left(t\over m^2\right)^2
+ {\lam^3+3 p_2 \lam + p_3\over6}\left(t\over m^2\right)^3 ,
  \label{eq:fplus}
\end{equation}
with $p_2$ and $p_3$ as given in table \ref{tab:p2p3}.
\begin{table}[htb]
\begin{center}
\begin{tabular}{lcccc}
\hline
                    & Vector (\eq{eq:fplus})   & Scalar (\eq{eq:fzeroa})   \\
\hline
 $p_2 \times 10^4$  &  $5.84\pm0.93$ & $4.16 \pm 0.50$ \\
 $p_3 \times 10^4$  &  $0.30\pm0.02$ & $0.27 \pm 0.01$ \\
\hline
\end{tabular}
\end{center}
\caption{Constants appearing in dispersive representations of vector and scalar form factors.}
\label{tab:p2p3}
\end{table}

The pion spectrum in $K_{\mu3}$ decay has also been measured recently
 \cite{KTeV:ff,NA48:ff2,istra2}. As discussed in \cite{Fra07:Kaon}, there is
 no sensitivity to $\lambda''_0$. All authors have fitted their data using a
 linear scalar FF:
\begin{equation}
\fzhat =1+\lamo {t\over m^2}.
  \label{eq:fzero}
\end{equation}
 Because of the strong correlation between $\lambda_0'$ and $\lambda_0''$,
 use of the linear rather than the quadratic parameterization gives a value
 for $\lambda_0$ that is greater than $\lambda_0'$ by an amount equal to
 about 3.5 times the value of $\lambda_0''$. To clarify this situation,
 it is necessary to obtain a form for \fzhat\ with at least $t$ and $t^2$ terms
 but with only one parameter.

 The Callan-Treiman relation \cite{ct} fixes the
 value of scalar FF at $t=\Delta_{K\pi}$ (the so-called Callan-Treiman point) to the
 ratio of the pseudoscalar
 decay constants $f_K/f_\pi$. This relation is slightly modified by SU(2)-breaking
 corrections\cite{dct}:
\begin{equation}
\tilde{f}_0(\Delta_{K\pi})=\frac{f_K}{f_\pi}\:{1\over f(0)}+\Delta_{CT},
  \label{eq:ct}
\end{equation}
where $\Delta_{CT}$ is of the order of $10^{-3}$.
 A recent parametrization for the scalar FF \cite{stern}
 allows the constraint given by the Callan-Treiman relation to be exploited.
 It is a twice-subtracted representation of the FF at $t=\Delta_{K\pi}$ and $t=0$:
\begin{equation}
\fzhat =\exp\left(\frac{t}{\Delta_{K\pi}} \log (C-G(t))\right)
  \label{eq:stern}
\end{equation}
 such that $C=\tilde{f}_0(\Delta_{K\pi})$ and $\tilde{f}_0(0) = 1$.
 $G(t)$ is derived from $K\pi$ scattering data.
As suggested in \cite{stern}, a good approximation to \eq{eq:stern} is
\begin{equation}
\fzhat =1+\lamo {t\over m^2}  + {\lamo^2+p_2\over 2} \left({t\over m^2}\right)^2
+ {\lamo^3+3 p_2 \lamo+ p_3\over 6}\left({t\over m^2}\right)^3 ,
  \label{eq:fzeroa}
\end{equation}
with $p_2$ and $p_3$ as given in table \ref{tab:p2p3}. The Taylor expansion gives
$\log C= \lamo \Delta_{K\pi}/m^2_\pi + (0.0398 \pm 0.0041)$.
\Eq{eq:fzeroa} is quite similar to the result in  \cite{pich2}.

To measure $K_{e3}$ form-factor slopes,
we start from the same sample
of \kl\ decays to charged particles used to measure the main \kl\
BRs. We impose additional, loose kinematic cuts and make use of
ToF information from the calorimeter clusters associated to the daughter
tracks to obtain better particle identification.
The result is a high-purity sample of 2 million $\kl\to\pi e\nu$ decays.
Within this sample, the identification of the electron and pion tracks
is certain, so that the momentum transfer $t$ can be safely evaluated
from the momenta of the \kl\ and the daughter tracks.
We obtain the vector form-factor slopes from binned log-likelihood
fits to the $t$ distribution.
Using the quadratic parameterization of \eq{eq:ff2}, we
obtain~\cite{KLOE+06:Ke3FF}
$\lambda_+' = \SN{(\VSS{25.5}{1.5}{1.0})}{-3}$ and
$\lambda_+'' = \SN{(\VSS{1.4}{0.7}{0.4})}{-3}$,
where the total errors are correlated with $\rho = -0.95$.
Using the pole parameterization of \eq{eq:pole}, we obtain
$M_V = \VSS{870}{6}{7}$~MeV.
Evaluation of the phase space integral for $\kl\to\pi e\nu$ decays
gives $0.15470\pm0.00042$ using the values of $\lambda_+'$ and
$\lambda_+''$ from the first fit and $0.15486\pm0.00033$ using
the value of $M_V$ from the second; these results differ by \ab0.1\%, while
both fits give $\chi^2$ probabilities of $\le$92\%.
The results we obtain using quadratic and pole fits are manifestly
consistent.

The measurement of the vector and scalar form-factor slopes
using $\kl\to\pi\mu\nu$ decays is more complicated.
First, there are two form factors to consider, and since all information
about the structure of these form factors is contained in the
distribution of pion energy (or equivalently, $t$),
the correlations between form-factor slope
parameters are very large. In particular, it is not possible to measure
$\lambda_0''$ for any conceivable level of experimental statistics
\cite{Fra07:Kaon}.
Second, at KLOE energies, clean and efficient $\pi/\mu$ separation is
much more difficult
to obtain than good $\pi/e$ separation.
However, the form-factor slopes may be obtained from fits to the
distribution of the neutrino energy $E_\nu$, rather than to the
distribution in $t$.
$E_\nu$ is simply the missing momentum in the
$\kl\to\pi\mu\nu$ decay evaluated in the \kl\ rest frame, and
requires no $\pi/\mu$ assignment to calculate.
A price is paid in statistical sensitivity: the $E_\nu$ distribution is
related to the $t$ distribution via an integration over the pion energy,
and so the statistical errors on the form-factor slope parameters
will be 2--3 times larger when the $E_\nu$ distribution, rather than
the $t$ distribution, is fit (assuming that the fit parameters
are $\lambda_+'$, $\lambda_+''$, and $\lambda_0$).

\begin{figure}[htb]
\begin{center}
\includegraphics[width=0.5\textwidth]{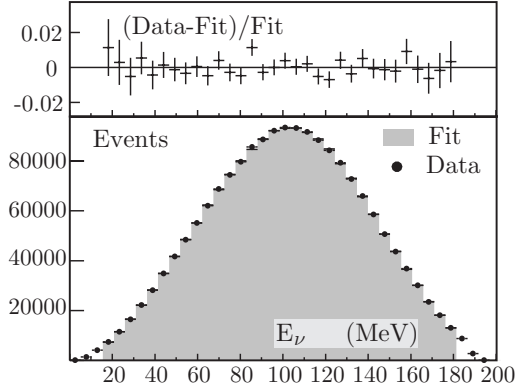}
\caption{Residuals of the fit (top plot) and $E_\nu$ distribution for data
  events superimposed on the fit result (bottom plot).}
\label{fig:fit_res}
\end{center}
\end{figure}

\noindent
 About 1.8 million decays were accepted.
 We first fit the data using equations \ref{eq:ff2} and \ref{eq:fzero} for the vector
 and scalar FFs respectively.
The result of this fit is shown in \fig{fig:fit_res}.

We obtain~\cite{kloe_ffkmu3}:
\begin{eqnarray}
\eqalign{\lamp &= (22.3\pm 9.8_{\rm{stat}}\pm 3.7_{\rm{syst}}) \times10^{-3}\cr
\noalign{\vglue-1mm}
\lampp &= (4.8\pm 4.9_{\rm{stat}}\pm 1.6_{\rm{syst}}) \times10^{-3}\cr
\noalign{\vglue-1mm}
\lambda_0 &= (9.1\pm 5.9_{\rm{stat}}\pm 2.6_{\rm{syst}}) \times10^{-3}\cr}\kern1cm
\pmatrix{ 1 & -0.97 & 0.81\cr   &     1 & -0.91\cr   &       & 1\cr}
\label{eq:ffquadke3}
\end{eqnarray}

with $\chi^2/{\rm dof}=19/29$, and correlation coefficients as given in the matrix.
Improved accuracy is obtained by combining the above results
 with those from our $K_{Le3}$ analysis.
We then find:
\begin{equation}
\eqalign{\lamp &= (25.6\pm 1.5_{\rm{stat}}\pm 0.9_{\rm{syst}}) \times10^{-3}\cr
\noalign{\vglue-1mm}
\lampp &= (1.5\pm 0.7_{\rm{stat}}\pm 0.4_{\rm{syst}}) \times10^{-3}\cr
\noalign{\vglue-1mm}
\lamo &= (15.4\pm 1.8_{\rm{stat}}\pm 1.3_{\rm{syst}}) \times10^{-3}\cr}\kern1cm
\pmatrix{ 1 & -0.95 & 0.29\cr   &     1 & -0.38\cr   &       & 1\cr}
\label{eq:ffquad}
\end{equation}
with $\chi^2/{\rm dof}=2.3/2$ and the correlations given in the matrix on the right.
Finally,
the same combination of $K_{e3}$ and $K_{\mu 3}$ results has been performed
to take advantage of the recent parameterizations of the
FFs based on dispersive representations (eqs. \ref{eq:sternv} and \ref{eq:stern}).
We perform a fit to the values obtained for \lamp, \lampp, and \lamo\
that makes use of the total error matrix as described above, and
the constraints implied by equations \ref{eq:fplus} and \ref{eq:fzeroa}.
Thus, the vector and scalar FFs are each described by
a single parameter.
Dropping the ``$\;'\;$'' notations, we find:
\begin{equation}\eqalign{%
\lam &= (25.7\pm 0.4_{\rm{stat}}\pm 0.4_{\rm{syst}}\pm 0.2_{\rm {param}}) \times10^{-3}\cr
\lamo &=(14.0\pm 1.6_{\rm{stat}}\pm 1.3_{\rm{syst}}\pm 0.2_{\rm {param}}) \times10^{-3}\cr}
\label{eq:ffdisp}
\end{equation}
with $\chi^2/{\rm dof}=2.6/3$ and a total correlation coefficient of
$-0.26$. The uncertainties arising from the choice of parameterization for
the vector and scalar FFs are given explicitly.

The values of the phase space integrals for $K_{\ell3}$ decays are listed
in table \ref{tab:ikl}, for both determinations
of the FF slopes (equations \ref{eq:ffquad} and \ref{eq:ffdisp}).
\begin{table}[htb]
\begin{center}
\begin{tabular}{lcccc}
\hline
Parametrization & $I(K^0_{e3})$ &$I(K^0_{\mu3})$ & $I(K^+_{e3})$ & $I(K^+_{\mu3})$     \\
\hline
 \eq{eq:ffquad} &  0.15483(40) &  0.10271(52) & 0.15919(41) & 0.10568(54) \\
 \eq{eq:ffdisp} &  0.15477(35) &  0.10262(47) & 0.15913(36) & 0.10559(48) \\
\hline
\end{tabular}
\end{center}
\caption{Phase space integrals for $K_{\ell3}$ decays, obtained using FF slopes as determined
with quadratic/linear (first row) and dispersive (second row) parameterizations of the FFs.}
\label{tab:ikl}
\end{table}
We note that the use of the dispersive parameterization changes the value of the
phase space integrals by at most $\sim 0.09\%$ with respect to what
obtained using quadratic and linear parametrizations for vector and scalar FFs, respectively. This reflects
into a systematic uncertainty from FF modelling of less than one per mil on the determination of \fVus,
which is evaluated using the phase space factors from the FF dispersive parameterization.

Finally, from the Callan-Treiman relation, \eq{eq:ct}, we compute
$\fo  = 0.967\pm0.025$ using $\fkfp = 1.189\pm 0.007$ from a recent lattice
 calculation~\cite{hpqcdukqcd07:fkfp}, and $\Delta_{CT} = (-3.5 \pm 8.0)\times 10^{-3}$.
Our value for \fo,
 although with a rather large error, is in agreement with the present best lattice determination,
$\fo  = 0.9644\pm0.0049$~\cite{rbcukqcd07:f0}.
\subsubsection{Radiative $K_{Le3}$ decay}
\label{sec:kle3g}
Two different processes contribute to photon emission in kaon decays: inner
bremsstrahlung (IB) and direct emission (DE). DE is radiation from intermediate
hadronic states, and is sensitive to hadron structure. The relevant kinematic
variables for the study of radiation in $K_{l3}$ decays are \Estar, the
energy of the radiated photon, and \qstar, its angle with respect to the
lepton momentum in the kaon rest frame. The IB amplitude diverges for
$\Estar \to 0$. For $K_{e3}$, for which $m_e \approx 0$, the IB spectrum in
\qstar\ is peaked near zero as well. The IB and DE amplitudes interfere. The
contribution to the width from IB-DE interference is 1\% or less of the purely
IB contribution; the purely DE contribution is negligible.
To disentangle the IB and DE
components, we measure the double differential spectrum
$d^2\Gamma/d \Estar d \qstar$ and compare the result with expectations from
Monte Carlo generators.
In the ChPT treatment of \cite{ke3gtheo}, the photon spectrum is approximated
by
\begin{equation}
\frac{d\Gamma}{d\Estar} \simeq \frac{d\Gamma_{\rm IB}}{d\Estar} + \X f(\Estar).
\label{eq:xdef}
\end{equation}
The DE contributions are summarized in the function $f(\Estar)$, which represents the
deviation from a pure IB spectrum. All information on the strength of the DE terms
is contained in the parameter \X.  Theoretical predictions on this
 parameter suffer by large uncertainties, due to the poor knowledge of ChPT
low energy constants. In contrast, a fit to the \Estar-\qstar\ spectrum allows
us to measure for the first time a value for \X. From the fit, we extract also
\begin{equation}
R \equiv \frac{\Gamma (K_{e3\gam}; \Estar>30\MeV, \qstar > 20^{\circ})}{K_{e3(\gam)}}.
\label{eq:Rdef}
\end{equation}
 The value of this ratio has been computed at $\mathcal{O} (p^6)$ in ChPT, leading to the prediction  $R = (0.963 \pm 0.006 \X \pm 0.010)\times 10^{-2}$.

The experimental distribution in (\Estar, \qstar) has been fit using the sum of
four independently normalized MC distributions: the distributions for $K_{e3\gam}$
events from IB satisfying (not satisfying) the kinematic cuts
 $\Estar>30\MeV$ and $\qstar>20^{\circ}$, the distribution corresponding
to the
function $f(\Estar)$, and the physical background from $\kl \to \pic$ and
$\kl \to \pi \mu \nu$ events.
Using a sample of 9\,000 $K_{e3\gam}$ events, and 3.5 million $K_{e3(\gam)}$ for
normalization, we find~\cite{ke3g}:
\begin{eqnarray}
 & R &  =  (0.924 \pm 0.023_{\rm stat} \pm 0.016_{\rm syst})\times 10^{-2} \nonumber \\
 &\X &  =  -2.3 \pm 1.3_{\rm stat} \pm 1.4_{\rm syst}.
\label{eq:ke3gres}
\end{eqnarray}
The dependence of R on \X, predicted by ChPT, can be used to further
constrain the possible values of R and \X\ from our measurement. The constraint is
applied via a fit, which gives $R = (0.944 \pm 0.014)\times 10^{-2}$ and
$\X = -2.8 \pm 1.8$, with $\chi^2/{\rm ndf}  = 0.64/1$ ($P = 42\%$). The
resulting value of $R$ may be compared with the value quoted in
\cite{ke3gtheo}, calculated for $\X = -1.2 \pm 0.4$.

\subsubsection{\kl\ $\rightarrow \pi^+\pi^-$}
\label{sec:klpp}
 $CP$ violation was discovered in 1964
 through the observation of the decay \kl \to \pic ~\cite{CCFT}.
 The value of BR($\kl\to \pic$) is known today with high accuracy from the
 results of many experiments \cite{pdg06}.
 In the SM, $CP$ violation is naturally accommodated
 by a phase in the quark mixing matrix \cite{Cab63,KM73}.
 BR(\kl \to \pic), together with the well known values of BR(\ks \to \pic),
$\tau_{S}$, and $\tau_{L}$, determines the modulus of the amplitude ratio
 \begin{equation}
|\eta_{+-}|=\sqrt{\Gamma(\kl \to \pic)/\Gamma(\ks \to \pic)},
\label{eq:eta+-}
\end{equation}
which is
parameterized as $\eta_{+-}=\epsilon+\epsilon'$. Here
$\epsilon$ quantifies the \kl\ $CP$ impurity and
$\epsilon^{\prime}$ is due to a direct $CP$-violating term.
Since we know that $\epsilon^{\prime} \sim 10^{-3} \epsilon $~\cite{pdg06},
it follows that a measurement of $|\eta_{+-}|$ can be directly
compared with the SM prediction for $\epsilon$.
We obtain a precise determination of $\BR(\kl \to \pic)$ by
measuring the ratio  $\BR(\kl \to \pic)/\BR(\kl \to \muiii)$. This approach
is particularly
convenient, since the values of the tagging efficiencies for
the  \kl\to\pic\ and $\kl \to \pi\mu \nu$ decays are very similar, and the
related systematic uncertainties do cancel in the ratio.
For this analysis, we used the same data sample as for \kl\ BRs determination
(sect.~\ref{sec:brl}), and our value of $\BR(\kl \to \muiii)$ has been used
to extract $\BR(\kl \to \pic)$\footnote{Statistical correlations have been
taken into account in the determination of the measurement error.}.

Given a $\ks \to \pic$ tagging decay, we select events with two tracks with opposite charge
belonging to the same vertex along the \kl\ line of flight. A fiducial volume in the DC
is chosen to count \kl\ decays.
The best variable to select $\kl \to \pic$ decays is
$\sqrt{E^2_{miss}+|\vec{p}_{miss}|^2}$, where
 the missing 4-momentum is evaluated using the tag information for \kl, and DC for
\kl\ decay particles, which are assumed to be pions.
We count \pic\ events by fitting the $\sqrt{E^2_{miss}+|\vec{p}_{miss}|^2}$ distribution,
and using MC for the signal and the background shapes.
\begin{figure}[htb]
\centering
\includegraphics[width=0.4\textwidth]{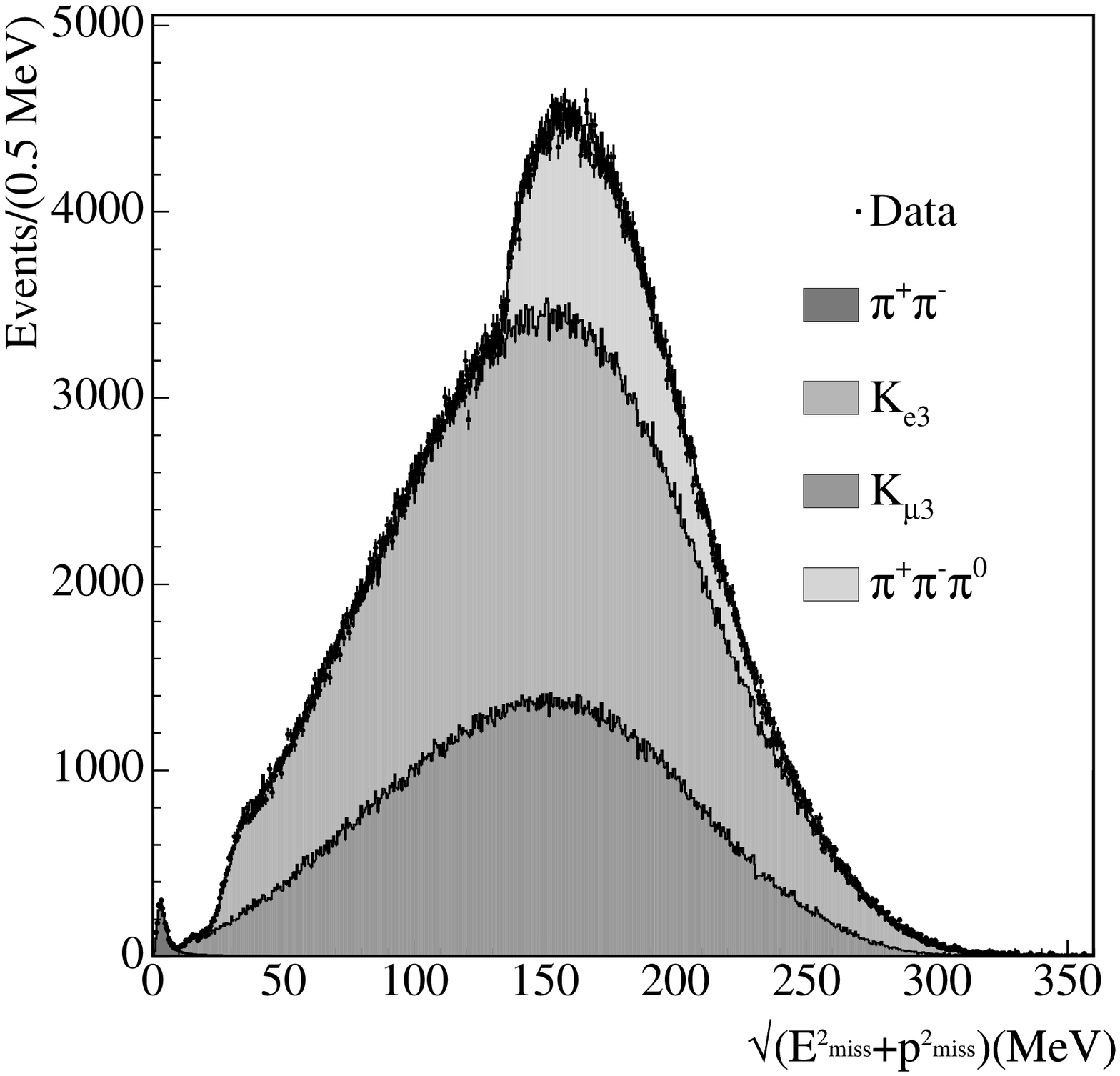}\kern1cm
\includegraphics[width=0.4\textwidth]{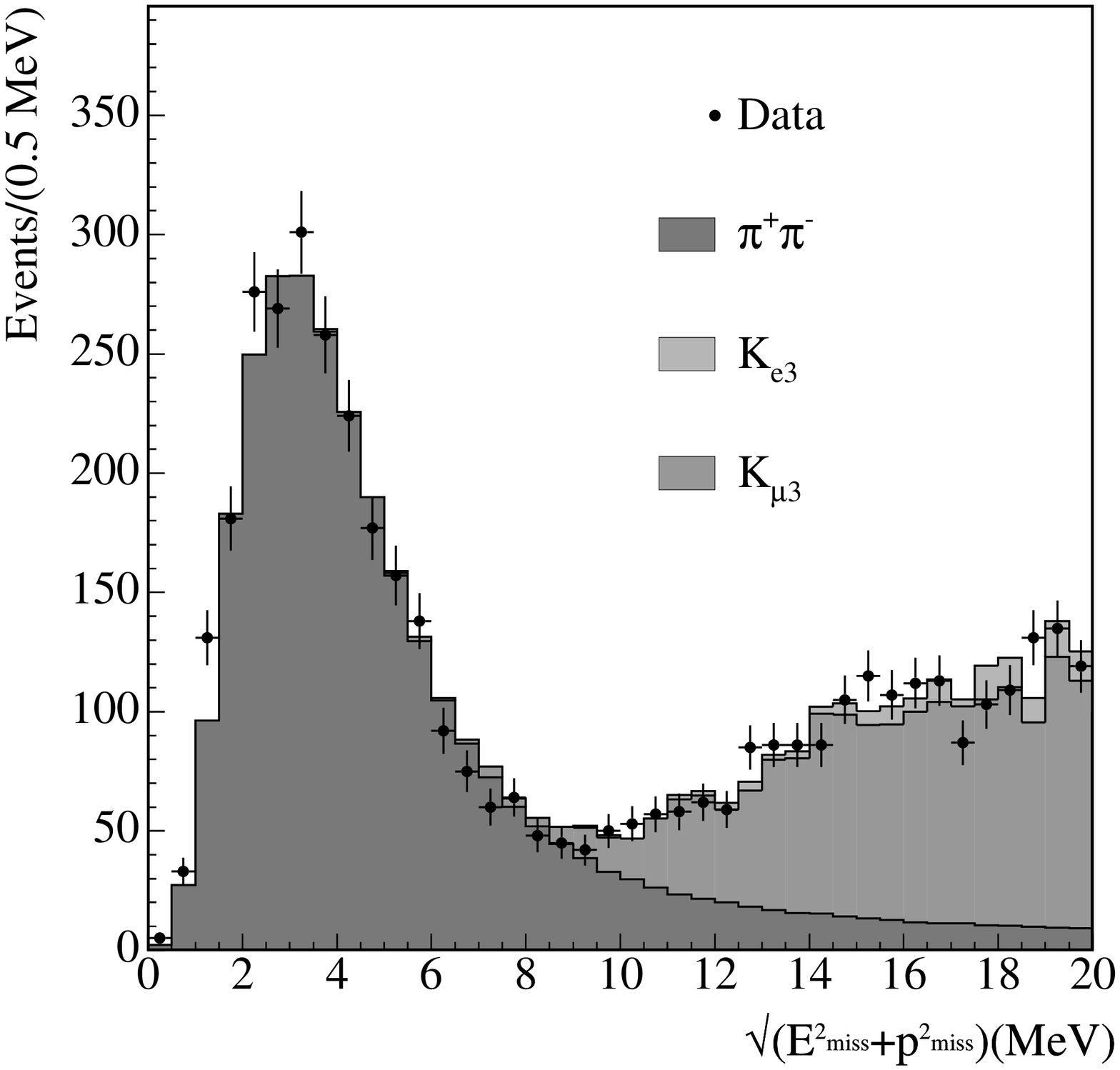}
\caption{Distribution of $\sqrt{E^2_{miss}+|\vec{p}_{miss}|^2}$, with fit to
MC distributions for different decay channels. Left, full fit range; right, signal region.}
\label{fig:klpp}
\end{figure}
The result of this fit is shown for a fraction of the whole data sample
in \fig{fig:klpp};
the signal region is expanded in the same figure, right.
The systematic uncertainty in the result is dominated by distortions introduced by MC in the
simulation
of signal and background shapes, which are particularly sensitive to momentum resolution.
We obtain~\cite{klpipi}:
\begin{equation}
BR(\kl\ \to\ \pic) = (1.964 \pm 0.012_{\rm stat} \pm 0.017_{\rm syst})\times 10^{-3}
\label{eq:klpp}
\end{equation}
 This branching ratio measurement is fully inclusive of
final-state radiation, and includes both the inner bremsstrahlung and the CP-conserving
direct emission (DE) components.
Using our measurements of BR($\ks\to \pic$) and $\tau_{L}$,
  the value of  $\tau_{S}$ from the
 PDG\cite{pdg06}, and subtracting the contribution of the $CP$-conserving DE
 process~\cite{Ramberg} from the inclusive measurement of BR($\kl\to \pic$), we obtain
 $|\eta_{+-}| = (2.219 \pm 0.013)\times 10^{-3}$.
Finally, using the world average of
$\rm{Re}(\epsilon'/\epsilon)=(1.67 \pm 0.26)\times 10^{-3}$
 and assuming $\arg\epsilon'=\arg\epsilon$, we obtain
$|\epsilon| = (2.216 \pm 0.013)\times 10^{-3}$.
 The value of $|\epsilon|$ can also be
 predicted from the measurement of the $CP$-conserving observables
$\Delta M(B_d)$, $\Delta M(B_s)$,
 $V_{ub}$, and  $V_{cb}$.
This is particularly interesting to test the mechanism of the $CP$ violation
in the SM.
 For this purpose, we use the prediction
 $|\epsilon| = (2.875 \pm 0.455)\times 10^{-3}$ obtained in \cite{UTfit}.
 No significant deviation from the SM prediction is
 observed. Notice however that, due to the large uncertainties on the computation
 of the hadronic matrix element corresponding to the \PK-\PKbar\ mixing,
 the theoretical error is much larger than the experimental one.
\subsubsection{\kl $\rightarrow \gamma \gamma $}
\label{sec:klggover3p0}
\begin{figure}[htb]
\centering
\includegraphics[width=0.5\textwidth]{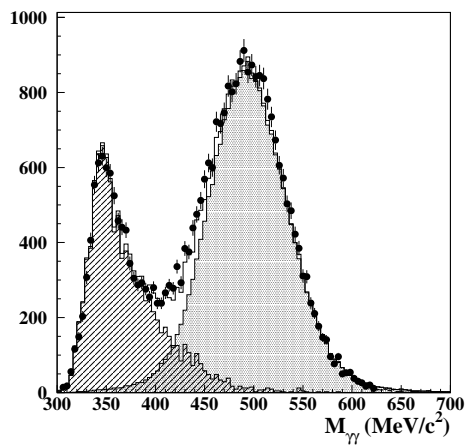}\kern1cm
\caption{Distribution of $M_{\gam\gam}$ after the selection cuts; dots are
data, histograms are MC shapes for signal ($M_{\gam\gam}\sim m_K$) and
background ($M_{\gam\gam}<m_K$).}
\label{fig:klgg}
\end{figure}

The $\kl \to \gam\gam$ decay provides interesting tests of ChPT.
This is because in the $SU(3)$ limit it takes contribution from
$\mathcal{O}(p^6)$ terms only~\cite{klggtheo}. Moreover,
a precise measurement of $\kl \to \gam \gam$ is needed to compute
the absorptive part of the $\kl \to \mu^+\mu^-$ decay rate. This
could be used to constrain the dispersive part of the same decay,
which in turn is related to the parameter $V_{td}$ of the CKM matrix.
From $\sim 300$~pb$^{-1}$ of integrated luminosity,
we measure the ratio
$\Gamma(\kl \to \gam\gam) / \Gamma(\kl \to 3 \po)$.
Using $\ks \to \pic$ as a tagging decay,
we select the signal by requiring two energetic photons belonging to
a single vertex along the \kl\ line of flight. A good rejection power over the
$\kl \to 3\po$ background is achieved by profiting of the two-body decay
kinematics. After the selection, we count the signal by fitting the
two-photon invariant mass, $M_{\gam\gam}$, with a combination of MC
distribution for signal and background (\fig{fig:klgg}).
From the previous fit, we count $22\,185 \pm 170$ signal events, with
efficiency $\sim 81\%$.
Finally, we obtain~\cite{klgg}
\begin{equation}
\frac{\Gamma(\kl \to \gam\gam)}{\Gamma(\kl \to 3 \po)} =
(2.79 \pm 0.02_{\rm stat} \pm 0.02_{\rm syst})\times 10^{-3}.
\label{eq:klgg}
\end{equation}
Using our result for $\BR(\kl\ \to 3 \po)$ decay,
a decay width $\Gamma$(\kl\ $\rightarrow \gamma\gamma)= (7.2 \pm 0.1)\times
10^{-12}$ eV is evaluated. This value is in reasonable agreement with
$\mathcal{O}(p^6)$ ChPT predictions,
provided the value of the pseudoscalar mixing angle
is close to our measurement of
$\theta_P =\varphi_{P}-\arctan \sqrt 2 =
(-13.3\pm0.3_{\text{stat}}\pm0.7_{\text{sys}}\pm0.6_{\text{th}})^{\circ}$
(section~\ref{sec:etaprimemix}).
%
%
\subsection{\kpm\ decays}
\label{sec:kchdec}
The charged kaons, \kpm, decay mostly into $\mu^\pm\nu$ ($K_{\mu2}$) and
$\pi^\pm\po$ ($K_{\pi2}$), about 60\% and 20\% of all decays respectively,
and the semileptonic modes account for about 8\%.
In the following the measurement of the main branching ratios of \kpm\ done
at KLOE will be reported. These are absolute branching ratios determined
tagging with two-body decays, easily identified as peaks in the p$^{\ast}$
distribution (sect.~\ref{sec:kpmbeam}). A residual dependency of the tagging
criteria on the decay mode of the tagged kaon is still present and
it is accounted for in the final branching ratio evaluation.
The few per mil accuracy on
BR($K_{\mu2}$) and BR($K_{\pi2}$) has been obtained tagging with \km\ and
using \kp\ for signal search, neglecting corrections to the BRs from
nuclear interactions (NI) of the kaon ($\sigma_{NI}(K^+) \sim
\sigma_{NI}(K^-) / 10^{2}$). All the measurements are inclusive of
final-state radiation. The results on BR($K^\pm_{\ell3}$)
and \kpm\ lifetime $\tau_{\pm}$ (\eq{eq:taukpm}) have been used to
evaluate \Vus\ (sect.~\ref{sec:fvus}) and,
together with the results from neutral kaons, to test lepton universality (sect.~\ref{sec:univer}).
From the BR($K^+_{\mu2}$) and $\tau_{\pm}$ values we have measured
$\Vus/\Vud$ (sect.~\ref{sec:vudvus}) and combining this result with the
values of \Vus\ and \Vud\ we have tested CKM unitarity (sect.~\ref{sec:ckm}).
Leptonic two-body decays $K^+_{\mu2}$ and $K^+_{e2}$ allow us to put bounds on New
Physics expected in Super-Symmetric extensions of the SM
(secs. \ref{sec:ke2} and \ref{sec:newphys}).

\subsubsection{$K^+\rightarrow\mu^+\nu(\gamma)$}
\label{sec:kmu2}
\begin{figure}[htb]
\begin{center}
\includegraphics[width=0.98\textwidth]{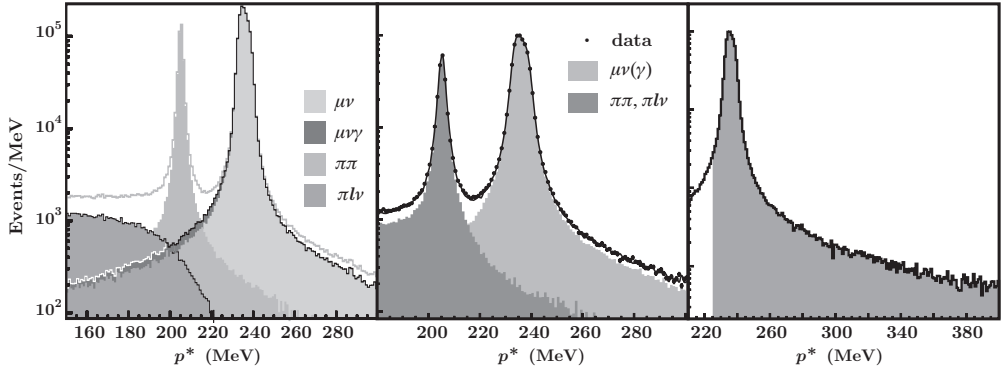}
\caption{Left. Monte Carlo spectra of $p^*$ for \kp\ decays, showing contributions from various channels. Center. Distribution of $p^*$. Right. Distribution of $p^*$ after background subtraction. The shaded area is used to count $\kp\to\mu^+\nu$ events.}
\label{kcmu}
\end{center}
\end{figure}

The KLOE measurement of this absolute BR \cite{kmu2}
is based on the use of $\km\to\mu^-\bar{\nu}$ decays for event tagging.
%
%
This choice minimizes possible interference in track reconstruction
and cluster association of the tagging \km\ and the tagged \kp\
decays.
The large number of $K_{\mu 2}$ decays allows for a statistical precision of \ab0.1\%,
while setting aside a generous sample for systematic studies.
For all the tagged events, the search for a positive kaon moving outwards
from the IP in the DC with momentum $70< p_{K^\pm}<130$ Mev/c is performed.
Then kaon decay vertices in
the DC volume, $40 < R_V < 150$ cm, are selected.
The number of signal counts is extracted from the distribution of $p^\ast$,
 the momentum of the charged decay particle evaluated in the kaon reference frame and in the
pion mass hypothesis, between $p^*_{min}$ = 225 MeV/c and  $p^*_{Max}$ = 400 MeV/c.
This distribution is shown in \fig{kcmu} left.
The spectrum shows the contamination, for a total amount of 2\%,
from $K^+_{\pi 2}$ and  $K^{+} \to \pi^0 l^{+}\nu$  events.
%
%
%
All the background sources have one neutral pion
in the final state and therefore can be identified by the observation
 of two photon's clusters
of equal time and compatible with the $\pi^0$ mass.
This selection provides the sample to evaluate
on data the p* distribution of the background.
Also the distribution for the signal events can be obtained from data
control sample selected using EMC information only.
This distribution is used together with the shape of the
background sources to fit the overall spectrum and to perform
background subtraction, \fig{kcmu}, center.
\noindent
\Fig{kcmu} right shows the spectrum after background
subtraction.
The branching ratio is then obtained dividing the signal counts by
the number of tagged events and correcting for the reconstruction and selection efficiency.
%
%
The efficiency has been determined directly on data using a
control sample of $K_{\mu2}$ events selected exploiting their
signature in the EMC.
The control sample is constituted by events with $K^-_{\mu2(\gamma)}$ providing
the tag and $K^+_{\mu2(\gamma)}$ selected using EMC information only.
This criterium is largely independent from the selection procedure based
on DC information that have been used for obtaining the signal sample.

%
%

In a sample of four million tagged events,
KLOE finds \ab865,000 signal events
 giving:
\begin{equation}
{\rm BR}(\kp\to\mu^+\nu(\gam)) = 0.6366\pm0.0009_{\rm stat}\pm0.0015_{\rm
  syst}.
\label{eq:kpmkmu2}
\end{equation}
This measurement is fully inclusive of
final-state radiation and has a 0.27\% uncertainty.

\subsubsection{$K^+\rightarrow\pi^+\pi^0(\gamma)$}
\label{sec:kpi2}
\begin{figure}[htb]
\centering
\includegraphics[width=0.4\textwidth]{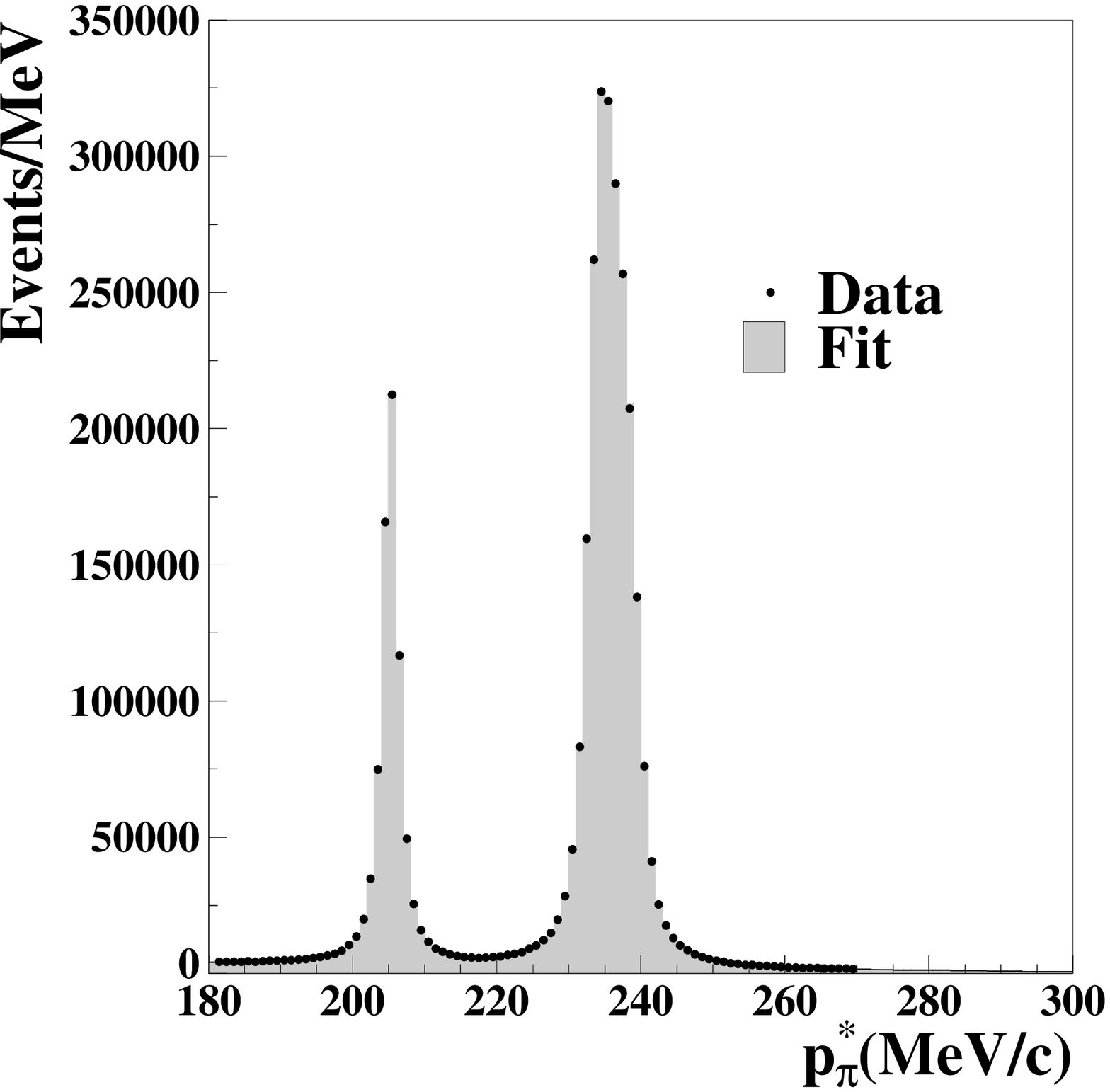}\kern1cm
\includegraphics[width=0.37\textwidth]{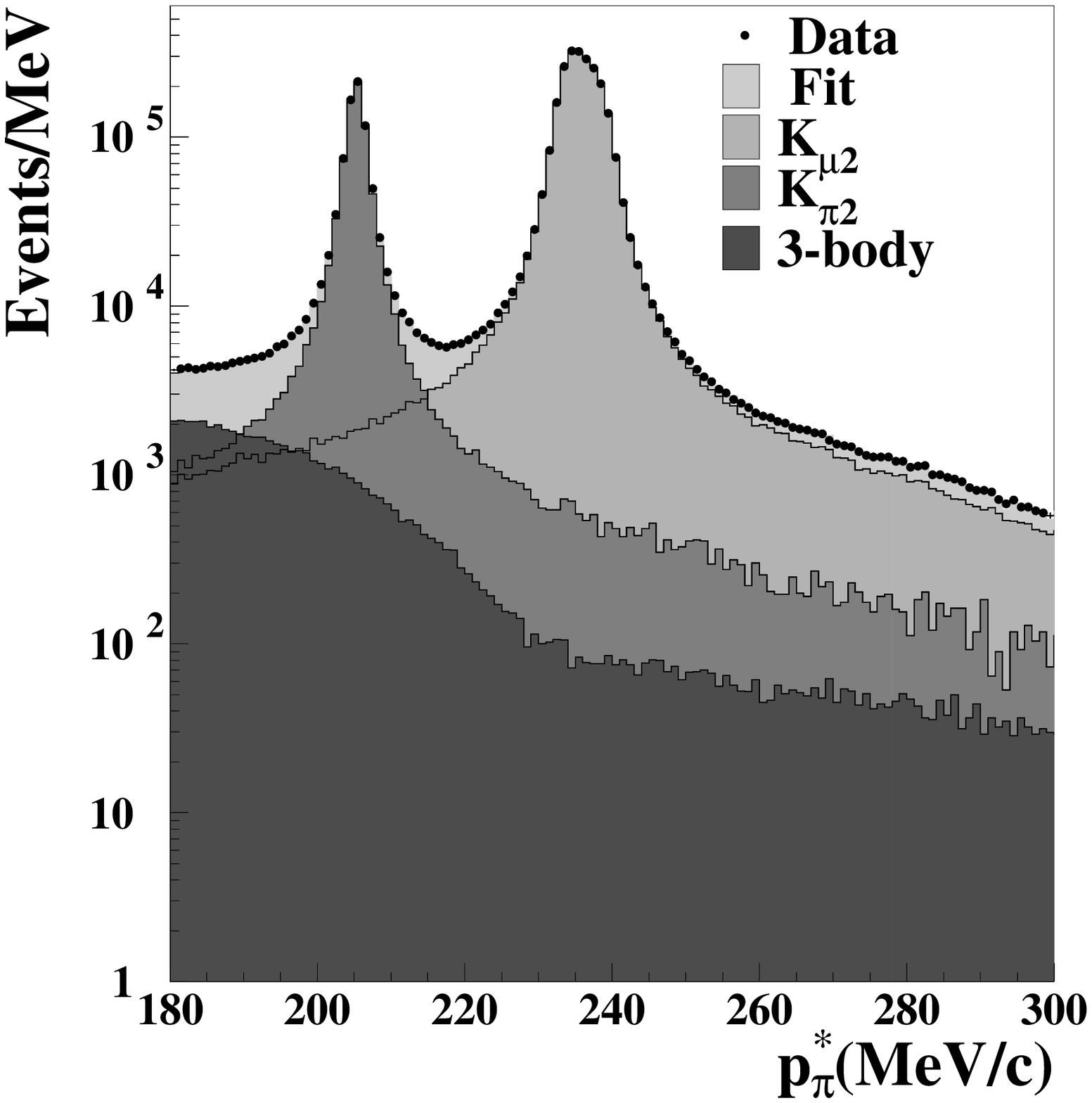}
\caption{Fit of the $p^\ast$ distribution. Left: black dots are Data and
grey histogram is the fit output. Right: the three contributions used to fit the Data
are shown: $K_{\mu 2}$, $K_{\pi 2}$ and {\it three-body} decays.}
\label{fig:fit_pstar}
\end{figure}

The measurement of the absolute BR of this decay,
inclusive of radiation, is performed tagging with both $K^-_{\mu2}$ and
 $K^-_{\pi2}$ decays, providing a pure $K^+$ beam
for signal search.
The selection of $K^+_{\pi2}(\gamma)$ decays uses DC information only,
with $K^+$ and its decay vertex selected
as done in the $K^+\rightarrow\mu^+\nu(\gamma)$ analysis.
Loose cuts on $p^{\ast}$ and on the momentum difference between the
kaon and the charged secondary track are applied to reject $K\to 3 \pi$
decays and \kpm\ split tracks.
The signal count is then extracted from the fit
of the $p^{\ast}$ distribution in the window starting from
$p^{\ast}_{cut}$=180 MeV/c (see \fig{fig:fit_pstar}).
This spectrum exhibits two peaks, the first at about $236$ MeV/c from
$K^+_{\mu2}$ decays
and the second at about $205$ MeV/c from $K^+_{\pi2}$ decays.
 Lower $p^{\ast}$ values are due to three body decays.
The momenta of the charged secondaries produced in the kaon decay
have been evaluated in the kaon rest frame using the pion mass hypothesis.
Therefore the $K_{\pi 2}$ peak
appears to be symmetric while the $K_{\mu 2}$ one is asymmetric due
 to the incorrect mass
hypothesis used.

The fit to the $p^{\ast}$ distribution
\noindent
is done using the following three contributions: $K_{\mu 2}$, $K_{\pi 2}$
and three-body decays.
For $K_{\mu 2}$ and $K_{\pi 2}$ components we use shapes obtained from data control samples
selected using EMC information only. For the contribution from three-body
decays we use the MC distribution.
\Fig{fig:fit_pstar} left shows the result of the fit of the
$p^\ast$ distribution performed on the $K^-_{\mu2}$-tagged data sample.
\Fig{fig:fit_pstar} right shows the three contributions: $K_{\pi2}$, $K_{\mu2}$ and  three-body decays.
Using a total number of 12,113,686 $K^-_{\mu2}$-tagged events
we obtain $818,347 \pm 1,912$ signal counts.
From the sample of 9,352,915 $K^-_{\pi2}$-tagged events we get
 $621,612 \pm 1,678$ signal counts.
The reconstruction and selection efficiency has been evaluated on data from
a control sample selected using EMC information only,
to avoid correlation with the DC driven sample selection.
Once a tagging $K^-_{\mu2}$ decay has been identified,
the control sample selection is given by \kp\ decays with a $\pi^0$ in the final state identified
via the reconstruction of $\pi^0\to\gamma\gamma$ decays.
Corrections to the efficiency accounting for possible distortions induced by the
control sample selection have been evaluated using MC simulation.
The final efficiency evaluation is strongly related to the charged kaon
lifetime $\tau_{\pm}$ via the geometrical acceptance.
The BR depends on $\tau_{\pm}$ as:
 BR/BR$^{(0)}$=1-0.0395 ns$^{-1}$ ($\tau_{\pm}$-$\tau_{\pm}^{(0)}$)
with reference value $\tau_{\pm}^{(0)}=12.385\pm0.024$~\cite{pdg06} used in the final BR evaluation.

The weighted average, accounting for correlations, of the absolute branching ratios obtained
using events tagged by $K^{-}_{\mu 2}$ and $K^{-}_{\pi 2}$ decays is a measurement with
4.6 per mil accuracy~\cite{kpmkp2}:
\begin{equation}
 BR (K^+\rightarrow \pi^+ \pi^0 (\gamma)) =
   0.2065 \pm 0.0005_{\rm stat} \pm 0.0008_{\rm syst}.
\label{eq:kpmkp2}
\end{equation}
The value reported is shifted by -1.3\% (\ab2$\sigma$) with respect to
the PDG06 fit value~\cite{pdg06} and has a 20\% improvement in the fractional accuracy.
\subsubsection{\kpm\ semileptonic decays}
\label{sec:kpml3}
\begin{figure}[htb]
\begin{center}
\includegraphics[width=0.5\textwidth]{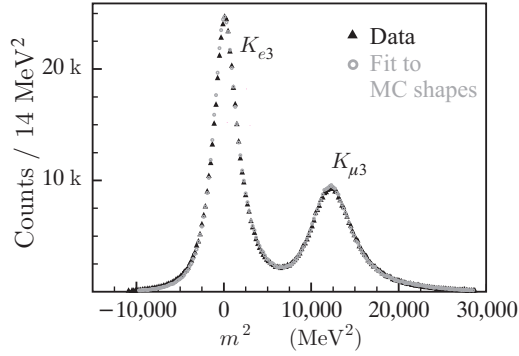}
\caption{Distribution of $m^2_{lept}$, for events selected as $K^\pm_{\ell3}$ decays. Also shown is a fit to MC distributions for signal and background.}
\label{kcsl}
\end{center}
\end{figure}

The values of BR($\kpm\to\po e^\pm\nu(\gam)$) and BR($\kpm\to\po\mu^\pm\nu(\gam)$) are
each determined from four independent measurements:
\kp\ and \km\ decays tagged by $K\to\mu\nu$ and $K\to\pi\po$. In the
analyzed data set about 60 million tag decays were identified.
The signal selection asks for a decay vertex in the DC volume.
After removal of $K_{\pi2}$  decays from the signal sample exploiting kinematics,
the \po\ is reconstructed from the two \gam s
which provide the kaon decay time $t^{decay}_{\pi^0}$.
This sample is composed mainly of semileptonic decays, residual two-body
and $K^{\pm} \to \pi^{\pm} \pi^{+}\pi^{-}$ decays.
To reject $K_{\pi 2}$ events with an early $\pi^{\pm} \to \mu^{\pm}\nu$ decay,
we cut on the lepton momentum evaluated  in the
center of mass of the $\pi^{\pm}$ using the muon mass hypothesis.
Then to isolate $K_{e3}$ and $K_{\mu3}$ decays, the lepton is
identified using a time of flight technique. Requiring the charged decay track to point
to an energy deposit in EMC, the kaon decay time is given
by $t^{decay}_{lept}=t_{lept}-\frac{L_{lept}}{p_{lept}c}\sqrt{p^{2}_{lept}+m^2_{lept}}$.
Measuring the time of the cluster associated to the charged decay particle
$t_{lept}$, its momentum $p_{lept}$ and track length $L_{lept}$,
we can evaluate $m^2_{lept}$ by imposing $t^{decay}_{\pi^0}=t^{decay}_{lept}$.
The $m^2_{lept}$ distribution is shown in \fig{kcsl}:
the $K_{e3}$ component shows as a narrow peak around zero
 while the $K_{\mu3}$ component is the peak around the $m^2_{\mu}$ value.
The signal count is extracted from a constrained likelihood fit to this distribution,
using a linear combination of $K_{e3}$ and $K_{\mu3}$ shapes and of background
sources all taken from MC simulation.
The fit result superimposed to the data distribution is also shown in \fig{kcsl}.
%
%
%
%
The reconstruction and selection efficiency has been measured using MC simulation and corrected for relevant
data/MC differences. Control samples have been selected to measure tracking
and calorimeter clustering efficiency as a function of a suitable set of variables to perform data/MC
comparisons.

The BRs depend on the value used for $\tau_\pm$ as
\begin{equation}
\mbox{BR}(\kpm \to f) / \mbox{BR}^{(0)}(\kpm\to f) = 1 - 0.0364~\mbox{ns}^{-1}\:\left(\tau_\pm
  - \tau_\pm^{(0)}\right).
\label{eq:fvdepch}
\end{equation}
Using the current world average value of the
$K^{\pm}$ lifetime $\tau_{\pm}^{(0)}$
and averaging over the available samples,
we get\cite{kpmkl3}:
\begin{eqnarray}
{\rm BR^{(0)}}(K^\pm_{e3}) & = & (0.04965\pm0.00038_{\rm stat}\pm0.00037_{\rm syst})  \nonumber\\
{\rm BR^{(0)}}(K^\pm_{\mu3}) & = & (0.03233\pm0.00029_{\rm stat}\pm0.00026_{\rm syst}).
\label{eq:kpml3}
\end{eqnarray}
These results, completely inclusive of final-state radiation,
 have a fractional accuracy of 1.1$\%$ for the $K_{e3}$
and of 1.2$\%$ for the $K_{\mu3}$ decays. The total errors are
correlated with 0.627 and do not include any contribution from the
uncertainty on $\tau_{\pm}$.
Inserting in \eq{eq:fvdepch} the KLOE result for $\tau_\pm$ from
\eq{eq:taukpm}, we obtain the values of \kpm\ semileptonic BRs reported
 in table \ref{tab:kpm}, which have been used for our evaluation of \Vus.
\begin{table}
\begin{center}
\begin{tabular}{lcccc}
\hline
Parameter & Value & \multicolumn{3}{c}{Correlation coefficients} \\
\hline
BR$(K_{e3})$     &  0.04972(53)   &   1     &         &    \\
BR$(K_{\mu3})$   &  0.03237(39)   & $+0.63$ &   1     &    \\
$\tau_\pm$      & 12.347(30)~ns  & $-0.10$ & $-0.09$ &  1 \\
\hline
\end{tabular}
\end{center}
\caption{KLOE measurements of \kpm\ semileptonic decays and lifetime.}
\label{tab:kpm}
\end{table}
%
\subsubsection{$K^{\pm}\rightarrow\pi^{\pm}\pi^0\pi^0$($\tau'$)}
\label{sec:kpmtaup}
This measurement is based on a sample of $\sim 3.3 \times 10^{7}$ tagged
 \kpm\ \cite{plbtauprime}, using $K^{\pm}_{\mu 2}$ and $K^{\pm}_{\pi 2}$ decays.
The signal selection requires a decay vertex in the
DC volume and at least four energy deposits in EMC, on-time with respect to
 the decay vertex, and $p^{\ast}< 135$ MeV.
Finally we require the sum of the energies of the most energetic
photons to be  $\sum E_{i} < 450 $ MeV.
The background sources are $K^{\pm}_{\pi 2}$,$K^{\pm}_{e3}$ and
radiative decays like $K^{\pm}\rightarrow\pi^{\pm}\pi^0 \gamma$,
in which a spurious cluster in EMC is paired with the cluster of the
radiated photon mimicking the second $\pi^{0}$.
The total background fraction has been estimated from MC to be less than
 1\%. The overall efficiency is given by the product of the
efficiency to reconstruct  the kaon track, the track of the charged secondary and the
decay vertex, times the efficiency of having at least 4 clusters fulfilling
 the energy and timing requirements. Each component has been measured
 separately for each tag and charge. Data control samples from EMC
 selection have been used for the evaluation of efficiencies involving DC
 variables and viceversa. Corrections evaluated using MC simulation account
 for possible contamination of the control samples.

We get 52253 and 30798 $\tau'$ decays from 1.9925$\times 10^{7}$ events
tagged by $K^{\pm}_{\mu 2}$ and 1.2753$\times 10^{7}$ tagged by
$K^{\pm}_{\pi 2}$ decays respectively. The final branching is:
\begin{eqnarray}
 {\rm BR} (K^{\pm}\rightarrow\pi^{\pm}\pi^0\pi^0) = (1.763 \pm 0.013_{\rm stat} \pm 0.022_{\rm syst})\%
\label{eq:kpmtaup}
\end{eqnarray}
%
\subsubsection{$R_{\rm K}=$\rk }
\label{sec:ke2}
For this measurement
we decided to perform a ``direct search'' for \ken\ and \kmn\ decays
without tagging, to keep the statistical uncertainty on the number of \ken\ counts below 1\%.
%
%
The presence of a one-prong decay vertex in the DC volume with a secondary
charged decay track with relatively high momentum (180$\div$270~MeV)
is required.
To distinguish \ken\ and \kmn\ decays from the background we use the lepton
mass squared $M_\mathrm{lep}^2$ obtained from the momenta of the kaon and the
charged decay particle, assuming zero neutrino mass. This provides a clean
identification of the \kmn\ sample, while to improve background rejection in
selecting \ken\ decays we need to use the information from the EMC.
A particle identification based on the
asymmetry of the energy deposits
 between the EMC planes,
on the spread $E_\mathrm{\rm R.M.S.}$ of these energy deposits and on the
position of the plane with the maximum energy release is used.

\begin{figure}[htb]
\begin{center}
    \includegraphics[width=0.36\textwidth]{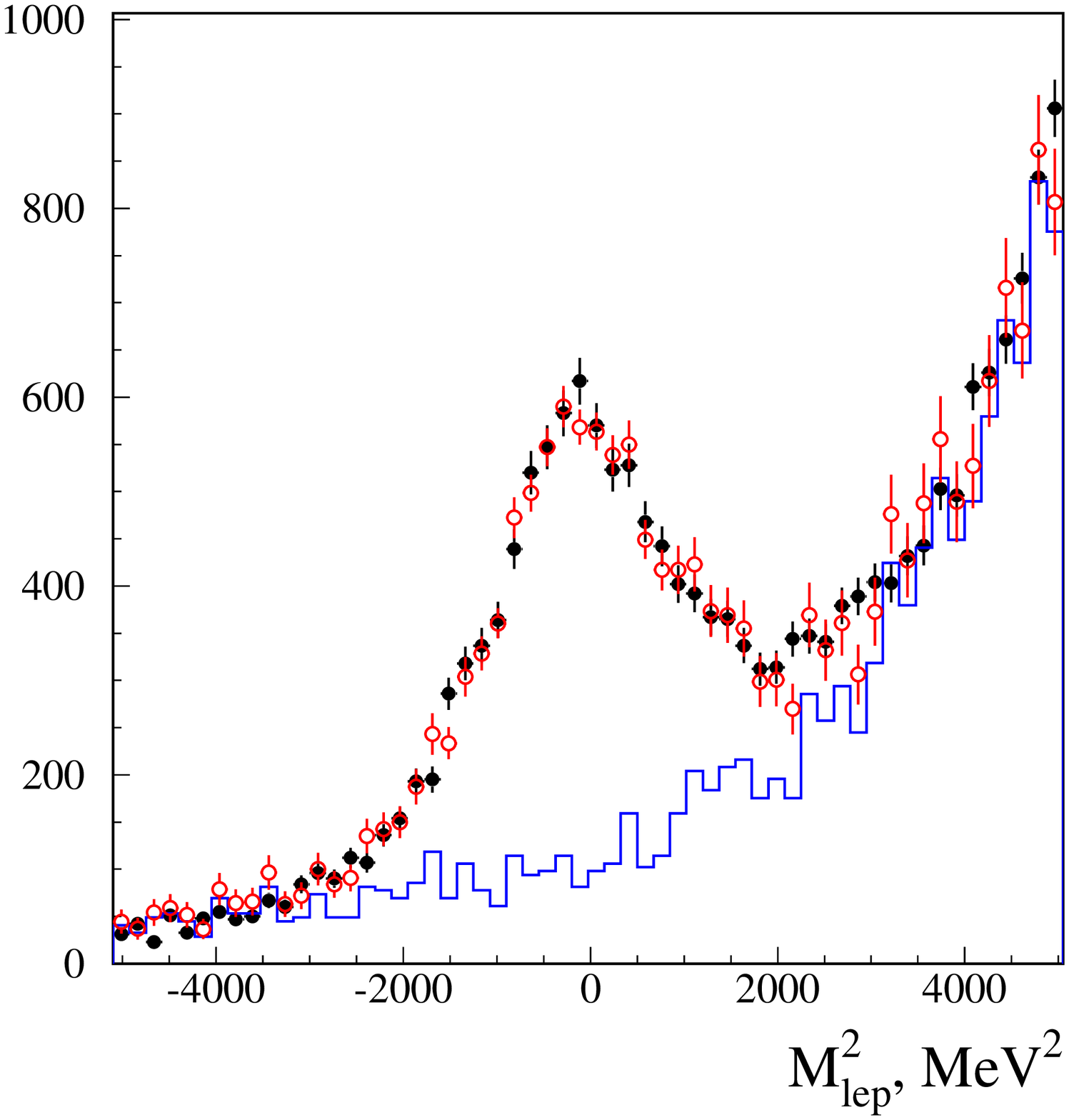}\kern1cm
     \includegraphics[width=0.4\textwidth]{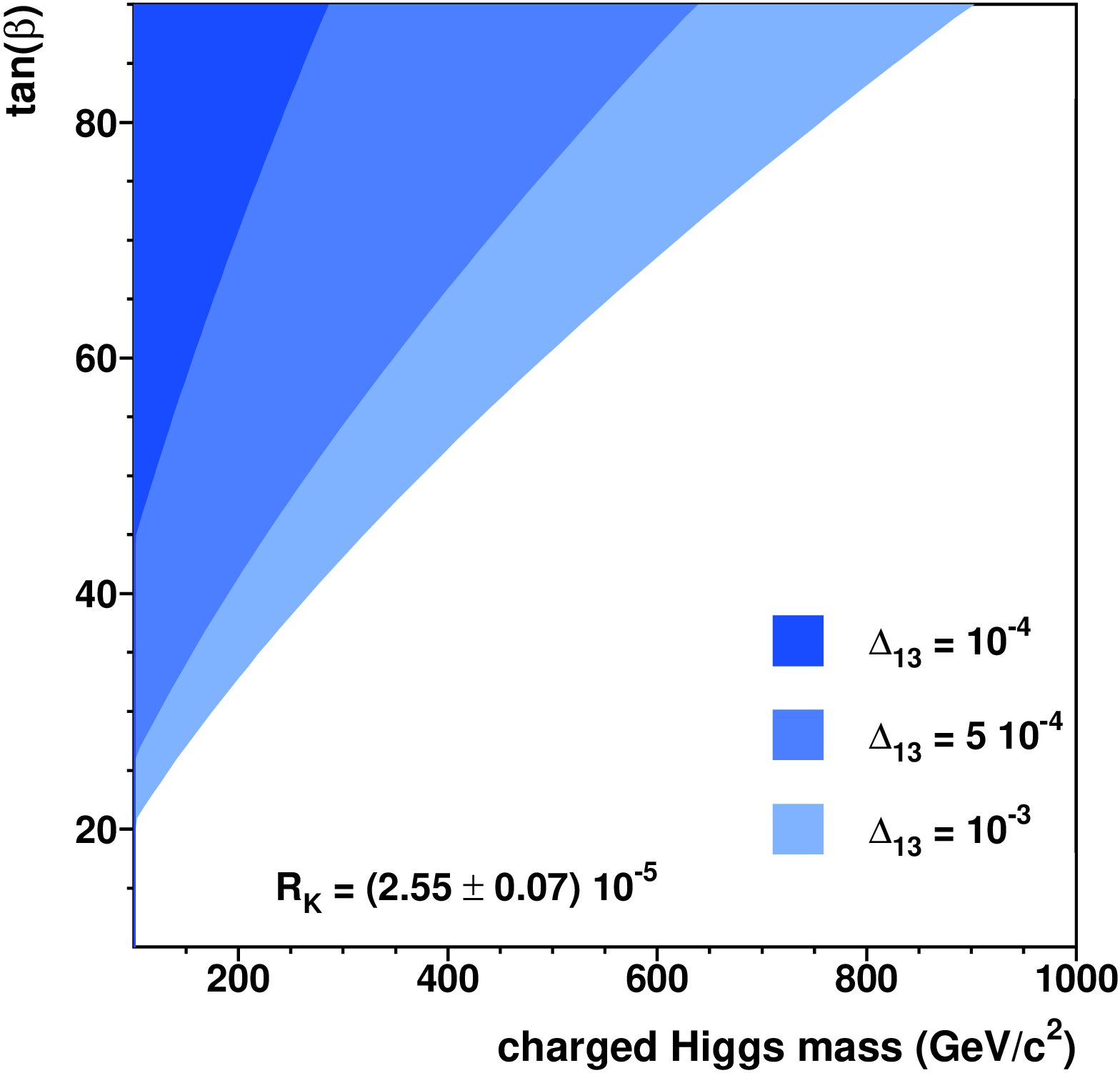}
\caption{Left: Distribution of the squared lepton mass of the
         charged secondary track $M^2_\mathrm{lep}$:
         filled dots are data, open dots the result from the
         maximum-likelihood fit.
         Right: Exclusion limits at $95\%$ CL on $\tan \beta$ and the
         charged Higgs mass $m_{H^\pm}$ from $R_{\rm K}$
         for different values of $\Delta_{13}$.}
\label{ke2:plot}
\end{center}
\end{figure}

The number of signal events is extracted from
a likelihood fit to the two-dimensional $E_\mathrm{\rm R.M.S.}$ vs
$M_\mathrm{lep}^2$ distribution, using MC shapes for signal and
background. The shape for $K\to e\nu(\gamma)$ have been evaluated
considering both the Inner Bremsstrahlung (IB) and Direct Emission (DE) contributions.
For signal count only events with photon energy in the kaon rest frame
$E^{\ast}_{\gamma}<20$~MeV have been considered.
With this choice the contribution from DE is negligible and we measure the IB contribution,
the only one entering in the SM prediction of $R_{\rm K}$.
The number of signal events obtained from the fit is $N_{Ke2}=8090\pm156$
and the projection of the fit result on the $M_\mathrm{lep}^2$ axes is compared to data in
 \fig{ke2:plot} left.
The number of \kmn\ events in the same data set is extracted from
a similar fit to the $M_\mathrm{lep}^2$ distribution,
where no PID cuts are applied. The fraction of
background events under the muon peak is estimated from MC to be less than one per mil.
The number of \kmn\ events in the same data set is 499\,251\,584$\pm$35403.
Using the number of observed \ken\ and \kmn\ events, we get the preliminary
result \cite{ke2}:
\begin{equation}
R_{\rm K} = (2.55\pm0.05\pm0.05)\times10^{-5}.
\end{equation}
This value is compatible within errors with the very precise SM prediction
$R_{\rm K} = (2.472\pm0.001)\times10^{-5}$.
Recently it has been pointed out that, in a SUSY framework, sizeable violations of
lepton universality can be expected in $K_{l2}$ decays \cite{masiero} from
the couplings to the charged Higgs boson $H^+$.
With this Yukawa coupling, the dominant non-SM contribution
 to $R_{\rm K}$ modifies this ratio to:
\be
R_{\rm K}^{\text{LFV}} \approx R_{\rm K}^{\text{SM}} \left[ 1 + \left( \frac{m_K^4}{m_{H^\pm}^4} \right) \left( \frac{m_\tau^2}{m_e^2} \right) |\Delta_{13}|^2 \tan^6 \beta \right].
\label{eqn:susy}
\ee
With the lepton flavor violating term $\Delta_{13}$ of the order
 of $10^{-4}-10^{-3}$, as expected from neutrino mixing, and
moderately large values of $\tan \beta$ and $m_{H^\pm}$, SUSY contributions may therefore
enhance $R_{\rm K}$ up to a few percent.
\Fig{ke2:plot} right shows the
strong constraints on $\tan
 \beta$ and $m_{H^\pm}$ given by the KLOE $R_{\rm K}$ result.
For a moderate value of $\Delta_{13} \approx 5 \times 10^{-4}$, the region
$\tan \beta > 50$ is excluded for charged Higgs masses up to 1000~GeV/$c^2$ at 95\% CL.
\subsubsection{ Fit to \kpm\ BRs.}
\label{sec:kpmbrfit}
We fit the six largest \kpm\ BRs and the lifetime $\tau$
using our measurements of $\tau_\pm$ (\eq{eq:taukpm}), $\BR(K^+_{\pi2})$
(\eq{eq:kpmkp2}),
$\BR(K^+_{\mu2})$ (\eq{eq:kpmkmu2}), $\BR(K^\pm_{\rm l3})$ (\eq{eq:kpml3}) and
$\BR(K^{\pm}\to\pi^{\pm}\pi^0\pi^0)$ (\eq{eq:kpmtaup}), with their dependence on $\tau_\pm$, together with
$\BR(\kpm\to\pi^\pm\pi^+\pi^-)=0.0550\pm0.0010$ from the
PDG04 average\footnote{PDG '06 gives the result of their constrained fit but not the average of the data}
~\cite{pdg04}, with the sum of the BRs constrained to unity.
The fit results~\cite{kpmkp2}, with  $\chi^2/{\rm ndf}=0.59/1$ (CL=44\%),
are shown in Table~\ref{tab:kpmbrfit}.
\begin{table}[htb]
\begin{tabular}{llccccccc}
\hline
Parameter & Value & \multicolumn{6}{c}{Correlation coefficients} \\
\hline
$\BR(K_{\mu2})$          & 0.6376(12)       &         &         &         &         &         &            \\
$\BR(K_{\pi2})$          & 0.2071(9)        & $+0.48$ &         &         &         &         &            \\
$\BR(\pi^\pm\pi^+\pi^-)$ & 0.0553(9)        & $-0.48$ & $+0.21$ &         &         &         &            \\
$\BR(K_{e3})$            & 0.0498(5)        & $+0.37$ & $-0.13$ & $+0.16$ &         &         &            \\
$\BR(K_{\mu3})$          & 0.0324(4)        & $+0.34$ & $-0.12$ & $+0.15$ & $+0.58$ &         &            \\
$\BR(\pi^\pm\pi^0\pi^0)$ & 0.01765(25)      & $-0.11$ & $+0.05$ & $-0.05$ & $+0.04$ & $+0.04$ &            \\
$\tau_\pm$ (ns)             & 12.344(29)      & $-0.15$ & $-0.21$ & $-0.07$ & $-0.06$ & $-0.05$ & $-0.015$   \\
\hline
\end{tabular}
\caption{Results of the fit to \kpm\ BRs.}
\label{tab:kpmbrfit}
\end{table}
We can also evaluate $\BR(\kpm\to\pi^\pm\pi^+\pi^-)$ by using our measurements of the above listed BRs and
imposing the constraint $\sum \BR(\kpm\to f)=1$. With $\BR(K^+_{\mu2})=0.63660\pm0.00175$,
$\BR(K^{\pm}\to\pi^{\pm}\pi^0\pi^0)= 0.01763\pm0.00025$ and
$\BR(K^+_{\pi2})= 0.20681\pm0.00094$, $\BR(K^\pm_{e3})= 0.04972\pm0.00053$ and $\BR(K^\pm_{\mu3})=0.03237\pm0.00039$,
evaluated at $\tau_\pm$ equal to our measured value 12.347$\pm$0.030 ns,
we get $\BR(\kpm\to\pi^\pm\pi^+\pi^-)= 0.0568\pm0.0022$. This result is in agreement with the PDG04 average.
\section{$V_{us}$ and lepton universality}
\label{kloevus}
\subsection{Introduction}
\label{sec:vusintro}
While much emphasis is placed on the search for New Physics,
we still lack precise information on the validity of certain
aspects of the SM itself.
One of the postulates of the SM is the form of the
Lagrangian for the
charged-current weak interactions, which can be written as
\begin{equation}
{\cal L}_{\rm cc} = {g\over\sqrt2}W_\alpha^+(\rlap{\bf U}\raise2.2ex\hbox to.6em{\hrulefill}\kern.3em_L\,{\bf V}_{\rm CKM}\gamma^\alpha\,{\bf D}_L+\bar e_L\gamma^\alpha\nu_{e\,L}+\bar \mu_L\gamma^\alpha\nu_{\mu\,L}+\bar \tau_L\gamma^\alpha\nu_{\tau\,L})\ +\ \mbox{h. c.}
\end{equation}
Two important properties are evident: there is only one coupling constant
for leptons and quarks, and quarks are mixed by the Cabibbo-Kobayashi-Maskawa
matrix~\cite{Cab63,KM73}, ${\bf V}_{\rm CKM}$, which must be unitary.
The 4-fermion Fermi coupling constant $G_F$ is related to the gauge coupling $g$ by $G_{\rm F}=g^2/(4\sqrt2\,m_W^2)$.

Precise measurements of leptonic and semileptonic kaon decay rates can
provide information about lepton universality. Combined
with results from nuclear $\beta$ decay and pion decays, such measurements also
provide information about the unitarity of the mixing matrix. Ultimately they
tell us whether quarks and leptons do indeed carry the same
weak charge.
The universality of electron and muon interactions can be tested by
measuring the ratio $\Gamma(K\to\pi \mu \nu)/\Gamma(K\to\pi e \nu)$.
The partial rates $\Gamma(K\to\pi e \nu)$ and $\Gamma(K\to\pi \mu \nu)$
provide measurements of $g^4\abs{\Vus}^2$, which, combined with
$g^4\abs{\Vud}^2$ from nuclear $\beta$ decay and the muon decay rate,
test the unitarity condition
$\abs{\Vud}^2+\abs{\Vus}^2+\abs{V_{ub}}^2=1$.
We recall that in 1983 it was already known that $\abs{V_{ub}}^2<4\times10^{-5}$ \cite{Klopfenstein:1983nz}
and today \abs{V_{ub}}\up2\ab\pt1.5,-5, \cite{pdg06}. We will therefore ignore \abs{V_{ub}}\up2 in the following.
The ratio $\Gamma(K\to\mu \nu)/\Gamma(\pi\to\mu \nu)$ provides an
independent measurement of $\abs{\Vus}^2/\abs{\Vud}^2$.

To perform these tests at a meaningful level of accuracy,
radiative effects must be carefully corrected for.
Strong-interaction effects also introduce form factors, which must be
calculated from first principles, or measured whenever possible.
Usually, corrections for SU(2)- and SU(3)-breaking must also be applied.
Recently, advances in lattice calculations have begun to catch up with
experimental progress.
\subsection{Via semileptonic kaon decays }
\label{sec:vuskl3}
%
Assuming lepton universality the muon decay rate provides the value for the Fermi constant,
$G_F$=\pt1.16637(1),-5, GeV\up{-2}, \cite{pdg06}.
The semileptonic decay rates, fully inclusive of radiation, are given by
\begin{equation}
\Gamma(K_{\ell3(\gamma)}) =
\frac{C_K^2 G_F^2 m_K^5}{192\pi^3}\,S_{\rm EW}\,\abs{\Vus}^2\,f_+(0)^2\,
I_{K,\ell}\:(1 + \delta_K^{SU(2)} + \delta_{K\ell}^{\rm EM})^2.
\label{eq:Vus}
\end{equation}
In the above expression, the index $K$ denotes $\ko\to\pi^\pm$ and
$\kpm\to\po$
transitions, for which $C_K^2 =1$ and 1/2, respectively. $m_K$ is the appropriate kaon mass,
and $S_{\rm EW}$ is the universal short-distance electroweak correction \cite{Sir82:SEW}
\footnote{Often in literature the numerical factor 192 is replaced by 768
  for which the phase space is normalized to 1 for all final particles
  masses vanishing \cite{FranzMoul06}. The phase space integrals listed in sect.~\ref{sec:ff} are evaluated accordingly to
  the chosen convention.}.
Conventionally, $f_+(0) \equiv f_+^{K^0\pi^-}(0)$, and the mode
dependence is encoded in the $\delta$ terms:
the long-distance
electromagnetic (EM) corrections, which depend on the meson charges and lepton masses,
and the SU(2)-breaking corrections, which depend on the kaon charge \cite{C+02:Kl3rad}.
$I_{K,\ell}\:$ is the integral of the Dalitz plot density over the physical
region and includes $|\tilde f_{+,\,0}(t)|^2$ (sect.~\ref{sec:ff}). $I_{K,\ell}$ does not account for
 radiative effects, which are included in $\delta_{K\ell}^{\rm EM}$.

The experimental inputs to \eq{eq:Vus} are the semileptonic decay
rates (tables \ref{tab:klbr2}, \ref{tab:ksdata}, \ref{tab:kpm})
and the phase space integrals (table \ref{tab:ikl}), which have been discussed in previous sections.
$SU(2)-$breaking and $EM$ corrections which are used to extract \fVus\ are summarized in
table~\ref{tab:radcorr}. The $SU(2)-$breaking correction is evaluated with ChPT to $O(p^4)$, as described
in~\cite{C+02:Kl3rad}.
The long distance $EM$ corrections to the full inclusive decay rate are evaluated with ChPT to
$O(e^2p^2)$~\cite{C+02:Kl3rad}, and using low-energy constants from \cite{moussallam}. The quoted results
have been evaluated recently~\cite{neufeld}, and include
for the first time the $K_{\mu 3}$ channels for both neutral and charged kaons.
\begin{table}[htb]
\begin{center}
\begin{tabular}{lcc}
\hline
 Channel         & $\delta_K^{SU(2)}$  & $\delta_{K\ell}^{EM}$ \\
\hline
$K^0_{e3}$       &          0          &   0.57(15)\%       \\
$K^0_{\mu3}$     &          0          &   0.80(15)\%       \\
$K^{\pm}_{e3}$   &         2.36(22)\%  &   0.08(15)\%       \\
$K^{\pm}_{\mu3}$ &         2.36(22)\%  &   0.05(15)\%       \\
\hline
\end{tabular}
\caption{Summary of $SU(2)-$breaking and $EM$ corrections used to extract \fVus.}
\label{tab:radcorr}
\end{center}
\end{table}
%
%
%
\subsubsection{\fVus\ and \Vus }
\label{sec:fvus}
Using all of the experimental and theoretical inputs discussed above, the values of \fVus\ have been obtained
for $K_{Le3}$, $K_{L\mu3}$, $K_{Se3}$, $K^{\pm}_{e3}$, and $K^{\pm}_{\mu3}$ decay modes, as shown in table \ref{tab:vusf} and
in \fig{fig:f0vus}. It is worth noting that the only external experimental input to
this analysis is the \ks\ lifetime. All other experimental inputs are KLOE results.
\begin{table}[htb]
\begin{center}
\begin{tabular}{lcccccc}
\hline
Channel & \fVus & \multicolumn{5}{c}{Correlation coefficients}  \\
\hline
 $K_{Le3}$        & 0.2155(7)  & 1    &       &      &      &    \\
 $K_{L\mu3}$      & 0.2167(9)  & 0.28 &  1    &      &      &    \\
 $K_{Se3}$        & 0.2153(14) & 0.16 &  0.08 & 1    &      &    \\
 $K^{\pm}_{e3}$   & 0.2152(13) & 0.07 &  0.01 & 0.04 &  1   &    \\
 $K^{\pm}_{\mu3}$ & 0.2132(15) & 0.01 &  0.18 & 0.01 & 0.67 & 1  \\
\hline
\end{tabular}
\end{center}
\caption{KLOE results for \fVus.}
\label{tab:vusf}
\end{table}
\begin{figure}[htb]
\centering
\figb f0vus;4;
\caption{KLOE results for \fVus.\label{fig:f0vus}}
\end{figure}

The five different determinations have been averaged, taking into account all known correlations. We find~\cite{vusklo}

\begin{equation}
   \fVus = 0.2157 \pm 0.0006,
\label{eq:f0vusresult}
\end{equation}
with $\chi^2/{\rm ndf} = 7.0/4$ ( $13\%$).
To evaluate the reliability of the $SU(2)-$breaking correction, a comparison is made
between separate averages of \fVus\ for the neutral and the charged channels, which are  $0.2159(6)$ and
 $0.2145(13)$. With correlations taken into account, these values
agree within $1.1\sigma$. Alternatively, an experimental estimate of
$\delta_K^{SU(2)}$ is obtained by
comparing the neutral result
with the charged one evaluated without correcting for $SU(2)-$breaking.
 We get $\delta_{exp}^{SU(2)} = 1.67(62)\%$, which is in agreement with
the value estimated from theory (table \ref{tab:radcorr}).

\noindent
The previous  determination of \fVus\, with
$2.8$ per mil fractional accuracy, can be used together with
the \fo\ value from  \cite{L&R},
$\fo = 0.961(8)$, to extract $\Vus = 0.2245 \pm 0.0020$.
However, lattice evaluations of \fo\ are rapidly improving in precision. For example, the RBC and UKQCD
Collaborations have recently obtained $\fo  = 0.9644(49)$ from a lattice calculation with $2+1$ flavors of
dynamical domain-wall fermions \cite{rbcukqcd07:f0}. Using their value of \fo, our $K_{l3}$ results give
$\Vus = 0.2237(13)$.

\subsubsection{Lepton universality}
\label{sec:univer}
Comparison of the values of \fVus\ for $K_{e3}$ and $K_{\mu3 }$ modes provides a test of lepton
universality. Specifically,
\begin{equation}
r_{\mu e} = \frac{\left(\fVus\right)^2_{\mu3, exp} }{\left(\fVus\right)^2_{e3, exp}} =
\frac{\Gamma_{\mu 3}}{\Gamma_{e3}}\;
\frac{I_{e3}(1+\delta_{Ke})^2}{I_{\mu3}(1+\delta_{K\mu})^2},
\label{eq:leptuniv}
\end{equation}
where $\delta_{Kl}$ is shorthand for $\delta_K^{SU(2)} + \delta_{K\ell}^{\rm EM}$.
By comparison with \eq{eq:Vus} and given $G_{\rm F}=g^2/(4\sqrt2\,m_W^2)$,
$r_{\mu e}$ is equal to the ratio $g^2_{\mu}/g^2_{e}$, with
$g_{\ell}$ the coupling strength at the $W \rightarrow \ell \nu$ vertex. In the SM, $r_{\mu e}=1$.
Averaging between charged and neutral modes, we find
\begin{equation}
r_{\mu e} =  1.000 \pm 0.008.
\label{eq:unilep}
\end{equation}
\noindent
This has to be compared with the sensitivity obtained in $\pi \rightarrow \ell \nu$ decays,
\break $(r_{\mu e})_{\pi} =  1.0042(33)$~\cite{piuniv}, and in $\tau$ leptonic decays,
$(r_{\mu e})_{\tau} =  1.000(4)$~\cite{pdg06}.
\subsection{Via $K\rightarrow\mu\nu$ decay}
\label{sec:vudvus}
High-precision lattice quantum chromodynamics (QCD) results have
recently become available and are
rapidly improving. The availability of precise values for the
pion- and kaon-decay constants $f_\pi$ and $f_K$ allows use of a relation
between $\Gamma(K_{\mu2})/\Gamma(\pi_{\mu2})$ and $\abs{\Vus}^2/\abs{\Vud}^2$,
with the advantage that lattice-scale uncertainties and radiative corrections
largely cancel out in the ratio \cite{Mar04:fKpi}:
\begin{equation}
{\Gamma(K_{\mu2(\gamma)})\over\Gamma(\pi_{\mu2(\gamma)})}=%
{|V_{us}|^2\over|V_{ud}|^2}\;{f_K^2\over f_\pi^2}\;%
{m_K\left(1-m^2_\mu/m^2_K\right)^2\over m_\pi\left(1-m^2_\mu/m^2_\pi\right)^2}\x(0.9930\pm0.0035),
\label{eq:fkfp}
\end{equation}
where the uncertainty in the numerical factor is dominantly from
structure-dependent corrections and may be improved.
Thus, it could very well
be that the abundant decays of pions and kaons to $\mu\nu$ ultimately
give the most accurate determination of the ratio of \abs{\Vus} to
\abs{\Vud}.
This ratio can be combined with direct measurements of \abs{\Vud} to obtain
\abs{\Vus}.
From our measurements of BR$(K_{\mu 2})$ (\eq{eq:kpmkmu2}) and
$\tau_{\pm}$ (\eq{eq:taukpm}), and using $\Gamma(\pi_{\mu2})$ from \cite{pdg06}, we
evaluate:
\begin{equation}
 \fVusVud = 0.2766 \pm 0.0009 .
\label{eq:fvusvud}
\end{equation}
Using the recent lattice determination of \fkfp\ from the HPQCD/UKQCD
collaboration, $\fkfp =1.189(7)$~\cite{hpqcdukqcd07:fkfp}, we finally
obtain $\Vus/\Vud = 0.2326 \pm 0.0015$.

\subsection{Bounds on New Physics from $K \rightarrow \mu \nu$ decay}
\label{sec:newphys}
\begin{figure}[htb]
\begin{center}
\includegraphics[width=0.4\textwidth]{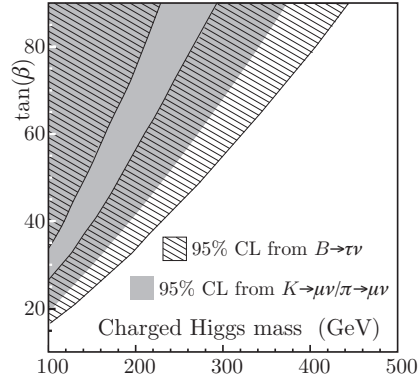}
\caption{Excluded region in the $m_{H^+} - \tan\beta$ plane by the measurement
$R_{\ell23}$; the region excluded by $B \rightarrow \tau \nu$ is also indicated.\label{fig:higgs}}
\end{center}
\end{figure}

A particularly interesting observable is the ratio of the \Vus\ values
obtained from helicity suppressed and helicity allowed modes:
$R_{\ell23} = \left| \Vus(K_{\ell2})/\Vus(K_{\ell3}) \times V_{ud}(0^+\to0^+)/V_{ud}(\pi_{\mu2}) \right|$. This ratio, which is equal to 1 in the SM, would be affected
 by the presence of scalar density or extra right-handed currents. A scalar current due
to a charged Higgs $H^+$ exchange is expected to lower the value of $R_{\ell23}$, which becomes \cite{rl23}:
\begin{equation}
R_{\ell23} = \left| 1 - \frac{m^2_{K^+}}{m^2_{H^+}}\left( 1 -  \frac{m^2_{\pi^+}}{m^2_{K^+}}\right)
\frac{\tan^2\beta}{1+0.01 \tan\beta} \right|,
\label{eq:rl23}
\end{equation}
with $\tan \beta$ the ratio of the two Higgs vacuum expectation values in the MSSM.
In addition, in this scenario both $0^+ \rightarrow 0^+$ nuclear beta decays and $K_{\ell3}$ are not affected,
and the unitarity constraint for this modes can be applied.
To evaluate $R_{\ell23}$, we fit our experimental data on $K_{\mu2}$ and $K_{\ell3}$ decays, using as external
inputs the most recent lattice
determinations of \fo\ ~\cite{rbcukqcd07:f0} and \fkfp
~\cite{hpqcdukqcd07:fkfp}, the value of \Vud\ from \cite{vud07ht},
and $\Vud^2 + \Vus(K_{l3})^2 = 1$ as a constraint.
We obtain
\begin{equation}
R_{\ell23} = 1.008 \pm 0.008,
\end{equation}
which is one $\sigma$ above the SM prediction.
This measurement can be used to set bounds on the charged Higgs mass and $\tan \beta$.
\Fig{fig:higgs} shows the region excluded at 95\% CL in the $m_{H^+}-\tan\beta$ plane.

The measurement of BR$(B \rightarrow \tau \nu)$~\cite{btaunu} can also be used to
set bounds on the $m_{H^+}-\tan\beta$ plane, which are shown in \fig{fig:higgs}.
While the $B\rightarrow \tau \nu$ can exclude quite an extensive region of
this plane, there is an uncovered region
 corresponding to the change of sign of the correction.
This region is fully covered by our result.
\subsection{Test of CKM}
\label{sec:ckm}
%
\begin{figure}[htb]
\centering
\includegraphics[width=0.5\textwidth]{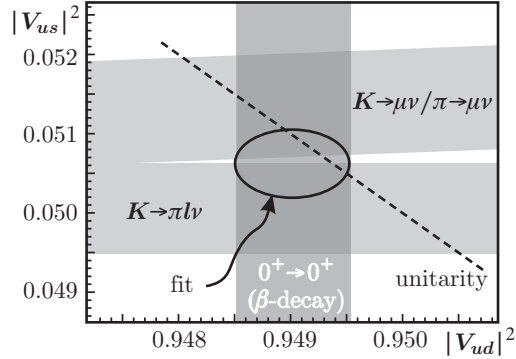}
\caption{KLOE results for $|\Vus|^2$ and $|\Vus/\Vud|^2$, together with $|\Vud|^2$ from $\beta$-decay measurements. The ellipse is the $1\sigma$ contour from the fit. The unitarity constraint is illustrated by the dashed line.
\label{fig:vusvud}}
\end{figure}

To test the unitarity of the quark mixing matrix, we combine
all the information from our measurements on
$K_{\mu2}$, $K_{e3}$, $K_{\mu3}$, together with superallowed
$0^+ \to 0^+$ nuclear $\beta$ decays (\fig{fig:vusvud}).
The best estimate of  $|\Vus|^2$ and $|\Vud|^2$ can be obtained from a
fit to our results $\Vus = 0.2237(13)$
and $\Vus/\Vud  = 0.2326(15)$, together with
$\Vud =0.97418(26)$ \cite{vud07ht}.
The fit gives $|\Vus|^2 = 0.0506(4)$ and $|\Vud|^2 =0.9490(5)$, with $\chi^2$/ndf = 2.34/1
(13\%) and a correlation coefficient of 3\%.
The values obtained confirm the unitarity of the CKM quark mixing matrix as
applied to the first row.
We find
\begin{equation}
|\Vus|^2+|\Vud|^2-1=-0.0004\pm0.0007\quad(\ab0.6\sigma)
\label{eq:ckmunitarity}
\end{equation}
In a more conventional form, the results of the fit are:
\begin{equation}
\eqalign{|V_{us}|&=0.2249\pm0.0010\cr
         |V_{ud}|&=0.97417\pm0.00026.\cr}
\end{equation}
Imposing unitarity as a constraint, $|\Vus|^2+|\Vud|^2=1$, does not improve the accuracy.
One should also keep in mind that while lattice results for $f_+(0)$ and $f_K/f_\pi$ appear to be converging and
are quoted with small errors there is still a rather large spread between different calculations. If we were to
use instead $f_+(0)=0.961\pm0.008$ as computed in \cite{L&R} and still preferred by many authors,
 we find $|V_{us}|=0.2258\pm0.0012$ which is less precise but satisfies more closely unitarity.

\section{Quantum interferometry}
\label{sec:qinterf}
\subsubsection{Quantum coherence}
\label{sec:qcohe}
As stated in sect.~\ref{sec:ksbeam}, the neutral kaon pair from $\f\to\ko\kob$
are produced in a pure quantum state with $J^{PC}$=1\up{--}. We evolve the initial state in time, project to any two
possible final states $f_1$ and $f_2$, take the modulus
squared, and integrate over all $t_1$ and $t_2$ for fixed $\Delta t=t_1-t_2$
to obtain (for $\Delta t > 0$, with $\Gamma\equiv\Gamma_L+\Gamma_S$):
\begin{eqnarray}
I_{f_1,\,f_2}(\Delta t) & = & {1\over2\Gamma}|\bra f_1;\ks\rangle\bra f_2;\ks\rangle|^2 \x
\Big[|\eta_1|^2e^{-\Gamma_L\Delta t}+|\eta_2|^2e^{-\Gamma_S\Delta t} \nonumber \\
& - & 2(1-\zeta)|\eta_1||\eta_2|
e^{-(\Gamma_L+\Gamma_S)\Delta t/2}\cos(\Delta m\Delta t+\f_2-\f_1)\Big].
\label{eq:two}
\end{eqnarray}
The last term is due to interference between the decays to states $f_1$ and
$f_2$ and $\zeta$=0 in quantum mechanics. Fits to the $\Delta t$
distribution provide measurements of the magnitudes and phases of the
parameters $\eta_i = \bra f_i;\kl\rangle/\bra f_i;\ks\rangle$,
as well as of the \kl-\ks\ mass difference $\Delta m$ and the decay widths $\Gamma_L$ and $\Gamma_S$.

\begin{figure}[htb]
\centering
\includegraphics[width=0.55\textwidth]{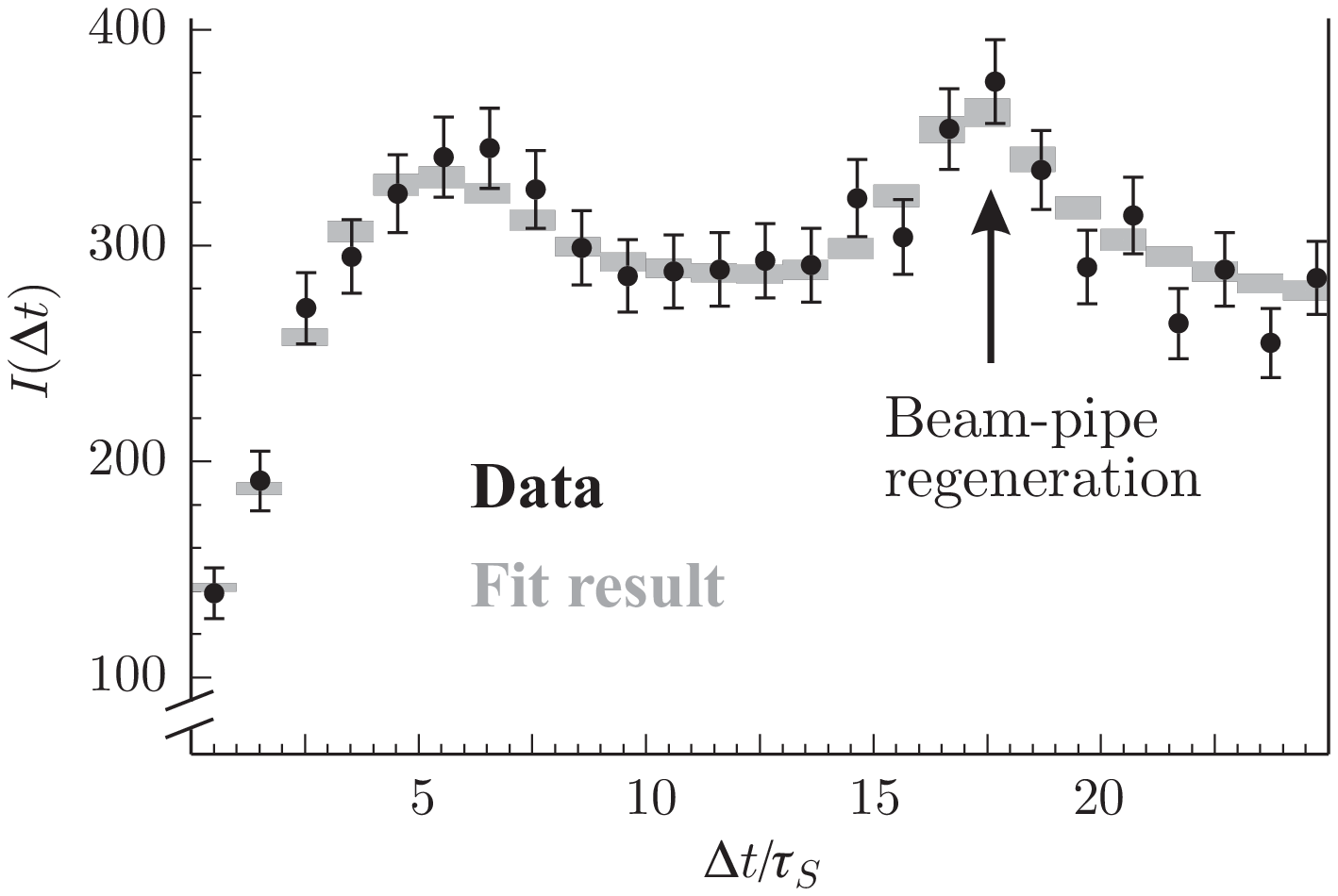}
\caption{$\Delta t$ distribution for $\f\to\ks\kl\to\pic\pic$ events.}
\label{fig:interf}
\end{figure}

Such fits also allow tests of fundamental properties of quantum mechanics. For example, the persistence of quantum-mechanical coherence can be tested by choosing $f_1 = f_2$. In this case, because of the antisymmetry of the initial state and the symmetry of the final state, there should be no events with $\Delta t = 0$.
Using $\sim 300$~pb$^{-1}$ of integrated luminosity, we published an analysis
of the $\Delta t$ distribution for $\ks\kl\to\pic\pic$ events
that establishes the feasibility of such tests.
The $\Delta t$ distribution is fit with a function of the form of \eq{eq:two},
including the experimental resolution and the peak from
$\kl\to\ks$ regeneration in the beam pipe. The results are shown in
\fig{fig:interf}.
Observation of $\zeta\neq0$ would imply loss of quantum coherence. The
value of $\zeta$ depends on the basis, \ko-\kob\ or \ks-\kl\, used in the
analysis. Using the \ko-\kob\ basis we find~\cite{interf}
\begin{equation}
\zeta =\pt(0.10\pm0.21_{\rm stat}\pm0.04_{\rm syst}),-5,
\label{eq:qmtest}
\end{equation}
which means no violation of quantum mechanics.
The statistical error has been already reduced by a factor of two
analyzing $\sim 1$~fb$^{-1}$ more of integrated luminosity.
\subsubsection{Quantum interference and quantum gravity}
\label{sec:qginterf}
In the context of a hypothetical
quantum gravity theory, $CPT$ violation effects might occur in
correlated neutral kaon states~\cite{mavro1,mavro2}, where
the resulting loss of particle-antiparticle identity could induce a
breakdown of the correlation of state
imposed by Bose statistics.
As a result, the initial state can be parametrized as:
\begin{eqnarray}
|i \rangle   &=&   \frac{1}{\sqrt{2}}   \left(    |K^0, \vec{p} \rangle
|\bar{K}^0, -\vec{p}\rangle -|\bar{K}^0,\vec{p}\rangle |K^0, -\vec{p}\rangle
\right) \nonumber \\
             &+& \frac{\omega}{\sqrt{2}} \left( |K^0, \vec{p}\rangle |\bar{K}^0, -\vec{p}\rangle + |\bar{K}^0,\vec{p}\rangle |K^0,-\vec{p}\rangle \right) ~,
\label{eq:statea}
\end{eqnarray}
where $\omega$ is a complex parameter describing a completely novel $CPT$
violation phenomenon, not included in previous analyses. Its order of magnitude
could be at most $|\omega| \sim \left [(m^2_K/m_{\rm{Planck}})/\Delta
   \Gamma \right ]^{1/2} \sim 10^{-3}$
with $\Delta \Gamma = \Gamma_S -\Gamma_L$.
Inserting the modified initial state in the expression of the
$\ks\kl\to\pic\pic$ decay intensity, and fitting the $\Delta t$ distribution
as in the case of the previous analysis, we
obtained the first measurement of the complex parameter $\omega$ \cite{interf}:
\begin{equation}
\Re(\omega) =\left( 1.1^{+8.7}_{-5.3} \pm 0.9 \right)\times{10^{-4}}\qquad
\Im(\omega) =\left( 3.4^{+4.8}_{-5.0} \pm 0.6 \right)\times{10^{-4}}.
\label{eq:omega}
\end{equation}
This can be translated into an upper limit on the absolute value,
$\omega < 2.1\times 10^{-3}$ at 95\%~CL.
Also in this case
the statistical error has been reduced by a factor of two
analyzing $\sim 1$~fb$^{-1}$ more of integrated luminosity, thus reaching
the interesting Planck's scale region.

\section{Test of CPT}
\label{sec:cpt}
\subsection{$CPT$ SM}
\label{sec:cptsm}
The three discrete symmetries of quantum mechanics, charge conjugation ($C$), parity ($P$) and time reversal ($T$) are known
to be violated in nature, both singly and in pairs. Only $CPT$ appears to be an exact symmetry of nature. Exact $CPT$ invariance
 holds in quantum field theory which assumes Lorentz invariance (flat space), locality and unitarity~\cite{CPT}. These assumptions
could be violated at very large energy scales, where quantum gravity cannot be ignored~\cite{CPTV}. Testing the validity of $CPT$
invariance probes the most fundamental assumptions of our present understanding of particles and their interactions.
The neutral kaon system offers unique possibilities for this study.

Within the Wigner-Weisskopf approximation, the time evolution
of the neutral kaon system is described by
\begin{equation}
i {\partial \over \partial  t} \Psi(t) = H\Psi(t) = (M-{i\over 2}\Gamma)\Psi(t)~,
\label{LOY}
\end{equation}
where $M$ and  $\Gamma$ are $2\times 2$ time-independent
Hermitian matrices and $\Psi(t)$ is a two-component
state vector in the $\kzero$--$\kzerob$ space.
Denoting by $m_{ij}$ and $\Gamma_{ij}$ the elements
of $M$ and  $\Gamma$ in the $\kzero$--$\kzerob$ basis,
$CPT$ invariance implies
\begin{equation}
m_{11} = m_{22} \quad ({\rm or}~\Mno=\Mnb) \qquad{\rm and}\qquad
\Gamma_{11} = \Gamma_{22} \quad ({\rm or}~\Gno=\Gnb)~.
\end{equation}
The eigenstates of \eq{LOY} can be written as
\begin{eqnarray}
\label{epsLS}
    K_{S,L} &=&
        \dis\frac{1}{\sqrt{2\left( 1 + |\epsilon_{S,L}|^2 \right)}}
                  \left[ \left(1 + \epsilon_{S,L}\right)K^0 \pm
                  \left( 1 - \epsilon_{S,L} \right) \bar{K}^0\right]~, \\
\epsilon_{S,L} &=&  \frac{
    -i \Im\left( m_{12}\right) -
            \frac{1}{2} \Im\left(\Gamma_{12}\right) \pm
    \frac{1}{2} \left[ \Mnb - \Mno -\frac{i}{2}
    \left( \Gnb -\Gno\right) \right]
                    }{
    m_L - m_S +i(\Gamma_S - \Gamma_L)/2
                    }
 \equiv  \epsilon \pm \delta \no,
\end{eqnarray}
such that $\delta=0$ in the limit of exact $CPT$ invariance.
Unitarity allows us to express the four entries of $\Gamma$ in terms
of appropriate combination of kaon decay amplitudes $\mathcal{A}_i$:
\begin{equation}
\Gamma_{ij} = \sum_f \mathcal{A}_i (f) \mathcal{A}_j(f)^*,\quad i,j = 1,2 = \kzero,\kzerob,
\end{equation}
where the sum runs over all the accessible final states. Using this decomposition
in \eq{epsLS} leads to  the Bell-Steinberger relation~\cite{BS:paper}: a link between
$\Re(\epsilon)$, $\Im(\delta)$ and the physical kaon decay amplitudes.
In particular, without any expansion in the $CPT$-conserving parameters
and neglecting only $\cal{O}(\epsilon)$ corrections to the coefficient of the
$CPT$-violating parameter $\delta$, we find
\begin{equation}
     \ \kern-2em\left({{\GAMS+\GAML}\over{\GAMS-\GAML}}+i\tan\phi_{\rm SW}\right)
      \left(\frac{\Re(\epsilon)}{1+|\epsilon|^2 }-i\Im(\delta) \right) =
      {1\over{\GAMS-\GAML}} \sum_f \cA_L(f) \cA^*_S(f),
\label{eqbsr}
\end{equation}
where $\cA_S(f)$ and $\cA_L(f)$ are the physical
\ks\ and \kl\ decay amplitudes, and $\phi_{\rm SW} = \arctan\left(\vv 2 (m_L -m_S)/(\GAMS-\GAML)\right)$.
The advantage of the neutral kaon
system is that only the $\pi\pi(\gamma)$, $\pi\pi\pi$ and $\pi\ell\nu$ decay modes give significant contributions
to the right-hand side of \eq{eqbsr}. The solution to the unitarity relation in \eq{eqbsr} is:
\begin{equation}
\left(
\begin{array}{l}
   \dis\frac{\Re(\epsilon)}{1+\left|\epsilon \right|^2 } \\
   \Im(\delta)
\end{array}
\right) = \frac{1}{N}
\left(
\begin{array}{cc}
   1+\kappa(1-2b) & (1-\kappa) \tan\phi_{\rm SW} \\
   (1-\kappa) \tan\phi_{\rm SW}  & -(1+\kappa)
\end{array}
\right)
\left(
\begin{array}{l}
\Sigma_i \Re(\alpha_i)   \\
\Sigma_i \Im(\alpha_i)
\end{array}
\right)~,
\label{eq:b-s_sol}
\end{equation}
where the $\alpha$ parameters are related to the product of \ks\ and \kl\ decay amplitudes to
 $\pi\pi(\gamma)$, $\pi\pi\pi$ and $\pi\ell\nu$ final states, $\kappa=\tau_S/\tau_L$,
$b=BR(\kl\to \pi \ell \nu)$, and
\begin{equation}
N = (1+\kappa)^2+(1-\kappa)^2  \tan^2\phi_{\rm SW} -2b \kappa(1+\kappa)~.
\end{equation}

\begin{figure}[htb]
\begin{center}
\includegraphics[width=0.7\textwidth]{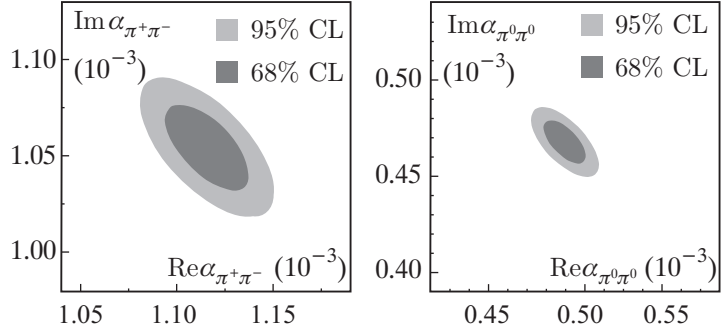}
\caption{Representation of $\alpha_{\pic}$ and $\alpha_{\pio}$ in the complex plane. In each case, the
two ellipses represent the 68\% and the 95\% CL contours.}
\label{fig:alphapipi}
\end{center}
\end{figure}

\begin{figure}[htb]
\begin{center}
\includegraphics[width=0.7\textwidth]{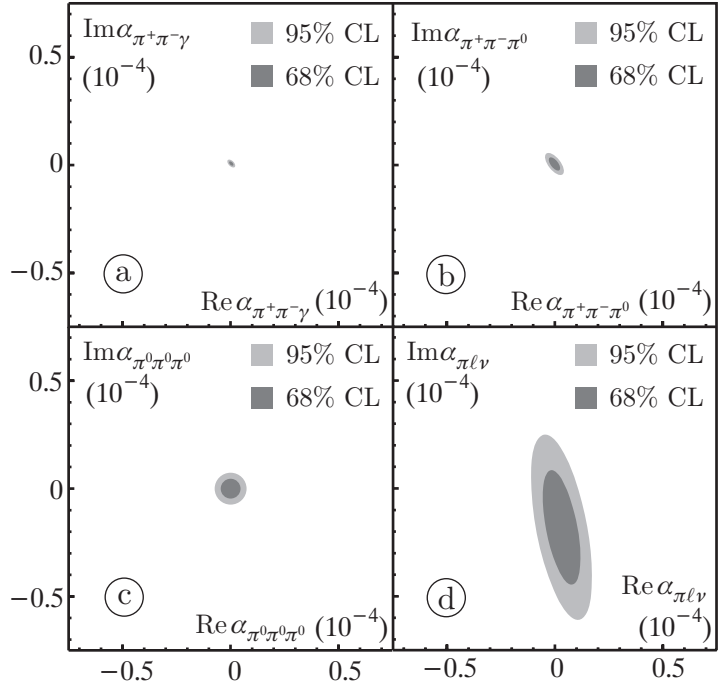}
\caption{Bounds of the $\alpha$ values for the three-body decays of \ks, \kl. Note that the same scale is
used for all plots.}
\label{fig:alpha3body}
\end{center}
\end{figure}

 Recently, we published improved results for the two parameters $\Re(\eps)$
and $\Im(\delta)$ using \eq{eqbsr} and our measurements of neutral kaon
decays~\cite{BSRpaper}.
Our analysis benefits in
particular from three measurements: i) the branching ratio for \kl\ decays to
\pic \cite{klpipi} (sect.~\ref{sec:klpp}), which is relevant for the
determination of $\Re(\eps)$; ii) the new upper limit on \BR(\ks \to \pio\po) \cite{ks3pi0}
(sect.~\ref{sec:ks3p0}), which is necessary to improve
the accuracy on $\Im(\delta)$; and iii) the measurement of the \ks\ semileptonic charge asymmetry
$A_S$~\cite{kslept} (sect.~\ref{sec:ksl3}), which
allows, for the first time, the complete determination of the direct contribution
from semileptonic channels, {\it without assuming unitarity}.
The experimental inputs to the determination of the $\alpha$ parameters of \eq{eq:b-s_sol} are
 all of the KLOE neutral kaon measurements (tables \ref{tab:klbr2} and \ref{tab:ksdata}),
 the values of $\tau_S$, $m_L - m_S$, $\phi_{+-}$, $\phi_{00}$ from
\cite{pdg06},  and the $\pic \gamma$ and $\pic\po$ amplitudes from \cite{E773} and
\cite{CPLEAR:p+p-p0}, respectively.  Using all of these results we obtain the values reported in
table \ref{tab:alpha}; \fig{fig:alphapipi} and \fig{fig:alpha3body} show the 68\% and 95\% CL
contours in the complex plane $\Re(\alpha_i) - \Im(\alpha_i)$.

 \begin{table}[htb]
\centering
\begin{tabular}{lc}
\hline
Parameter & Value   \\
\hline
$\alpha_{\pic}$           &  $((1.115 \pm 0.015)+i(1.055\pm0.015))\times 10^{-3}$  \\
$\alpha_{\pio}$           &  $((0.489 \pm 0.007)+i(0.468\pm0.007))\times 10^{-3}$  \\
$\alpha_{\pic\gamma DE}$   &  $((5 \pm 7)+i(6\pm7))\times 10^{-7}$  \\
$\alpha_{\pic\po}$        &  $((0 \pm 2)+i(0 \pm 2))\times 10^{-6}$  \\
$|\alpha_{\pio\po}|$      &  $ < 7 \times 10^{-6}$ at 95\% CL  \\
$\alpha_{\pi\ell\nu}$     &  $((0.3 \pm 0.6)+i(-1.8 \pm 1.8))\times 10^{-5}$  \\
\hline
\end{tabular}
\caption{Values of $\alpha$ parameters  of \eq{eq:b-s_sol} for all relevant kaon decay channels.}
\label{tab:alpha}
\end{table}

\begin{figure}[htb]
\begin{center}
\includegraphics[width=0.7\textwidth]{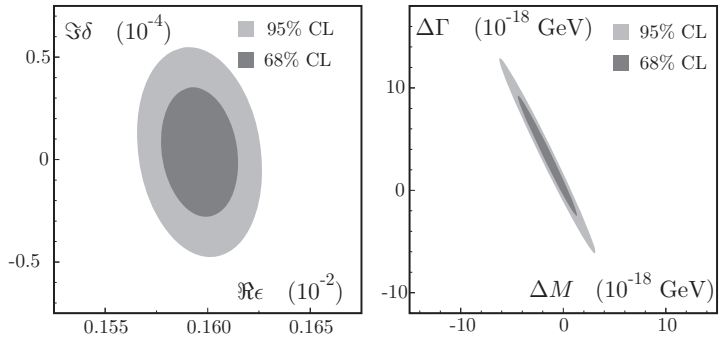}
\caption{Left: allowed region at 68\% and 95\% CL in the $\Re\eps$, $\Im\delta$ plane. Right: allowed
region at 68\% and 95\% CL in the $\Delta M, \Delta \Gamma$ plane.}
\label{fig:delm}
\end{center}
\end{figure}

Inserting all of the information in \eq{eq:b-s_sol} we finally obtain:
\begin{equation}
\Re(\eps) = (159.6 \pm 1.3)\times 10^{-5} , \qquad   \Im(\delta) = (0.4 \pm 2.1)\times  10^{-5}\, .
\label{eq:bellres}
\end{equation}
The allowed region in the ($\Re(\eps)$, $\Im(\delta)$) plane at 68\%~CL and 95\%~CL is shown in \fig{fig:delm}, left. Our results, \eq{eq:bellres}, improve by a factor of two the previous
best determinations.
The limits on $\Im (\delta)$ and $\Re(\epsilon)$ can be used to constrain the mass and width difference
between \kzero\ and \kzerob\ via
\begin{equation}
\delta = { { i(m_{\kzero} - m_{\kzerob}) + {1\over2} (\Gamma_{\kzero}-\Gamma_{\kzerob}) }
       \over{\GAMS-\GAML} }\cos\phi_{SW} e ^{i \phi_{SW}} [1+\mathcal{O}(\epsilon)].
\label{eq:belldelta}
\end{equation}
The allowed region in the $\Delta M = (m_{\kzero} - m_{\kzerob})$,
 $\Delta \Gamma = (\Gamma_{\kzero}-\Gamma_{\kzerob}) $ plane is shown in \fig{fig:delm}, right.
Since the total decay widths are dominated by long-distance dynamics, in models where $CPT$ invariance is a
pure short-distance phenomenon, it is useful to consider the limit  $\Gamma_{\kzero}=\Gamma_{\kzerob}$.
In this limit, neglecting $CPT-$violating effects in the decay amplitudes, we obtain the following
bound on the neutral kaon mass difference:
\begin{equation}
 -5.3  \times 10^{-19}~ {\rm GeV}< m_{\kzero} - m_{\kzerob}  < 6.3 \times 10^{-19}~ \rm{GeV \quad at~ 95 ~\% ~CL}\, .
\label{eq:belldm}
\end{equation}

\subsection{$CPT$ and Lorentz symmetry breaking}
\label{sec:LorentzB}
$CPT$ invariance holds
for any realistic Lorentz-invariant quantum field theory.
However a very general theoretical possibility for $CPT$ violation is
based on spontaneous breaking of Lorentz symmetry, as developed by
Kostelecky \cite{kost}, which
appears to be compatible with the basic
tenets of quantum field theory and retains the property of gauge
invariance and renormalizability (Standard Model Extensions, SME).
For neutral kaons, $CPT$ violation would manifest to lowest order only in the parameter $\delta$,
with dependence on the kaon 4-momentum:
\begin{eqnarray}
\label{eq:deltak}
\delta \approx i \sin \phi_{SW} e^{i \phi_{SW}} \gamma_K (\Delta a_0-
\vec{\beta_K}\cdot \Delta{\vec{a}})/\Delta m,
\end{eqnarray}
where $\gamma_K$ and $\vec{\beta_K}$ are the kaon boost factor and velocity in the observer frame, and $\Delta a_{\mu}$
are four $CPT$- and Lorentz-violating coefficients for the two valence quarks in the kaon.
Following \cite{kost}, the time dependence arising from the rotation of
the earth can be explicitly displayed in
\eq{eq:deltak} by choosing a three-dimensional basis ($\hat{X},\hat{Y},\hat{Z}$) in a non-rotating frame, with the $\hat{Z}$ axis along the earth's rotation axis, and a basis $(\hat{x},\hat{y},\hat{z})$ for the rotating (laboratory) frame.
The laboratory frame precesses around the earth's rotation axis $\hat{Z}$
at the sidereal frequency $\Omega$. Defining $\chi$ the angle between $\hat{Z}$ and the positron beam direction,
the $CPT$ violating parameter $\delta$ may then be expressed as:
\begin{eqnarray}
\label{eq:deltak3}
\delta (|\vec{p}|,\theta, t) &=& \frac{1}{2\pi}\int_0^{2\pi}\delta(\vec{p},t)d\phi\nonumber\\
&=& {\frac{i \sin \phi_{SW} e^{i \phi_{SW}}}{\Delta m}}
\gamma_K
\{
\Delta a_0 +\beta_K\Delta a_Z \cos\theta\cos\chi \nonumber\\
&&+\beta_K ( \Delta a_Y \sin\chi\cos\theta \sin\Omega t +\Delta a_X \sin\chi\cos\theta \cos\Omega t) \} ~,
\end{eqnarray}
The $\Delta a_0$ parameter can be measured through
the difference between \ks\ and \kl\ charge asymmetries, $A_S-A_L$.
Each asymmetry is integrated over the polar angle  $\theta$,
thus averaging to zero any possible contribution from the terms
proportional to $\cos\theta$ in \eq{eq:deltak3}, giving the relation:
\begin{eqnarray}
\label{eq:delta0}
A_S-A_L \simeq \left[{
\frac{4\Re\left( i \sin \phi_{SW} e^{i \phi_{SW}} \right)\gamma_K}{\Delta m}}  \right]
 \Delta a_0 ~.
\end{eqnarray}
From our measured value of $A_S$~\cite{kslept} (sect.~\ref{sec:ksl3}) and a preliminary evaluation of $A_L$,
we get $A_S-A_L=(-2\pm10)\times10^{-3}$, which gives as a
  preliminary evaluation of the $\Delta a_0$ parameter~\cite{testamoriond08}:
\begin{equation}
\Delta a_0 = (0.4\pm 1.8)\times 10^{-17}\hbox{ GeV}~.
\end{equation}
With the analysis of the full data sample, an
accuracy $\sigma(\Delta a_0) \sim 7 \times 10^{-18}\hbox{ GeV}$ could be reached.

The parameters $\Delta a_{X,Y,Z}$ can be measured using
$\phi \rightarrow K_S K_L \rightarrow\pi^+\pi^-,\pi^+\pi^-$ events, by fitting the decay intensity
$I\left(\pi^+\pi^-(\cos\theta>0),\pi^+\pi^-(\cos\theta<0);\Delta t\right)$,
where the two identical final states are distinguished by their forward or
backward emission.
A preliminary analysis, based on $\sim 1$~fb$^{-1}$, yields \cite{testamoriond08}:
\begin{eqnarray}
\Delta a_X &=& (-6.3 \pm 6.0)\times 10^{-18} \GeV \no \\
\Delta a_Y &=& ( 2.8 \pm 5.9)\times 10^{-18} \GeV \no \\
\Delta a_Z &=& ( 2.4 \pm 9.7)\times 10^{-18} \GeV,
\label{eq:deltaaz}
\end{eqnarray}
which represents the first determination of $\Delta a_Z$ to date.

\section{Light meson spectroscopy}
\label{sec:lightms}
\begin{figure}[htb]
\begin{center}
\epsfig{file=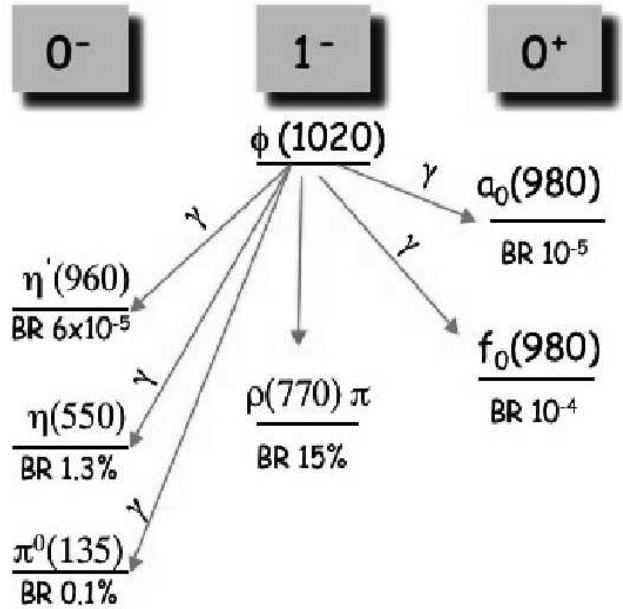,width=.6\textwidth,height=0.6\textwidth}
\end{center}
\caption{Graphical representation of $\phi$ meson decaying to lower lying meson states. }
\label{tab:decays}
\end{figure}

When running \DAF\ at the \f\ peak, the dominant decay channels
of the $\phi$ meson are $K\overline{K}$ pairs.
%
However, as shown in \fig{tab:decays}, the $\phi$ meson decays
with large rates also to  other interesting hadronic final states.
 In this chapter, we mainly discuss the
so called $\phi$ radiative decays i.e.
electromagnetic transitions ($\f\to\mbox{meson}+\gam$)
to other mesons, which result in a great profusion of lighter scalar
and pseudoscalar mesons. The transition rates are strongly dependent
on the wave function of the final-state meson. They also depend on its
flavor content, because the \f\ is a nearly pure $s\bar{s}$ state and
because there is no photon-gluon coupling. These radiative \f\ decays
are unique probes of meson properties and structure. This hadron source,
coupled with \klo\ 's great versatility, allows to perform many
hadronic experiments, with great precision. Moreover, we can
study  with high statistics the  properties of vector mesons.
Due to space constraint, we will limit ourselves to mentioning only
a handful of these measurements, just enough to give
a taste of hadron physics with \klo.
\subsection{$\eta$ meson}
\label{sec:eta}
The BR for the $\f\to\eta\gamma$ decay is 1.3\%. In 2.5~fb$^{-1}$ of KLOE data,
there are $\sim$ 100 million $\eta$ mesons which, for most of the final states,
are clearly identified by their recoil against a photon of $E = 363$~MeV.
The $\eta$'s decay predominantly into two photons or three pions.
The large production
rate and the clean $\eta$ signature made possible to search for rare or forbidden
$\eta$ decays even with reduced samples. A good example is provided by the decay
 $\eta\to\pic$ which, being the $\eta$ an isoscalar, violates both \P\ and \CP.
With the first 350 pb$^{-1}$, KLOE had searched for evidence of this decay  in
the $M_{\pi\pi}$ distribution of $e^+e^-\to\pic\gam$ events where the photon is
emitted at large polar angles ($\theta>45\deg$). No peak is observed in the
distribution of $M_{\pi\pi}$ in the vicinity of $m_\eta$. The corresponding
limit is ${\rm BR}(\eta\to\pic)\leq\pt1.3,-5,$ at 90\% CL \cite{etapp},
which improves the previous limit by a factor of 25.
Similarly, the decay $\eta\to3\gam$ violates \C .
KLOE has conducted a search for this  decay using 410~pb$^{-1}$ and has set
the  limit ${\rm BR}(\eta\to3\gam)\leq\pt1.6,-5,$ at 90\% CL, the most
stringent result obtained to date \cite{eta3g}. In the following subsections,
we will instead show precise measurement
on the $\eta$ meson such as the mass and the description of the dynamics
for the $\eta \to 3 \pi$ decays. We conclude by showing preliminary $BR$
measurements for very rare but observed decays which are still  in progress.
\subsubsection{$\eta$ mass}
\label{sec:etamass}
KLOE has performed the most precise determination to date of the $\eta$ mass,
on the basis of 17 millions of $\phi \to \eta \gamma, \eta \to \gamma \gamma$ decays.
The $\phi \to \eta \gamma$ events are selected by requiring at least three energy
deposits in the barrel with a polar angle
$\theta_{\gamma}: 50^{\circ} < \theta_{\gamma} < 130^{\circ}$,
 not associated to a charged track.
A kinematic fit imposing energy-momentum conservation and time of flight
of photons equal to the velocity of light is done for all 3 $\gamma$'s
combination of  N detected photons. The combination with the lowest
$\chi^2$ is chosen as a candidate event if $\chi^2 < 35$.
The inputs of the fit are the energy, the position and the time of the calorimeter clusters,
the mean position of the $e^+ e^-$ interaction point, ${\overline x_{\phi}}$,
and the total four-momentum, $p_{\phi}$,
of the colliding $e^+ e^-$ pair. The mean values of ${\overline x_{\phi}}$  and $p_{\phi}$ are
determined run by run using $e^+ e^- \to e^+ e^-$  events (almost 90000 events for each
run, allowing a very precise determination of the relevant parameters).

\begin{figure}
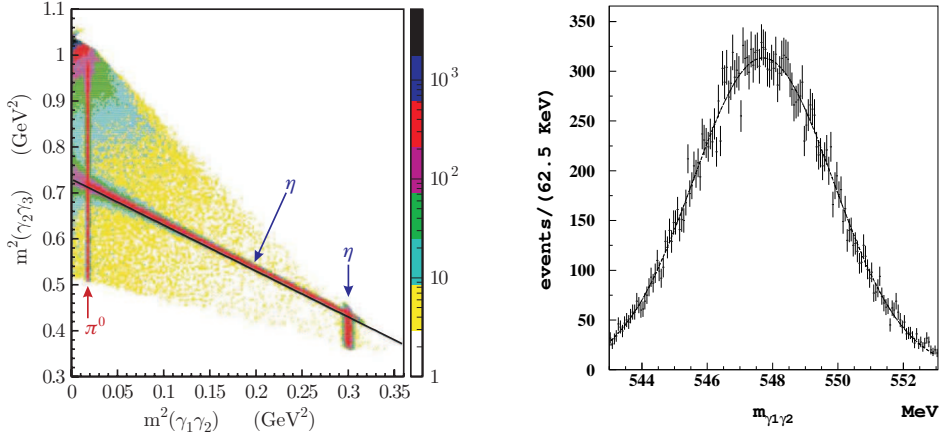

\cl{\figb dalit3a;6;\kern1cm\figb etapeak3;5.4 ;}
\caption{Left: Dalitz plot distribution of the selected 3$\gamma$ events.
 The photons are sorted according their energies $E_1 < E_2 < E_3$. Right:
 Invariant mass distribution of the $\gamma_1 \gamma_2$ pair in the $\eta$
region. The fitted
function is the continuous line.} \label{fig:dalitz}
\end{figure}

The events surviving the cuts are shown in the Dalitz plot, \fig{fig:dalitz}-left,
where three bands are clearly visible. The band at low $m^2_{\gamma \gamma}$ is given by
the $\phi \to \pi^0 \gamma, \; \pi^0 \to \gamma \gamma$, while the other two bands are
$\phi \to \eta \gamma, \; \eta \to \gamma \gamma$ events. Using the cuts indicated by the
black line shown in the Dalitz plot we select pure samples of $\eta, \; \pi^0 \to \gamma \gamma$
events. The resulting $m_{\gamma \gamma}$ spectrum for the $\eta$ (\fig{fig:dalitz}-right) is well
fitted  with a single gaussian of $\sigma \sim $ 2.1 MeV$/c^2$, thus showing that the kinematic
fit improves of a factor $\sim$ 20 the  mass resolution obtained by calorimetric reconstruction
only.

To determine the systematic error associated with the mass measurement, we have
evaluated the uncertainties on all the quantities used in the kinematic fit and
their effect on the fitted value. A sample of $e^+ e^- \to \pi^+ \pi^- \gamma$
events has been used to check the mean position of the interaction point and
the alignment of the calorimeter with respect to  the Drift Chamber.
The absolute energy scale of the calorimeter and the linearity of the
energy response was checked using both the $e^+e^- \to
e^+ e^- \gamma$ and the $\pi^+ \pi^- \gamma$ events. A linearity of
better than 2\% is found and the absolute scale results set to
better  than 1\%. Since the kinematic fit overconstrains the photon
energy with the cluster positions, the systematic uncertainties
related to the energy reconstruction are small and result in a $\eta$ mass error
of 4 keV for the scale and  4 keV for the linearity.
For the same reason, a larger effect can be expected if the calorimeter does not
have angular uniformity in $\phi$ or $\theta$. Misalignment of single modules in
the calorimeter has been checked resulting in a spread of about 10-15 keV on
the $\eta$ mass. While the chosen cut on $\chi^2$ has negligible effect,
the systematics due to the particular choice of the event
selection cut on the Dalitz plot, shown in \fig{fig:dalitz}-left,
was determined to contribute 12 keV to the error.
The measured value of the mass is, however, very sensitive to the center of mass energy
of the $\eta \gamma$ system. Due to initial state radiation emission (ISR), the available
center of mass energy  is a bit lower than the one of the $e^+ e^-$  beams, $W$,
measured using $e^+ e^- \to e^+ e^-$ events. A variation of 100 keV of the measured
mass value is obtained from the MC simulation which includes ISR emission. Due to the
large value of this correction, the $\eta$ mass has been determined
as a function of $W$ both in data and with the  MC simulation+ISR.
The resulting data-MC spread of 8 keV is assumed as systematic error.

All the above mentioned studies on systematics have been also done for the $\pi^0$ mass
using  the $\phi \to \pi^0 \gamma$ events and  for the ratio of the two masses
$r=m_{\eta}/m_{\pi^0}$.
The mass scale is set by the knowledge of the absolute scale of $W$ determined
using  the CMD2 $m_{\phi}$ value as shown in the $K$ mass measurement section.
We obtain \cite{ketamass}:
\begin{equation}
   m_{\pi^{0}} = (134.906 \pm 0.012_{\rm stat} \pm 0.048_{\rm syst}) \quad
   \mbox{MeV}
   \label{EQ2}
\end{equation}
\begin{equation}
   m_{\eta} = (547.874 \pm 0.007_{\rm stat} \pm 0.031_{\rm syst}) \quad
   \mbox{MeV}
   \label{EQ3}
\end{equation}
the  $\pi^{0}$ mass value is in agreement with the world
average~\cite{pdg06} within 1.4 $\sigma$.

As a check of this result, we use the measured ratio
\begin{equation}
   r = \frac{m_{\eta}}{m_{\pi^{0}}} = 4.0610 \pm 0.0004_{\rm stat} \pm 0.0014_{\rm syst}
   \label{EQ4}
\end{equation}
and the world average value of the $\pi^{0}$ mass, $m_{\pi^{0}} =
(134.9766 \pm 0.0006)$ MeV to derive
$m_{\eta} = (548.14 \pm 0.05_{\mathrm{stat}} \pm 0.19_{\rm syst})$ MeV,
consistent with the results quoted above although affected by a
larger systematic error. This is due to the soft energy spectrum
of the photons from $\pi^0$ which results in a worse position resolution.

While being the most accurate result today,
our measurement of the $\eta$ mass is in good agreement with the
recent determinations based on $\eta$ decays \cite{NA48,CLEO}.
Averaging those mass values
and our result we obtain $m_{\eta} = 547.851 \pm 0.025$ MeV, a value
different by $\sim$ 10 $\sigma$ from the average of the measurements
done studying the production of the $\eta$ meson at threshold in
nuclear reactions \cite{GEM}.
\subsubsection{Dynamics of the decays $\eta\rightarrow3\pi$}
\label{sec:eta3pi}
The decay $\eta\to3\pi$ violates isospin invariance. Electromagnetic
contributions to the process are very small and the decay is induced
dominantly by the strong interaction via the $u,d$ mass difference. The
$\eta\to3\pi$ decay is therefore an ideal laboratory for testing Chiral
Perturbation Theory, ChPT.
We have studied the dynamics of the decay $\eta\rightarrow\pi^+\pi^-\pi^0$ using about 1.4 million of such events.
From a fit to the Dalitz plot density distribution we made precise determinations of the parameters that characterize the decay amplitude.
In addition, we have also obtained the most stringent tests on \C-violation
from left-right, quadrant and sextant integrated asymmetries. \cite{etapicpo}

\begin{figure}[htb]
\begin{center}
\includegraphics[width=0.6\textwidth]{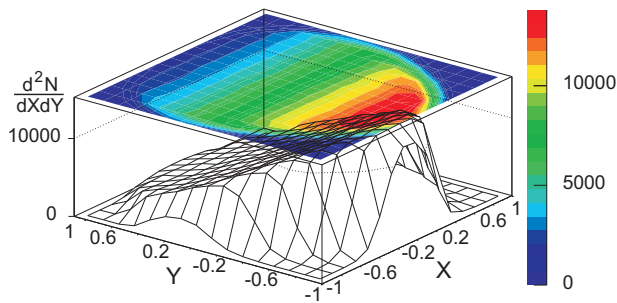}
\end{center}
\caption{Dalitz-plot distribution for the whole
    data sample. The plot contains $1.34$ millions of events in $256$
    bins.}
  \label{dalitz_plot}
\end{figure}

A three body decay\footnote{Both $\eta$ and $\pi$ are spinless, therefore
  there is no preferred direction.} is fully described by two variables. We
can choose two of the pion energies ($E_+,E_-,E_0$) in the $\eta$ rest
frame, two of the three combinations of the two-pion masses squared ($m^2_{+-},\ m^2_{-0},\ m^2_{0+}$) also called ($s,t,u$). Note that $E_+$ is linear in $m^2_{-0}$ and so on, cyclically. The Dalitz variables, $X,Y$ are linear combinations of the pion energies:
\begin{equation}\eqalign{X&= \sqrt3\,{E_+-E_-\over Q} = {\sqrt3\over 2m_\eta Q}\,(u-t)\cr
 Y& = 3\,{E_0-m_0\over Q}-1 = {3\over2m_\eta Q}\left[(m_\eta-m_{\po})^2-s\right]-1\cr}
\label{defXY}
\end{equation}

where $Q$ is the reaction ``$Q$-value''.
The decay amplitude is given in \cite{BijGa02} as:
\begin{equation}
A(s,t,u)=\frac{1}{\Delta^2}\frac{m_K^2}{m_\pi^2}
\left(m_\pi^2-m_K^2\right)\frac{\Ma(s,t,u)}{3 \sqrt{3} F^{2}_{\pi}}
\label{ampli}
\end{equation}
where $\Delta^2\equiv(m_s^2-\widehat m^2)/(m_d^2-m_u^2)$ and
$\widehat m=1/2(m_u+m_d)$ is the average $u, d$ quark mass. $F_\pi = 92.4$
MeV is the pion decay constant and $\Ma(s,t,u)$ the amplitude
we would like to know.
From \eq{ampli} it follows that the decay rate for \etappp\ is proportional to $\Delta^{-4}$.
The transition \etatrepi\ would be therefore very sensitive to $\Delta$ if the amplitude \Ma\ is known.
At lowest order in ChPT:
\begin{equation}
\Ma(s,t,u) =\frac{3 s-4 m_{\pi}^{2}}{m_{\eta}^{2}-m_{\pi}^{2}}.
\label{ma}
\end{equation}
and the decay width at leading order is $\Gamma^{\rm lo}\left(\etappp\right)=66$
eV~\cite{BijGa02} to be compared with the measured width 295 \plm 16 eV\cite{pdg06}.
\noindent
A one-loop calculation within conventional ChPT
improves  considerably the prediction giving $\Gamma^{\rm nlo}\left(\etappp\right)\simeq 167 \pm 50
\;\textrm{eV}$
 but is still far from the experimental value. Higher order corrections
help but do not yet bring agreement with measurements of both total rate and Dalitz plot slopes. Good agreement is found combining ChPT with a non perturbative coupled channels
approach using the Bethe Salpeter equation \cite{Borasoy}.
Therefore  a precision study of the \etatrepi\ Dalitz plot, DP, is highly desirable. The amplitude squared
is expanded around $X=Y=0$  in power of $X$ and $Y$
\begin{equation}
 |A(X,Y)|^{2}= 1+aY + bY^{2} + c X + dX^{2} + eXY + fY^{3} + \dots
\label{eq:amp_tra}
\end{equation}
The parameters ($a,b,c,d,e,f,...$ ) can be obtained from a fit to the observed DP density, see \Fig{dalitz_plot} and should be computed by the theory.
Any odd power of $X$ in $A(X,Y)$ implies violation of charge conjugation.

The event selection consists of a pre-selection, the requirement of
one charged vertex inside a small region together with three photons and a total photon energy
below 800 MeV. A constrained kinematic fit imposing four-momentum conservation and correct
time of arrival for the photons is then applied and only events with  a $\chi^2$ probability
greater than 1\% are kept for further analysis. We finally required that
the recoil photon energy lies between 320 and 400 MeV and we add two further constraints on
the sum of charged pion energies and on the invariant mass of the two softest photons.
All selection criteria and efficiencies were checked by data and Monte Carlo.
The overall selection efficiency, taking into account all the Data-MC
corrections is found to be $\epsilon = ( 33.4~ \pm~ 0.2 )$\%.
The expected background, obtained from MC simulation is as low as 0.3\%.
After the background subtraction the observed number of events in the Dalitz
plot is $N_{obs} = 1.337 \pm 0.001$ million events.
%
The results, from fitting the matrix element of Dalitz-plot as shown in
\eq{eq:amp_tra}, including the statistical uncertainties coming from the
fit and the estimate of systematics are:

$a = -1.090 \pm 0.005 (stat) ^{+ 0.008}_{- 0.019} (syst)$;

$b = 0.124 \pm 0.006 (stat) \pm 0.010 (syst)$;

$d = 0.057 \pm 0.006 (stat) ^{+ 0.007}_{- 0.016} (syst)$;

$f = 0.14 \pm 0.01 (stat) \pm 0.02 (syst)$.

Our fit raises several points unexpected from previous, low statistics analyses.
In particular: (i) the fitted value for the quadratic slope in $Y$, b, is almost
one half of simple Current Algebra predictions ($b = a^{2}/4$), indicating that
higher order corrections are probably necessary, (ii)
the quadratic term in $X$, d, is unambiguously found different from  zero,
and (iii) the same applies for the large cubic term in $Y$.

We have also fitted the Dalitz plot with a different parametrization
of the event density which takes into account the final state $\pi$-$\pi$ rescattering.
Since strong interactions are expected to mix
the two isospin $I = 1$ final states of the \etatrepi~ decay, it is possible to
introduce a unique rescattering matrix $R$ which mixes the corresponding
I=1 decay amplitudes \cite{D'Amb_Isi}.
Thus, we have made a fit using the alternative parametrization of the
Dalitz plot density distribution,
from which it is possible to extract the Dalitz plot
slope $\alpha$ parametrizing the 3$\pi^0$ decay amplitude. The result
$\alpha = -0.038 ~\pm 0.002$
is in reasonable agreement with the direct measurement by KLOE of
$\alpha=-0.027\pm 0.004^{+0.004}_{-0.006}$ \cite{eta3pi0}.

\begin{figure}[htb]
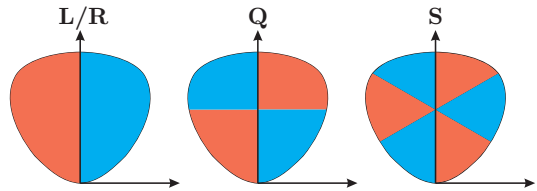

\cl{\figb etadp;7;}
\caption{Definition of the $\eta$ Dalitz plot asymmetry tests: left $A_{LR}$, center $A_Q$, right $A_S$}
\label{fig:etadp}
\end{figure}

While the polynomial fit of the Dalitz plot density gives valuable
information on the matrix element, some  specific integrated asymmetries as defined in
\fig{fig:etadp}  are
very sensitive in assessing the possible contributions to \C-violation
in amplitudes with fixed $\Delta I$.
In particular left-right asymmetry tests \C-violation with no
specific $\Delta I$ constraint;
quadrant asymmetry tests C violation in $\Delta I = 2$ and sextant
asymmetries tests C violation in $\Delta I = 1$.

In the following we present results on asymmetries which use four times the
statistics entering the Particle Data Group's fits and are the most stringent tests to date.
We first obtained from MC the efficiency for each region of the Dalitz plot
as the number of events reconstructed in a particular region divided by
the number of events generated in the same region, thus taking
resolution effects into account. Since no asymmetries were introduced in the Monte
Carlo, we have checked that the asymmetries estimated with our technique in Monte Carlo
are compatible with zero.  We then evaluated the
asymmetry on data by counting events in the regions, subtracting the background and
correcting for efficiency .
The measured asymmetries are as follows:

$A_{LR} = (0.09 \pm 0.10 (\rm stat.)^{+0.09}_{-0.14} (\rm syst.) )\times 10^{-2}$

$A_{Q}  = ( -0.05 \pm 0.10 (\rm stat.)^{+0.03}_{-0.05} (\rm syst.) )\times 10^{-2}$

$A_{S}  = (0.08 \pm 0.10 (\rm stat.)^{+0.08}_{-0.13} (\rm syst.) )\times 10^{-2}$

\subsubsection{$\eta\rightarrow\pi^0\gamma\gamma$}
\label{sec:etapi0gg}
The decay $\eta\to\po\gam\gam$ is particularly interesting in chiral
perturbation theory. There is no $\mathcal{O}(p^2)$ contribution,
and the $\mathcal{O}(p^4)$ contribution is small. At KLOE, the decay
can be reconstructed with full kinematic closure and without complications
from certain backgrounds present in fixed-target experiments, such as $\pi^-p\to\pio n$.
Using 450 pb$^{-1}$ of data, we obtain the preliminary
result ${\rm BR}(\eta\to\pi^0\gam\gam) = \pt(8.4\pm2.7\pm1.4),-5,$,
which is in agreement with several $\mathcal{O}(p^6)$
chiral perturbation theory calculations discussed in
a recent $\eta$ workshop \cite{wasa}. The final measurement
using the complete set of data is forthcoming in 2009.
\subsubsection{$\eta~\rightarrow~\pi^+~\pi^-~e^+~e^- $}
\label{sec:etappee}
\begin{figure}[htb]
    \begin{center}
      {\includegraphics[scale=0.29]{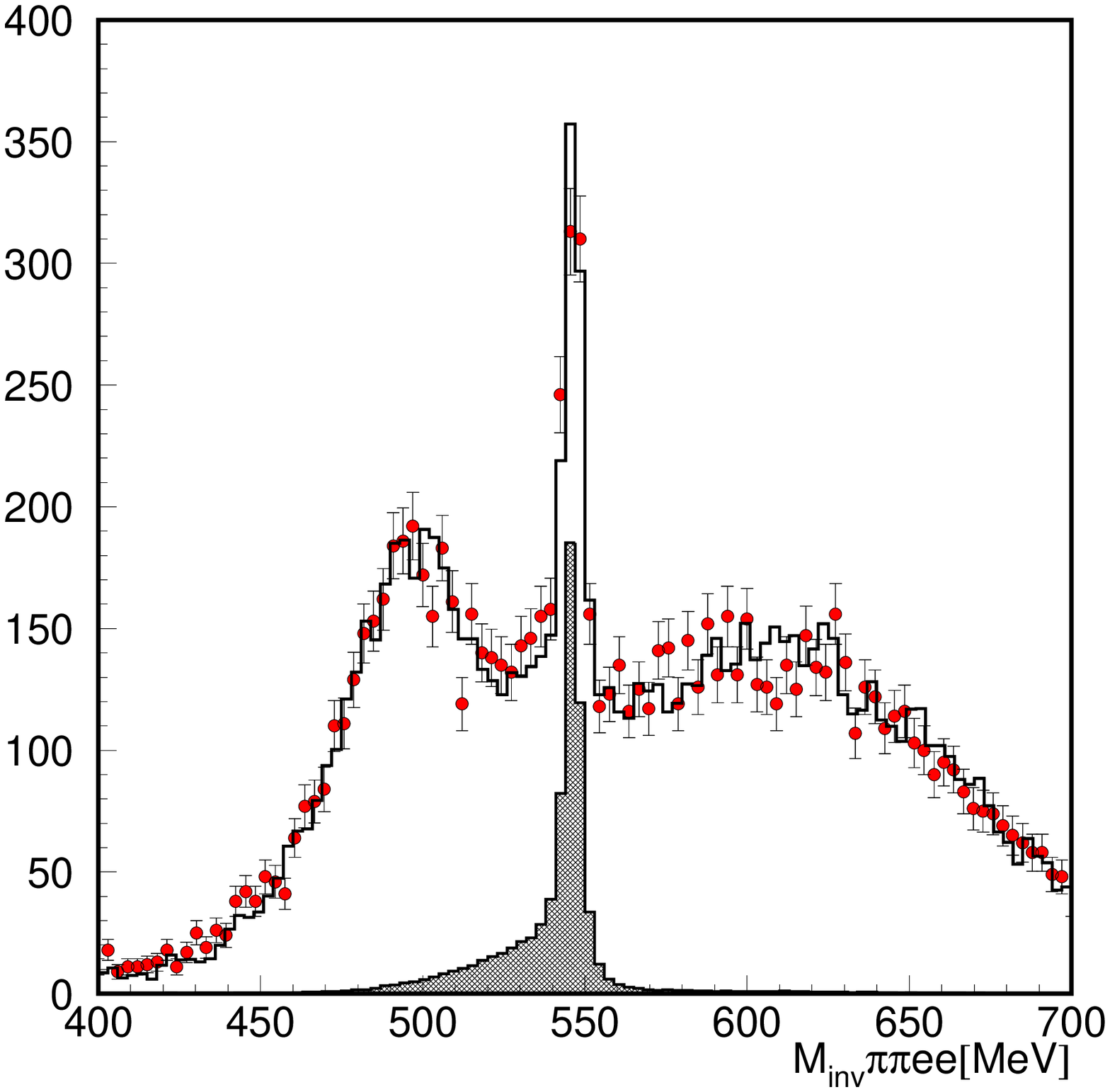}}\kern1cm
      {\includegraphics[scale=0.29]{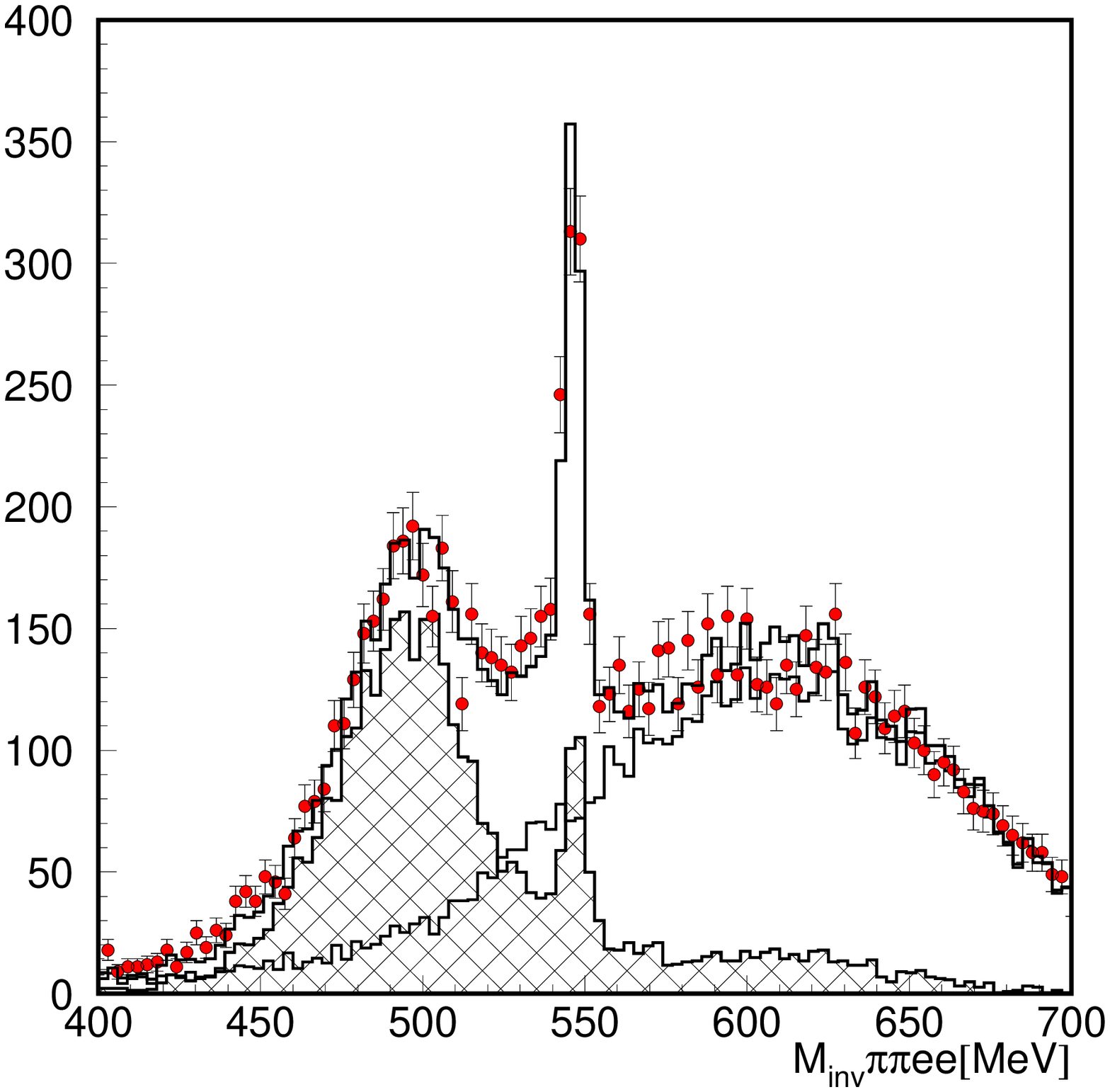}}
      \caption{Fit of the invariant mass of the four tracks selected.
        Dots, data;
        line, total MC.
    On the left signal MC is dark.
    On the right the hatch corresponds to $\phi \rightarrow \eta \gamma$ backgrounds;
        the continuous line is for all other sources of background.}
      \label{fig:minv4trks}
    \end{center}
\end{figure}

This reaction is interesting because it can test a particular
formulation \cite{Gao:2002gq} of CP violation beyond the SM,
through the measurement of the angular asymmetry between the pions and
electrons planes. The analysis is based on the reconstruction of the
invariant mass of four charged particles recoiling against a monochromatic
photon in $\phi \rightarrow \eta \gamma$ events.
The main background is represented by $\phi \rightarrow \pi^+\pi^- \pi^0$ events,
with either the $\pi^0$ undergoing a Dalitz decay or one of the two photons
from the $\pi^0$ decay converting in the beam pipe.
The same happens for $\phi \rightarrow  \eta \gamma$, with
$\eta \rightarrow \pi^+ \pi^- \pi^0$ with the $\pi^0$ Dalitz decay,
and for $\phi \rightarrow  \eta \gamma$  with
$\eta \rightarrow \pi^+ \pi^- \gamma$ events with photon conversion.
These backgrounds are well reduced by kinematic cuts.
Signal and background counts are extracted from the fit of the invariant
mass of the four tracks events, using MC spectra for the signal and
the various backgrounds, \fig{fig:minv4trks}.

Using a data sample of $\sim$622 pb$^{-1}$,
about 700 signal events have been observed while
previous experiments collected less than 20 events.
A preliminary estimate of the BR is $(24 \pm 2_{\rm stat} \pm 4_{\rm syst})
\times 10^{-5}$. Since approximately 3 times more statistics is available,
the CP violation test aforementioned will be accessible \cite{Robbie}.
\subsection{$\eta'$ meson}
\label{sec:etaprime}
The magnitude of ${\rm BR}(\f\to\eta'\gam)$
is a probe of the $s\bar{s}$ content of the $\eta'$. Furthermore
the ratio $R_{\phi} \equiv {\rm BR}(\f\to\eta'\gam)/{\rm BR}(\f\to\eta\gam)$
can be related to the pseudoscalar mixing angle
$\varphi_{\rm P}$ in the basis
$\{\ket u\bar u+d\bar d;/\sqrt2,\ \ket s\bar s;\}$,
offering an important point of comparison
for the description of the $\eta$-$\eta'$ mixing in extended
chiral perturbation theory \cite{Fel00}.
\subsubsection {$\eta'$ decays}
\label{sec:etaprimedec}
At \klo\ , the $\f\to\eta'\gam$ decay is identified either by the
final state with $\eta$ and  charged pions, such as
$\eta'\to\eta\pic$ with $\eta\to 3\po$,
or by the $\eta$ and neutral pions final state, $\eta'\to\eta\pio$ with $\eta\to\pic\po$.
The $\f\to\eta\gam$ decay in the channel where $\eta\to\pic\po$ which,
as we have learnt from the previous section, is practically background free,
is used as the normalization process.

To select $\phi\rightarrow \eta^{\prime}\gamma$ events, we require,
in addition to seven prompt photons, a vertex in a cylindrical region
around the interaction point formed by two oppositely charged tracks.
Then we perform a kinematic fit requiring energy-momentum conservation and
times and path lengths to be consistent with the speed of light for photon candidates.
From about 1.4 billion $\phi$ collected, the final number of $\eta$' events from
the two selected decay chains is, after background subtraction,
$N_{\eta^{\prime} \gamma} = 3405\pm61$.
The selection efficiency for detecting the events, evaluated from
MC simulations and checked extensively with data sub samples,  is
$\epsilon_{\eta^{\prime}}=(23.45\pm0.16)\%$.

We obtained $R_{\phi}=(4.77\pm0.09_{stat}\pm0.19_{sys})\x 10^{-3}$ from which we derive
$BR(\phi \to \eta^{\prime} \gamma)=(6.20\pm0.11_{stat}\pm0.25_{sys})\x 10^{-5}$.
Scant knowledge of intermediate branching ratios prevents us to benefit from having
twentyfold statistics increase over our previous BR determination \cite{etapr}.
\subsubsection {Pseudoscalar mixing angle}
\label{sec:etaprimemix}
The value of $R_{\phi}$ can be related to the pseudo-scalar mixing angle.
The $\eta-\eta^{\prime}$ system can be parametrized in terms of just an
angle only in the quark-flavor basis:
\begin{eqnarray*}
 |\eta> &=& \cos \varphi_P|q\bar{q}> + \sin \varphi_P |s\bar{s}> \\
 |\eta^{\prime}> &=& -\sin \varphi_P|q\bar{q}>  +\cos \varphi_P |s\bar{s}>
\end{eqnarray*}
where $|q\bar{q}>=(1/\sqrt{2})|u\bar{u}+d\bar{d}>$.
Using the approach of ref.\cite{Bramon,Rafael}, where the SU(3) breaking is taken
into account via constituent quark mass ratio $m_s/{\bar m}$, $R_{\phi}$ can be
parametrized as
\begin{equation}
R_{\phi} = \frac {{\rm BR}(\phi \rightarrow \eta^{\prime} \gamma)} {{\text
BR}(\phi \rightarrow \eta \gamma)} =  \cot^{2} \varphi_{P} \left(
1-\frac{m_s}{{\bar m}}\frac{C_{NS}}{C_{S}}\frac{\tan \varphi_V} {\sin2\varphi_{P}}\right )^2
\left( \frac{p_{\eta^{\prime}}}{p_{\eta}}  \right )^3
\label{R}
\end{equation}
where $\varphi_V=3.4^{\circ}$ is the mixing angle for vector mesons; $p_{\eta(\eta^{\prime})}$ is the
${\eta(\eta^{\prime})}$ momentum in the $\phi$ center of mass; the two
parameters $C_{NS}$ and $C_{S}$ take into account the OZI-rule effect
on the vector and pseudoscalar wave-function overlap, they are 0.91$\pm$0.05 and 0.89$\pm$0.07
respectively,
for m$_{s}$/$\bar{m}$=1.24$\pm$0.07 \cite{Rafael}. Then we obtain:

\begin{equation}
\phi_{P} = (41.4\pm0.3_{\text{stat}}\pm0.7_{\text{sys}}\pm0.6_{\text{th}})^{\circ}
\end{equation}
The theoretical uncertainty on the mixing angle has been evaluated from the
maximum variation induced
from the spread of the ratio $m_s/\bar{m}$, $C_{NS}$ and $C_{S}$ values.
In the traditional approach to the mixing, the $\eta-\eta^{\prime}$ meson are parametrized in the octet-singlet basis;
in this basis the value of the mixing angle becomes: $\theta_P =\varphi_{P}-\arctan \sqrt 2 =
(-13.3\pm0.3_{\text{stat}}\pm0.7_{\text{sys}}\pm0.6_{\text{th}})^{\circ}$.\\

\subsubsection {$\eta^{\prime}$ gluonium content}
\label{sec:etaprimegluo}

The QCD involves quanta, the gluons, which are expected to form bound states, and mix with neutral
mesons. While the $\eta$ meson is well understood as an $SU(3)-$flavor octet with a small quarkonium singlet admixture, the $\eta^{\prime}$ meson is a good candidate to have a sizeable gluonium content.
If we allow for a $\eta^{\prime}$ gluonium content, we have the following parametrization:
\begin{equation}
 |\eta^{\prime}> = X_{\eta^{\prime}}|q\bar{q}> + Y_{\eta^{\prime}}|s\bar{s}> + Z_{\eta^{\prime}}|gluon>
\end{equation}
where the $Z_{\eta^{\prime}}$ parameter takes into account a possible mixing with gluonium.
The normalization implies $X_{\eta^{\prime}}^2+Y_{\eta^{\prime}}^2+Z_{\eta^{\prime}}^2=1$ with
\begin{equation}
\eqalign{X_{\eta^{\prime}}&=\cos \phi_G \sin\varphi_P\cr
Y_{\eta^{\prime}}&=\cos\phi_G \cos\varphi_P\cr
Z_{\eta^{\prime}}&=\sin \phi_G\cr}
\label{parglue}
\end{equation}
where $\phi_G$ is the mixing angle for the gluonium contribution. A possible gluonium content of the
$\eta^{\prime}$ meson corresponds to a non zero value for $Z_{\eta^{\prime}}^2$, that implies
\begin{equation}
X_{\eta^{\prime}}^2+Y_{\eta^{\prime}}^2<1
\end{equation}
and \eq{R} becomes:
\begin{equation}
R_{\phi} = \cot^{2} \varphi_{P} \cos^{2} \phi_{G} \left(
1-\frac{m_s}{{\bar m}}\frac{C_{NS}}{C_{S}}\frac{\tan \varphi_V} {\sin2\varphi_{P}}\right )^2
\left( \frac{p_{\eta^{\prime}}}{p_{\eta}}  \right )^3.
\label{eqR}
\end{equation}

Combining our $R_{\phi}$ result with other experimental constraints, we estimate
the gluonium fractional content of $\eta^{\prime}$ meson as $Z^2 =
0.14\pm0.04$ and the mixing angle $\varphi_P = (39.7\pm0.7)^{\circ}$ \cite{didonato}.

%
%
\subsection{Scalar mesons, $f_0(980)$ and $a_0(980)$}
\label{sec:f0}
The composition of the scalar mesons with masses below 1~GeV,
$\sigma(600)$, $f_0(980)$ and $a_0(980)$, is not yet
well understood. \klo\ is well suited for this task,
since the radiative $\phi$ decays into two pseudoscalar
mesons are dominated by a scalar meson exchange,
$\phi \to S \gamma $, with $S= \sigma, f_0, a_0$, and we
have now collected literally millions of such decays.
A detailed proposal on how to elucidate the nature of these
mesons with  \klo\  was one of the first papers JLF had written,
after her arrival to LNF in 1991 \cite{JLF}.
Since then, theoretical models explaining how to describe these scalars
and how to calculate their decay rate
have  proliferated greatly. For this review, we only pick two
models which  are illustrated below.
The investigation on these mesons is experimentally carried out
by measuring branching ratios and
studying the  mass-spectra or the event density
in the Dalitz plot.

Branching Ratios: The branching ratio for the decays
$\f\to f_0\gam\to\pi\pi\gam$ and $\f\to a_0\gam\to\eta\po\gam$
are suppressed unless the $f_0$ and $a_0$ have significant
$s\bar{s}$ content. The BRs for these decays are estimated
to be  of $\sim 10^{-4}$ if the $f_0$ and $a_0$ are
$qq\bar{q}\bar{q}$ states, in which case they contain an
$s\bar{s}$ pair, or of $\sim 10^{-5}$ if the $f_0$
and $a_0$ are conventional $q\bar{q}$ states. If the $f_0$
and $a_0$ are $K\Kb$ molecules, the BR estimates are sensitive
to assumptions about the spatial extension of these states.

Invariant Mass distributions: Fits to the $\pi\pi$  ($\eta\pi^0$) invariant-mass distributions
can also shed  light on the nature of the $f_0$ ($a_0$) since they
provide an estimate of the couplings of these mesons
to the \f\ , $g_{\f S\gamma}$, and/or to the final-state particles.
The rate expression  for $\f\to S\gam$ E1-transitions
contains a factor $E_\gam^3$ , where $E_\gam$
is the energy of the radiated photon, as required by
considerations of phase space and gauge invariance.
As a result, the invariant-mass distributions are cut off above
$m_\f$ and develop a long tail toward lower mass values.
\begin{figure}[htb]
\begin{center}
\includegraphics[width=0.6\textwidth]{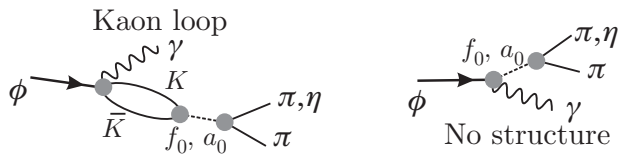}
\caption{Feynman diagrams used to describe the
$\pi\pi$ and $\eta\po$ mass spectra for the $\f\to S\gam$
decay in the kaon-loop (left) and No structure (right) models.}
\label{kloop}
\end{center}
\end{figure}
The fit results therefore strongly depend on the scalar meson
masses and widths and on the  assumed  specific model for the
decay mechanism. Because of the proximity of the $f_0$ and $a_0$
masses to the $K\Kb$ threshold, and because these mesons are
known to couple strongly to $K\Kb$, the kaon-loop model,
\fig{kloop} left, is often used \cite{AI89}. In this model,
the decay proceeds through a virtual $K^+K^-$ pair emitting
the photon and subsequently annihilating into a scalar.
This loop function damps the $E_\gam^3$ behavior. The
transition amplitude depends on the coupling of the
scalar meson both to the pseudoscalar meson in the final
state and to the $K^+K^-$ pair in the loop function. The propagator
include finite width corrections which are relevant close
to the $K\Kb$  threshold.
A second model, called no-structure ~\cite{IMP06}, sketched in
\fig{kloop} right, describes the process as a point-like
interaction where the dynamics of the scalar production is
absorbed in the $g_{\f S \gamma}$ coupling. The amplitude
is described by a Breit-Wigner propagator, with a mass dependent
width, which accounts for analytical continuation under
$\pi\pi$ and $K\Kb$
thresholds, and a complex polynomial describing a continuum background.

\subsubsection{$\phi\rightarrow\pi^0\pi^0\gamma$
and $\phi\rightarrow\eta\pi^0\gamma$ decays.}
\label{sec:phip0p0g}
The first KLOE studies of the  $\f\to\pio\gam$
\cite{f0} and $\f\to \eta\po\gam$ \cite{a0} decays
were performed with the 17~pb$^{-1}$ of data
collected in 2000. All BRs quoted therein are
in agreement with, but are much more precise, than earlier measurements.

The amplitudes contributing to $\f\to\pio\gam$ include $\f\to S\gam$,
with $S = f_0$ or $\sigma$, and $\f\to\rho^0\po$ with $\rho^0\to\po\gam$.
The final state was selected requiring five prompt $\gam$ in the event and
pairing the photons to get the best value for the pion masses. The energy
resolution was improved by performing a kinematic fit to the event.
In this final state, more than 50\% of the events were due to
the not-resonant $e^+e^- \to \omega \pi^0$ process,
with $\omega \to \pi^0\gamma$. This background
 was largely removed by directly cutting around $m_{\omega}$ on the
reconstructed $\pi \gam $ invariant mass.
After this rejection, the $M_{\pi\pi}$ distribution was
fit with a function including in the amplitude the terms
describing the process $\f\to S\gam$,  $\f\to\rho\pi$,
and their interference. We obtained a ${\rm BR}(\f\to\pio\gam)$
of \pt(1.09\pm0.03\pm0.05),-4,.

In analogy, for the $a_0$ case, the amplitudes contributing
to the process are $\f\to a_0\gam$ and
$\f\to\rho^0\po$, with $\rho^0\to\eta\gam$.
The not-resonant process $e^+e^- \to \omega \pi^0$
with $\omega \to \eta\gamma$,
constitutes a practically negligible background.
The final states corresponding to
two different $\eta$ decay channels were studied:
$\eta\to\gam\gam$ ($5\gam$ final state) and
$\eta\to\pic\po$ ($\pic5\gam$ final state).
The $M_{\eta\pi}$ distributions of both samples
were fit simultaneously with a function analogous to that discussed
above. The values obtained for ${\rm BR}(\f\to\eta\po\gam)$ were
\pt(8.51\pm0.51\pm0.57),-5, and \pt(7.96\pm0.60\pm0.40),-5,,
for the $5\gam$ and $\pic5\gam$ samples respectively.

The analyses of the $\f\to\pio\gam$ and $\f\to\eta\po\gam$ decays
based on a later data set, benefit of an increase in statistics
by a factor of \ab30 such that we could improve background removal
and analysis cuts.

\begin{figure}[htb]
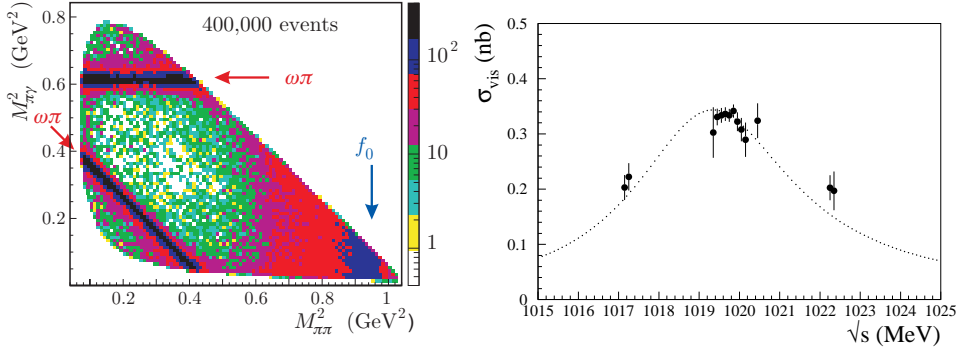

\figb dalitz_mass2_logz;6;
\figb ls_f0n_cimento;7.0;
 \caption{left, $M_{\pi_0\pi_0}$ Dalitz plot in logarithmic
    scale after background subtraction. The two bands in the low
    $M_{\pi\pi}^2$ region represent the $e^+ e^- \to\omega\pi^0$
    background. right, $\sigma_{e^+ e^- \to\pio\gam}$ near the $f_0$ mass region.}
    \label{p0p0g_dalitz}
\end{figure}

\begin{figure}[htb]
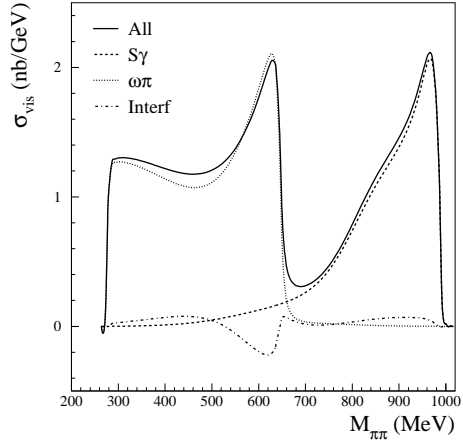

\centering\figb fit_all_bw_scatt1;6;
\caption{$M_{\pi_0\pi_0}$ spectrum coming from the KL fit result. }
   \label{fop0p0}
\end{figure}

For the $\pi^0\pi^0\gamma$ final state, the analysis is done without
cutting on the $\pi^0\gamma$ invariant mass, which could
bias the residual $M_{\pi\pi}$ spectra, and plotting all events
with same topology in the Dalitz plots shown in \fig{p0p0g_dalitz} left.
Two clear bands appear related to the not-resonant process
which constitutes the larger background to the scalar term
for $M_{\pi\pi}< 700$~MeV. Above this threshold, the $f_0(980)$
contribution is almost background free. We estimate with  theory
the event density on the Dalitz-plot by adding to the amplitudes
$\phi\to S\gamma$ and $\phi \to \rho \pi^0$ the complete VDM
description of the not-resonant process. The interferences
between all diagrams are  also taken into account~\cite{simona}.
We then fit the Dalitz plot density, by folding theory with reconstruction
efficiencies and with a smearing matrix describing the probability for an
event  to migrate from a bin to another one.
In particular, we fit data with (1), an improved k-loop model \cite{AK05},
and (2) with the no-structure model \cite{IMP06}.
In the improved k-loop model,the $f_0$ and $\sigma$
description are strongly coupled with the
energy dependence of the $\pi\pi$ scattering phase. Ten different parametrizations
are introduced. We find that,  without adding the $\sigma$ in the model,
it is impossible to adequately describe our data-set.
In \fig{fop0p0} the function
describing $M_{\pi\pi}$, as extracted by our best fit result for the k-loop
model, is shown. The contribution of each process is also
represented. A large coupling of the
$f_0(980)$ to kaons is obtained with large theory errors related to the
parametrization choice (see table \ref{f0coupling}). Discussing with
the authors of ref.~\cite{AK05}, some typos in the parametrization were found.
After correcting for it, a new preliminary set of values
is extracted which much reduced theory errors, referred as
{\em $\pi^o\pi^o\gamma$ update} in table \ref{f0coupling}.
All of the above indicates a not-conventional structure for the $f_0$ meson.
For the no-structure model, the situation is really different. The
$f_0$ coupling to kaons get reduced and the introduction of the
$\sigma$ does not substantially modify the results. However,
the physical interpretation gets somehow hidden from the introduction
of the continuum background in the parametrization.

\begin{table}[htb]
\vskip 0.1 in
\renewcommand{\arraystretch}{1.7}
{
\begin{tabular}{@{}cccc@{}}
\hline
Parameter & $\ \ \  \ \ \pi^+\pi^-\gamma \ \ \ \  \ $  & $\pi^0\pi^0\gamma$ & $\ \ \pi^0\pi^0\gamma$ updates $\ \ $ \\
\hline
 $m_{f_0}$ (MeV)         & $980\div987$ &
$976.8\pm0.3_{\rm fit}\,^{+0.9}_{-0.6}\,_{\rm sys}+10.1_{\rm th}$ & $984.7\pm1.9_{\rm th}$ \\
  $g_{f_0K^+K^-}$ (GeV)         & $5.0\div6.3$ &
$3.76\pm0.04_{\rm fit}\,^{+0.15}_{-0.08}\,_{\rm sys}\,^{+1.16}_{-0.48}\,_{\rm th}$ & $3.97\pm0.43_{\rm th}$\\
  $g_{f_0\pi^+\pi^-}$ (GeV) & $3.0\div4.2$ &
$-1.43\pm0.01_{\rm fit}\,^{+0.01}_{-0.06}\,_{\rm sys}\,^{+0.03}_{-0.60}\,_{\rm th}$ & $-1.82\pm0.19_{\rm th}$ \\[1.2ex]
 $\frac{g_{f_0K^+K^-}^2}{g_{f_0\pi^+\pi^-}^2}$ & $2.2\div2.8$ &
$6.9\pm0.1_{\rm fit}\,^{+0.2}_{-0.1}\,_{\rm sys}\,^{+0.3}_{-3.9}\,_{\rm th}$ & ~ \\ 
\hline
  $g_{f_0K^+K^-}$ (GeV)         & $1.6\div2.3$ &
$0.40\pm0.04_{\rm fit}\,^{+0.62}_{-0.29}\,_{\rm sys}$ & ~ \\
  $g_{f_0\pi^+\pi^-}$ (GeV) & $0.9\div1.1$ &
$1.31\pm0.01_{\rm fit}\,^{+0.09}_{-0.03}\,_{\rm sys}$ & ~ \\
  $\frac{g_{f_0K^+K^-}^2}{g_{f_0\pi^+\pi^-}^2}$ & $2.6\div4.4$ &
$0.09\pm0.02_{\rm fit}\,^{+0.44}_{-0.08}\,_{\rm sys}$ & ~ \\
 $g_{\phi f_0\gamma}$ (GeV$^{-1}$) & $1.2\div2.0$ &
$2.61\pm0.02_{\rm fit}\,^{+0.31}_{-0.08}\,_{\rm sys}$ & ~ \\ 
\hline
\end{tabular}
}
\centering\caption{\it Comparison of $f_0$ parameters estimates between
k-loop and no-structure models. In the last column, we report the preliminary
update of the KL fit to the $\pi^0\pi^0\gamma$ final state.}
\label{f0coupling}
\end{table}

By integrating the scalar term contribution, we determine a value for
the BR which slightly depends on the theory model used in the fit. We
quote a BR to $\f\to\pio\gam$ \cite{simona} of:
$${\rm BR}(\f\to S\gam \to \pio\gam)=\pt\left(1.07^{+0.01}_{-0.04}{\rm(fit)}\ ^{+0.04}_{-0.02}{\rm(syst)}\ ^{+0.06}_{-0.05} {\rm (mod)}\right),-4,.$$

\begin{figure}[htb]
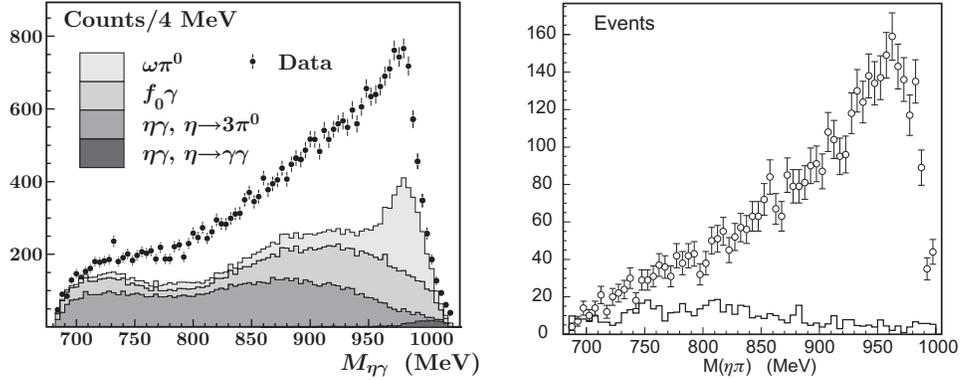

\centering\figb spettrotot;6.25; \kern.5cm \figb signfondo;5.75;
  \caption{Distribution of $M_{\eta\pi^0}$ for $\phi\to\eta\pi^0\gamma$ events. Right. $\pi^+\pi^-\pi^0\pi^0$ final state. Left. Five photon final state.  Points, data; solid line, fit result.}
    \label{aoetap}
\end{figure}

For the $\eta \pi^0 \gamma$ final state, same analysis strategy of the previous work was followed with the advantage of using a more realistic
detector response and simulation of the machine background in the
official KLOE Monte Carlo. The physics background was estimated after
event preselection by fitting MC shapes to data. At the end of the
analysis, we have a signal efficiency of $\sim$ 40\% (20\%) and a
residual background of 55\% (15\%) for the sample with
$\eta \to \gamma\gamma$ ($\eta \to \pi^+\pi^-\pi^0$). The absence
of a major source of interfering background allows to get the BR
directly from event counting. We get a preliminary value for the
${\rm BR}(\f\to\eta\po\gam)$ of
\pt(6.98 \pm0.10\pm0.23_{syst}),-5, and \pt(7.05\pm0.10\pm0.21),-5,,
for the $5\gam$ and $\pic5\gam$ samples respectively, which corresponds
to  a decrease of the central value with respect to our previous
measurement of $\sim$ 12\% with an improvement in precision better
than a factor 3.  In \fig{aoetap}, the $\eta\pi^0$ invariant
mass distribution, $M_{\eta\pi^0}$, is shown for both final states.
A simultaneous fit to the spectra is also performed to extract
the relevant $a_0$ coupling. Also for this meson, a large coupling
to kaons is found.

\subsubsection{\f\to$\pi^+\pi^-$\gam}
\label{sec:pipig}
KLOE has also published a study of the decay $\f\to f_0\gam\to\pic\gam$
\cite{f0ppg}. The $f_0$'s charged mode was extremely difficult to identify because only a small fraction of the $e^+e^-\to\pic\gam$ events involve the $f_0$; the principal contributions are from events in which the photon
is from ISR or FSR. The analysis is performed on the $M_{\pi\pi}$ distribution for events with large photon polar angle where the ISR contribution is much reduced. The $M_{\pi\pi}$ distribution was fit with a function composed of analytic expressions describing  the ISR, FSR, $\rho\pi$ contributions and two terms that describe the decay $\f\to S\gam\to\pi\pi\gam$ and its
interference with FSR which could be constructive or destructive.
\begin{figure}[htb]
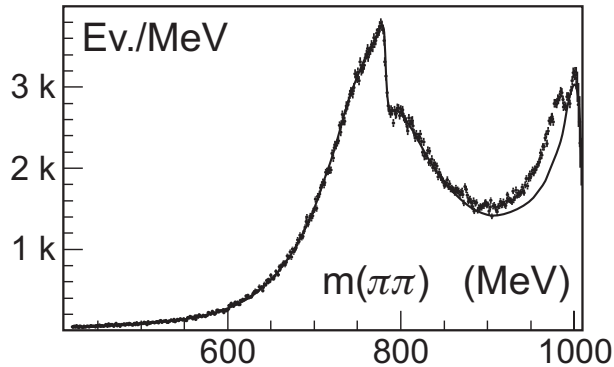

\centering\figb ppg-minv;8;
  \caption{Distribution of $M_{\pi^+\pi^-}$ spectra for $\phi\to\pi^+\pi^-\gamma$ events.}
    \label{fopippima}
\end{figure}

\Fig{fopippima} shows the $M_{\pi\pi}$ distribution,
with the result of the kaon-loop fit superimposed. The overall appearance of the distribution is dominated by the radiative return to the $\rho$; the $f_0$ appears as
the peak-like structure in the region 900--1000 MeV.
Both the kaon-loop and no-structure fits strongly prefer destructive interference between the $S\gam$ and FSR amplitudes.
The kaon-loop fit gives coupling values in reasonable agreement
with those obtained from the fit to the KLOE $\f\to\pio\gam$ data
discussed before. Integrating the appropriate terms of the fit functions gives values for ${\rm BR}(\f\to f_0\gam)$ in the neighborhood of \pt2,-4,.

Another important test in this final state is the
pion forward backward  asymmetry. The $\pi^+\pi^-$
pair has a different charge conjugation eigenvalue
depending on whether it is produced from FSR and $f_0(980)$
($C=+1$) or ISR ($C=-1$). An interference term
between two amplitudes of opposite charge conjugation
gives rise to $C$-odd terms that change sign by
the interchange of the two pions and results in an asymmetry $A_{\rm c}$:
$$
A_{\rm c}=\frac{N(\theta_{\pi^+}>90^{\circ})-N(\theta_{\pi^+}<90^{\circ})}
 {N(\theta_{\pi^+}>90^{\circ})+N(\theta_{\pi^+}<90^{\circ})}~.
$$
\begin{figure}[htb]
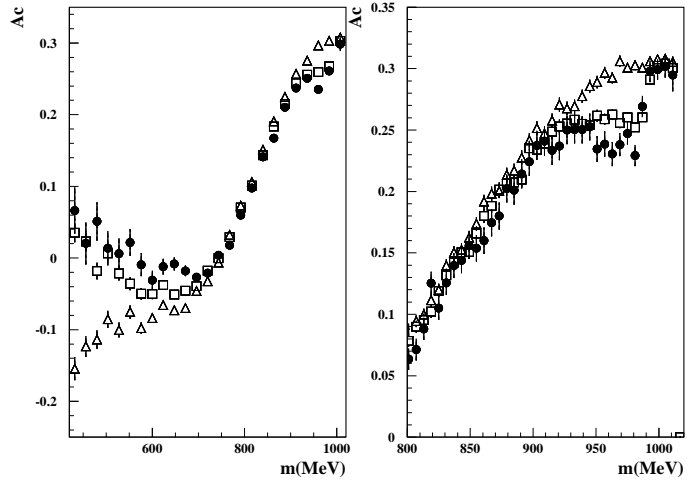

\centering\figb asim;9;
  \caption{Pion forward backward asymmetry distribution. Data: black dots;
    Simulation FSR+ISR: open triangles; Simulation FSR+ISR+ scalar(KL):
    open squares. Right: zoom of the $f_0$ mass region.}
    \label{fopippim}
\end{figure}

\Fig{fopippim} shows the comparison of the asymmetry
data with the expectations based on the
KL parametrization of the $f_0(980)$.~\cite{Czyz:2004nq,Pancheri:2006cp}
The $A_{\rm c}$ data distribution is reproduced only
by calculations including the $f_0(980)$.

\subsubsection{$\phi\rightarrow K\bar K\gamma$}
\label{sec:kkbarg}
A fundamental consistency check of the study of $\phi$ radiative
decays, as well as being another important tool for the investigation
of the $s$ quark content of $\f_0(980)$ and $a_0(980)$, is the measurement
of final states involving kaons.
The search of $\phi\to K\bar{K}\gamma$, which is expected to proceed
mainly through the $[f_0(980)+a_0(980)]\gamma$ intermediate state,
adds relevant information on the scalar meson structure.
Theoretical predictions for the BR spread over several orders
of magnitude, although the latest evaluations essentially concentrate
in the region of $10^{-8}$ (see \cite{KLOE_SKK} and references therein).
Experimentally, this decay has never been observed.
Using 1.4 fb$^{-1}$ collected at the $\phi$ resonance and an
equivalent MC statistics to model the background, the KLOE experiment
searched for this decay using the
$\phi\to\ks\ks\gamma\to\pip\pim\pip\pim\gamma$
decay channel \cite{KLOE_SKK}. The significant signal reduction
(24\% of the total rate) is compensated by a very clean event topology.
The MC signal has been generated according to phase-space and
radiative decay dynamics.
The selection cuts applied to reject the main background component
($\phi\to\ks\kl$ with an ISR or FSR photon) have been optimized
using MC only: no events survive the analysis cuts, while one
event is found in data. The resulting upper limit on the branching ratio of
$\phi\to\ks\ks\gamma<1.8\times 10^{-8}$ @ 90\% CL
rules out most of the theoretical predictions.
\subsection{Vector mesons}
\subsubsection{$\phi$ leptonic widths}
\label{sec:lepwidth}
The \f-leptonic widths provide information on the \f-structure and its production cross section in \epm annihilations. They are necessary for decay branching ratio measurements and estimates of the hadronic contribution to vacuum polarization. There is no direct measurement of the leptonic width.
The cross sections for processes\tot\ can be written as $\sigma_{ee} = \sigma_{ee,\phi}+ \sigma_{\rm int}$ and $\sigma_{\mu\mu} = \sigma_{\mu\mu,\phi} +\sigma_{\rm int}$, with $\sigma_{\rm int}$, the interference term, given by:
\begin{equation}
  \sigma_{\rm int} =\frac{3\alpha\Gamma_{\ell\ell}}{m_\phi}\:
  \frac{W^2-m_\phi^2}{(W^2-m_\phi^2)^2+W^2\Gamma_\phi^2}\:
  \int^{\cos\theta_{max}}_{\cos\theta_{min}} f_{\ell\ell}(\theta)~{\rm d}\cos\theta.
  \label{interfer}
\end{equation}
where W is the energy in the collision center of mass (CM), $\theta_{min}$ and $\theta_{max}$ define the acceptance in the polar angle $\theta$ (see later) and where $\Gamma_{\ell\ell}=\Gamma_{ee}$ for \baba\ and $\Gamma_{\ell\ell}=\sqrt{\Gamma_{ee}\Gamma_{\mu\mu}/\xi}$ for \mumu. The $\xi$ term takes into account for the phase space correction. KLOE used data collected at CM energies $W$ of 1017.2 and 1022.2 \mev, i.e. $m_{\phi}\pm\Gamma_\phi/2$, and at the $\phi$ peak, $W$=1019.7 \mev. For \mumu\ we measure the cross section. Since Bhabha scattering is dominated by the photon exchange amplitude, the interference term is best studied in the forward-backward asymmetry, $A_{FB}$, defined as
\begin{equation}
  A_{FB}= \frac{\sigma_F-\sigma_B}{\sigma_F+\sigma_B}.
  \label{asy}
\end{equation}
where $\sigma_F$ and $\sigma_B$ are the cross sections for events with electrons in the forward and backward hemispheres.

 We find $\Gamma_{ee}$=1.32\plm 0.05\plm 0.03\kev~and $\sqrt{\Gamma_{ee}\Gamma_{\mu\mu}}$=1.320\plm 0.018\plm 0.017\kev. These results, compatible with $\Gamma_{ee}$=$\Gamma_{\mu\mu}$, provide a precise test of lepton universality. Combining the two results gives $\Gamma_{\ell\ell}(\f)=1.320\pm0.023$ \kev \cite{llwidth}.

\subsubsection{$\phi \rightarrow \pi^+ \pi^- \pi^0$ }
\label{sec:pipipi0}
The decay of the $\phi$ meson to $\pi^+\pi^-\pi^0$, with a branching ratio
(BR) of \ab15.5\%, is dominated
by the $\rho\pi$ intermediate states \cite{Parrour}
$\rho^+\pi^-$, $\rho^-\pi^+$, and $\rho^0\pi^0$ with equal amplitudes.
CPT invariance requires equality of the masses and widths of
$\rho^+$ and $\rho^-$, while possible mass or width differences between
$\rho^0$ and $\rho^{\pm}$ are related to isospin-violating electromagnetic effects.
Additional contributions to $e^+ e^- \rightarrow \pi^+ \pi^- \pi^0$ are the so called ``direct term'',
$\phi \rightarrow \pi^+ \pi^- \pi^0$,
and $e^+ e^- \rightarrow  \omega \pi^0$, $\omega \rightarrow \pi^+ \pi^-$.
We use some 2 million of  $e^+ e^- \rightarrow \pi^+ \pi^- \pi^0$ events
to determine the masses $M(\rho^{\;+-0})$ and widths
$\Gamma(\rho^{\;+-0})$ of the three charge states
of the $\rho$-meson, from a fit to the Dalitz plot density distribution \cite{Bini03}.
%
%
If we define the variables $x=T^+\!-T^-$ and $y=T^0$, where $T^{\;+\;,\;-,\;0}$
are the kinetic energies
of the three pions in the center of
mass system (CM),
the Dalitz plot density distribution $D(x,y)$ is given by:
\begin{equation}
\label{eqdalitz}
D(x,y)\propto|\vec p^{\;*}_{+}\times\vec p^{\;*}_{-}|^2|A_{\rho\pi}+A_{\rm
  dir}+A_{\omega\pi}|^2 \nonumber
\end{equation}
where $\vec p^{\;*}_{\pm}$ are the $\pi^{\pm}$ momenta in the CM and
$A_{\rho\pi}$, $A_{\rm dir}$ and
$A_{\omega\pi}$ are the three amplitudes described above and containing the
dependence on the masses and widths of the $\rho$-mesons.
$\pi^+\pi^-\pi^0$ events are selected by asking for two
non-collinear tracks with opposite sign of curvature and polar angle $\theta>40^\circ$ which intersect the
interaction region. The acollinearity
cut ($\Delta\theta<175^\circ$) removes $e^+e^-\gamma$ events without incurring an
acceptance loss for the signal.
The missing mass, $M_{\rm miss}= \sqrt{(E_\phi-E_{\pi^+}-E_{\pi^-})^2-|\vec p_\phi- \vec
  p_{\pi^+}-\vec p_{\pi^-}|^2}$, where $E$ and $\vec p$ are laboratory energies and momenta,
is required to be within 20~MeV
of the $\pi^0$ mass. This corresponds to an effective energy cut of
$\leq20$ MeV on the total
energy radiated because of initial state radiation (ISR).
Two photons in the calorimeter are also required,
with opening angle in the $\pi^0$ rest frame $\cos\theta_{\gam\gam}<-0.98$.
The Dalitz plot variables $x$ and $y$ are evaluated using the measured momenta
  of the charged pions,
boosted to the center of mass system: $x=E_{+}^{*}-E_{-}^{*}$ and
  $y=E_\phi^{*}-E_{+}^{*}
-E_{-}^{*}-m_{\pi^0} =T_{\po}$.
$E_{\phi}$ and $\vec{p}_{\phi}$
are measured run by run using Bhabha scattering events.
The resolution on $x$ and $y$ is about 1~MeV over the full kinematical
range.
Three $\chi^2$ fits to the Dalitz plot density distribution have been
performed:
a) a fit assuming CPT and isospin invariance,
\ie\, $m_{\rho^0}=m_{\rho^+}=m_{\rho^-}$, $\Gamma_{\rho^0}=\Gamma_{\rho^+}=
\Gamma_{\rho^-}$; b) a fit assuming only CPT invariance \ie\ $m_{\rho^{+}}=
m_{\rho^-}$, $\Gamma_{\rho^+}=
\Gamma_{\rho^-}$; and finally c) without assumptions on masses and widths.
The results of the fits are shown in table \ref{param}.

\begin{table}[htb]
\begin{center}
\begin{tabular}{lcccc}
\hline
 parameter & fit(a) & fit(b) & fit(c) \\
\hline

$\chi^2$ [$p(\chi^2)$]&1939 [12\%]&1914 [21\%]&1902 [26\%] \\
$m_{\rho^0}$&                   &$775.9\pm0.5\pm0.5$&$775.9\pm0.6\pm0.5$\\
$m_{\rho^+}$&$775.8\pm0.5\pm0.3$&$775.5\pm0.5\pm0.4$&$776.3\pm0.6\pm0.7$\\
$m_{\rho^-}$&                   &                   &$774.8\pm0.6\pm0.4$\\
$\Gamma_{\rho^0}$&                   &$147.3\pm1.5\pm0.7$&$147.4\pm1.5\pm0.7$\\
$\Gamma_{\rho^+}$&$143.9\pm1.3\pm1.1$&$143.7\pm1.3\pm1.2$&$144.7\pm1.4\pm1.2$\\
$\Gamma_{\rho^-}$&                   &            &$142.9\pm1.3\pm1.4$ \\
\hline
\end{tabular}
\end{center}
\caption{Fit results. Masses and widths are in MeV. The amplitudes are
scaled in such a way that $a_{\rho}$=1.
$\chi^2$ values and probabilities of the fits are given in the first row.
}
\label{param}
\end{table}
The convergency of fit (a) shows that the
experimental distribution is consistent with CPT and isospin invariance.
The $\rho$ masses  and the widths
are close to recent results \cite{pdg06}.
The results obtained for mass and width differences, including correlations
between the parameters, are:
$m_{\rho^0}-m_{\rho^{\pm}}=0.4\pm0.7\pm0.6$
MeV,$m_{\rho^+}-m_{\rho^-}=1.5\pm0.8\pm0.7$ MeV,
$\Gamma_{\rho^0}-\Gamma_{\rho^{\pm}} = 3.6\pm1.8\pm1.7$ MeV,
$\Gamma_{\rho^+}-\Gamma_{\rho^-} = 1.8\pm2.0\pm0.5$ MeV.
Note that the neutral and charged $\rho$ masses are essentially equal within statistics.
%
%
\subsubsection{$\phi \rightarrow \omega \pi$}
\label{sec:omegap}
Using $\sim 600$ pb$^{-1}$ collected at center of mass energies between
1000 and 1030 MeV, the KLOE experiment studied the production cross
section of $e^+ e^- \to \pi\pim\pio$ and $ e^+ e^- \to\pio\gamma$
processes \cite{KLOE_wp}, which mainly proceed through the non-resonant
$\rho\to\omega\pi^{0}$ intermediate state.
The dependence of the cross section on the center of mass energy is
parametrized in the form $\sigma (W) = \sigma_0(W)
|1-Z\frac{m_\phi\Gamma_\phi}{D_\phi}|$, where $Z$ is the complex
interference parameter between the $\phi$ decay amplitude and the
non-resonant processes, whose bare cross section is $\sigma_0(W)$.
$m_\phi$, $\Gamma_\phi$ and $D_\phi$ represent the mass, the width
and the inverse propagator of the $\phi$ meson respectively. A model
independent parametrization is used to describe the increase of the
non-resonant cross section with $W$, which is linear in this energy
range: $\sigma_0(W) = \sigma_0+\sigma'(W-m_\phi)$.

\begin{figure}[htb]
\begin{center}
\includegraphics[width=7.25cm]{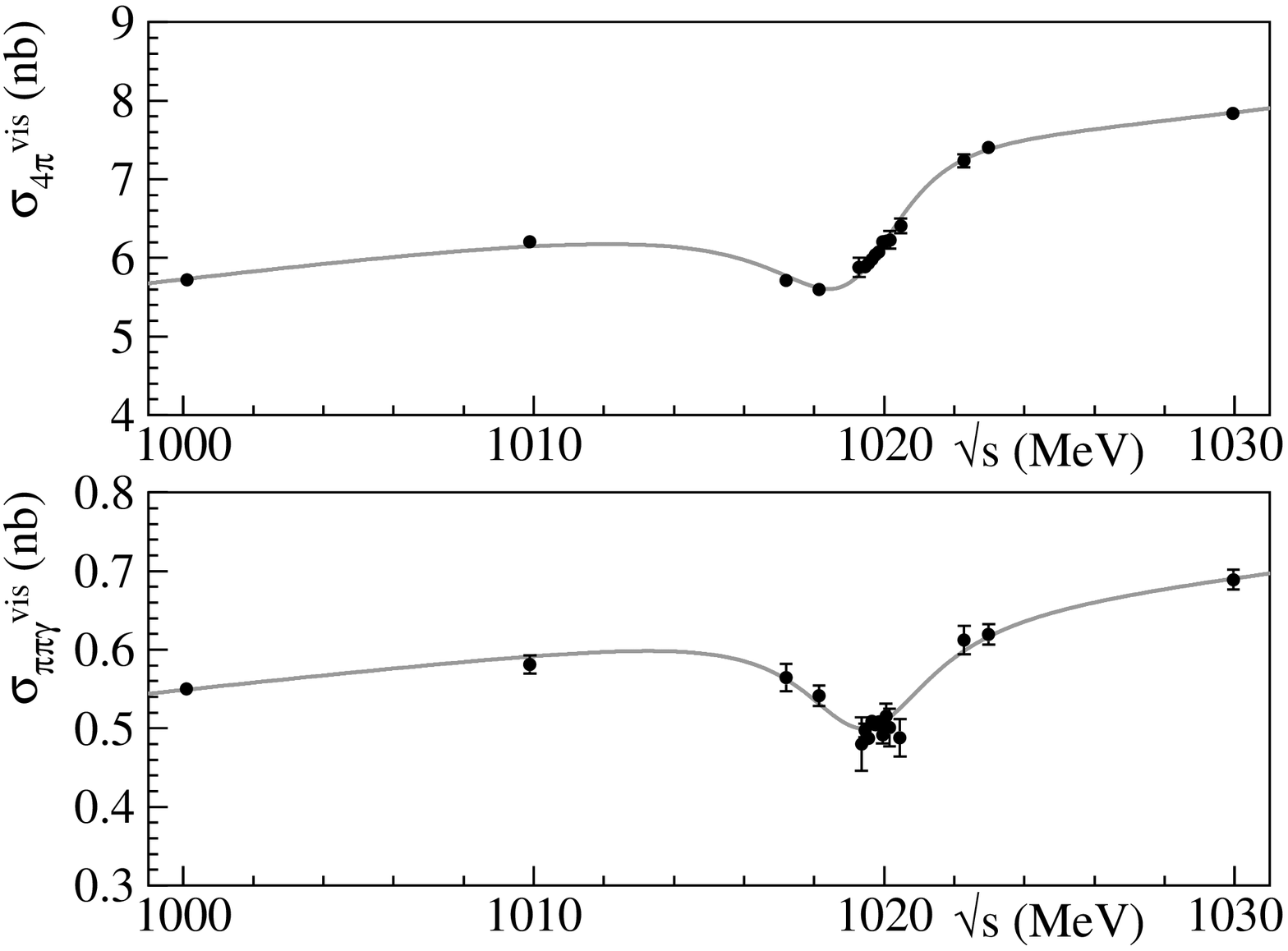} 
\includegraphics[width=5.4cm]{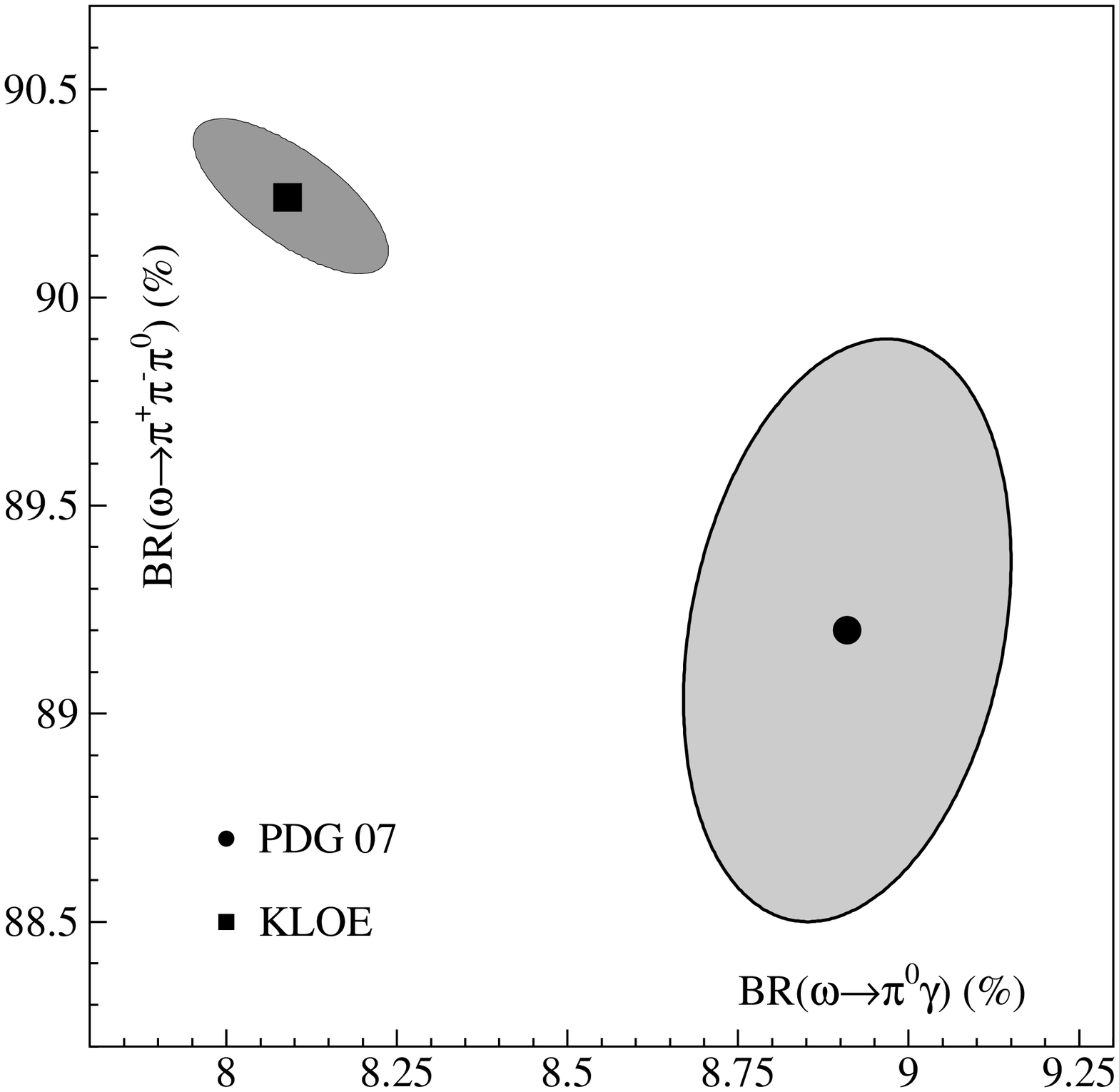}
\end{center}
\caption{Cross section fit results for the $e^+ e^- \to \pi\pim\pio$ (top-left) and
$ e^+ e^- \to\pio\gamma$ (bottom-left) channels. Black dots are data, solid line is the resulting fit function. Branching fraction for the two main $\omega$ decay
channels (right). The square is the KLOE fit result, while the dot is the
PDG constrained fit result. The shaded regions are the 68\% CL }
\label{fig:cross-br}
\end{figure}

In \fig{fig:cross-br} data points with the superimposed fit function are
shown for both channels.

By fitting the interference pattern for both final states, a
preliminary ratio
$\Gamma(\omega\to\po\gamma)/\Gamma(\omega\to\pip\pim\po) =   0.0897 \pm 0.0016$
is obtained.
Since these two final states represent the 98\% of the $\omega$
decay channels, this ratio and the sum of the existing BR
measurements on rarest decays are used to extract the main
$\omega$ branching fractions imposing unitarity.
BR$(\omega\to\pip\pim\po) = (\,90.24 \pm 0.18\,)\%$
and
BR$(\omega\to\po\gamma) = (\,8.09 \pm 0.14\,)\%$
are extracted, with a correlation of -71\%.
The parameters describing the $e^+ e^- \to\pip\pim\pio$ reaction
around $m_\phi$ allow to derive the branching fraction for
the OZI and G-parity violating decay $\phi\to\omega\po$, which is
found to be:
BR$(\phi\to\omega\po) = (\,4.4 \pm 0.6\,) \times 10^{-5}$.

\section{The hadronic cross section and $a_\mu$}
\label{sec:sighad}
\subsection{The muon anomaly $a_\mu$}
\label{sec:amu}
The observation of the anomalous magnetic moment of the electron
helped drive the development of quantum electrodynamics (QED).
The value of the muon anomaly, $a_\mu$,
is $(m_\mu/m_e)^2 \approx \mbox{40,000}$ times
more sensitive than that of the electron to high-mass states
in the polarization of the vacuum.
The Muon $g-2$ Collaboration (E821) at Brookhaven has used stored muons
to measure $a_\mu$ to 0.5~ppm \cite{E821+04}:
\begin{equation}
a_\mu = \frac {g_\mu - 2}{2} =
\pt116{,}592{,}080\pm60,-11,.
\end{equation}
The muon anomaly receives contributions from QED, weak, and hadronic
loops in the photon propagator, as well as from light-by-light scattering.
The lowest-order hadronic contribution is
$a_\mu^{\rm had} =$ \ab\pt{7000},{-11},, with an uncertainty of \ab\pt{60},{-11},.
To the extent that this uncertainty
can be reduced, the E821 measurement offers a potential probe
of new physics at TeV energy scales.

The low-energy contribution to $a_\mu^{\rm had}$
cannot be obtained from perturbative QCD.
However it can be calculated from
measurements of the cross section
for $e^+e^-$ annihilation into hadrons via the dispersion integral

\begin{figure}[htb]
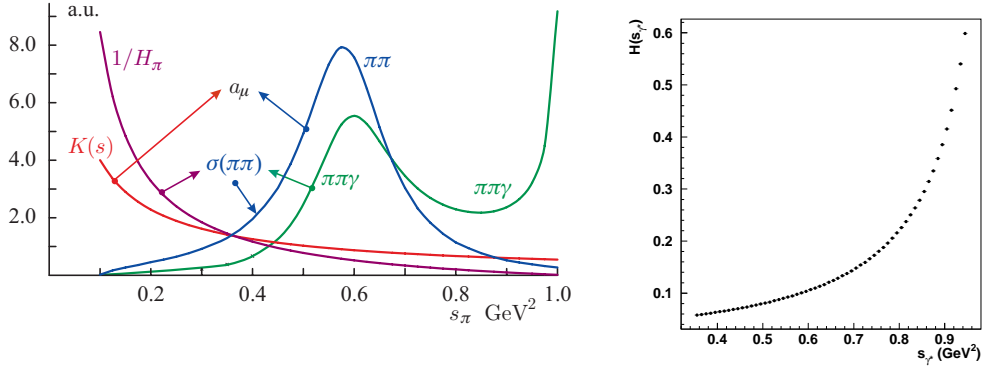

\centering\figb PipiGam1; 7.5;\kern.5cm\figb radfunction;5.5;
  \caption{left: connection between $\sigma(\pi\pi\gamma)$, H, $\sigma(\pi\pi)$, K and $a_{\mu\mu}$. Right, H, radiator function, $\sigma_{\pi\pi\gamma}$ with pion form factor set to one.}
    \label{hradiatorl}
\end{figure}

\begin{equation}
a_\mu^{\rm had} = \frac{1}{4\pi^3} \sum_f \int_{s_{\rm th}(f)}^\infty
       K(s)\, \sigma(e^+e^-\to f)\, ds,
\end{equation}
where $K(s)$, the QED kernel, is a monotonic function that varies
approximately as $1/s$, with $s$ the squared center-of-mass collision
energy, see \Fig{hradiatorl}, left. This amplifies the importance of the cross-section
measurements at low energy. Approximately two-thirds of the integral
is contributed by the process $e^+e^-\to\pic$ for $\sqrt s < 1$~GeV, i.e., in the vicinity of $m_\rho$.
Before the arrival of the KLOE results we discuss below, calculations
of $a_\mu^{\rm had}$ from world $e^+e^-$ data
(dominated at low energies by the CMD-2 measurement of \cite{CMD2})
led to a value of $a_\mu$ \ab$2.4\ \sigma$ lower \cite{DM04}
than the final value reported by E821.
\subsection{Measurement of $\sigma_{\pi^+\pi^-}$ and $a_\mu$ at KLOE}
\label{sec:sighadamu}
\DAF\ is highly optimized for running at $W = m_\f$
and no large variation of $W$ are accessible.
However, initial state radiation, ISR,  via the process $e^+e^-\to\pic\gam$ naturally provides access to hadronic states of lower mass.
The cross section $\sigma_{\pic}$ over the entire interval
from threshold to $m_\f$ is related to the
distribution of $s_\pi = M^2_{\pi\pi}$ for $\pic\gam$ events,
\begin{equation}
\sigma_{\pic}(s_\pi) = \frac{s_\pi}{H(s_\pi|s)}\,
\frac{d\sigma_{\pic\gam}}{ds_\pi}\Big|_{\rm ISR},
\label{eq:radiator}
\end{equation}
where $H$, the ``radiator function,'' describes the ISR spectrum.
Note the subscript ISR on the differential $\pic\gam$ cross section.
This is very important because the contribution from final state photon emission, FSR is of the same order as that from the ISR process unless one imposes special fiducial volume cuts to suppress it.

\begin{figure}[htb]
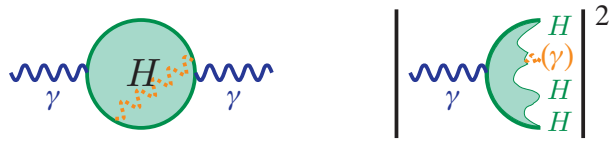

\centering\figb sig-h; 8;
  \caption{photon spectral function and cross section.}
    \label{spectralfunction}
\end{figure}

On the other hand, as is depicted in \fig{spectralfunction}
one must remember eventually to include FSR in the final evaluation of $a_\mu^{\rm had}$.
To correctly calculate $H$ and estimate the effects of FSR,
an accurate simulation of the $\sigma_{\pi\pi\gamma}$ is critical.
The radiator function used by KLOE to obtain $\sigma_{\pi\pi}$
is provided by the \texttt{Phokhara} code~\cite{Rodrigo:2001kf} by setting
the pion form factor $F_\pi(M_{\pi\pi})=1$, see \Fig{hradiatorl},
right. The \texttt{Phokhara} event generator has been continuously improved
by inclusion of next-to-leading-order ISR (two initial-state photons), leading-order FSR and ISR-FSR~\cite{PHOK3}.

Furthermore, $\sigma_{\pi\pi}$ has to be corrected for the running of the fine
structure constant~\cite{Jegerlehner:2006ju} (vacuum polarization) and for shifting from $M_{\pi\pi}$ to the virtual photon mass, $M_{\gamma^*}$ for those events
with both an initial and a final state radiated photon. The \texttt{Phokhara}
version of Ref.~\cite{Czyz:2005as} contains also these contributions and it
is used for the evaluation of the acceptance correction and for all
reconstruction efficiencies.
\subsubsection{Extraction of $a_\mu^{\pi\pi}$ and pion form factor}
\label{sec:amuextr}
The differential cross section of $e^+e^-\to\pi\pi\gamma$ as a function of
$M_{\pi\pi}$ is obtained by subtracting, bin by bin, the background events
$N_{bkg}$ from the number of observed events $N_{obs}$, correcting for
the selection efficiency, $\epsilon_{sel}(M_{\pi\pi}^2)$,
and finally dividing by the total luminosity obtained as described in sect.~\ref{sec:Lumi}:
\begin{equation}
\frac{\dd\sigma_{\pi\pi\gamma}}
{\dd M_{\pi\pi}^2} = \frac{N_{obs}-N_{bkg}}
{\Delta M_{\pi\pi}^2}\, \frac{1}{\epsilon_{sel}(M_{\pi\pi}^2)~ \mathcal{L}}~ ,
\label{eq:2}
\end{equation}
where our mass resolution allows us to have bins of
width $\Delta M_{\pi\pi}^2=0.01\GeV^2$.

The background content is found by fitting spectra
of selected data sample with a superposition
of Monte Carlo distributions describing the signal and
background sources. The only free parameters of such fits
are the relative weights of signal and backgrounds in data,
computed as a function of $M_{\pi\pi}$.

The "bare" cross section  $\sigma_{\pi\pi(\gamma)}^{bare}$ which is
obtained after applying the corrections written in \eq{eq:radiator},
inclusive of FSR, is used to determine $a_\mu^{\pi\pi}$:
\begin{equation}
\label{eq:3}
a_\mu^{\pi\pi} =
\frac{1}{4\pi^3}\int_{s_{low}}^{s_{up}}\dd s~
\sigma_{\pi\pi(\gamma)}^{bare}(s)\,K(s)~ ,
\end{equation}
where the lower and upper bounds of the spectrum depend on specific measurement configurations.

The measured $\sigma_{\pi\pi}$ is related to the pion form factor via the
relation $\sigma_{\pi\pi}(s)=\frac{\pi\alpha^2}{3s}\beta_\pi^3\left|F_\pi(s)\right|^2$,
where $s$ is the center-of-mass energy squared, $m_\pi$ is the pion mass
and $\beta_\pi=\sqrt{1-4m_\pi^2/s}$ is the pion velocity in the
center-of-mass frame. Since $\left|F_\pi(s)\right|^2$ is measured by many
experiments, it is customarily used for spectra comparisons.

\begin{figure}[htb]
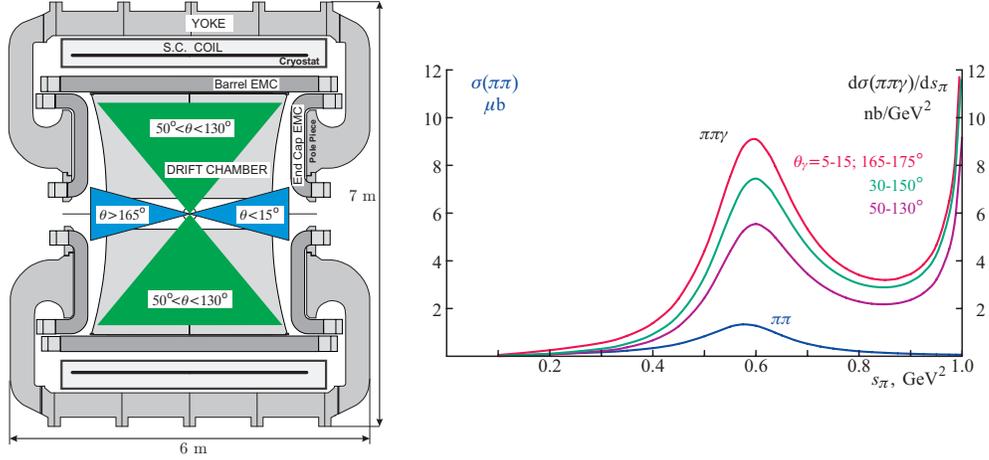

\centering\figb pfdetector_new2;5;\kern.5cm\figb PiPiGam;7.5;
\caption{Left: KLOE detector with the selection regions. Right: $\sigma(\pi\pi)$ and $d\sigma(\pi\pi\gamma)/ds_{\pi}$ as a function of $s_{\pi}$.}
\label{fig:1}
\end{figure}

Two different selection regions are used: the first one is named "Small
angle $\gamma$ selection'' (SA) since photons are required to be within a cone of
$\theta_\gamma<15^\circ$ around the beam line (narrow cones in \fig{fig:1},
left),  the second one is called "Large angle $\gamma$ selection" (LA) since
there should be at least one photon at a polar angle of $50^\circ<\theta_\gamma<130^\circ$ (large central
cones in \fig{fig:1}, left). In both cases the two charged pion tracks
should have $50^\circ<\theta_\pi<130^\circ$.

\vspace{-0.2cm}
\subsubsection{Small angle $\gamma$ selection and data analysis}
\label{sec:smallang}
At small photon polar angle $\theta_\gam$, ISR events are
vastly more abundant than FSR events. Requiring angular separation between
the pions and the photon further suppresses FSR \cite{BKM99}.
Thus, the first KLOE measurement \cite{sighad} was
performed using this event selection criterion.
Since ISR-photons are mostly collinear with the beam line, a high
statistics for the ISR signal events remains. On the other hand, a highly
energetic photon emitted at small angle forces the pions also to be at small
angles (and thus outside the selection cuts), resulting in a
kinematical suppression of events with $M^2_{\pi\pi}< 0.35\ {\rm GeV}^2$.

In this analysis the photon is not explicitly detected.
 Its direction is reconstructed from the tracks' momenta by closing
kinematics: $\vec{p}_\gamma\simeq\vec{p}_{miss}= \vec{p_{\phi}}-(\vec{p}_{\pi^+}
+\vec{p}_{\pi^-})$. The separation between pion and photon selection
regions greatly reduces the contamination from the resonant process
$e^+e^-\to \phi\to\pi^+\pi^-\pi^0$, in which the $\pi^0$ mimics the missing
momentum of the photon(s) and from the final state
radiation process $e^+e^-\to \pi^+\pi^-\gamma_\mathrm{FSR}$.

\begin{figure}[htb]
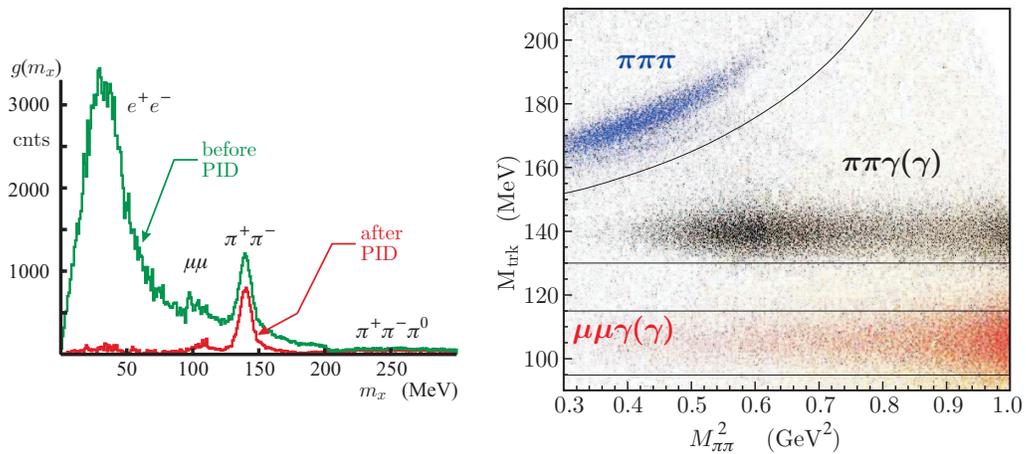

\centering\figb themass;6;\kern.5cm\figb trackmass_qq;7;
\caption{Left: $e+e-\gamma$ suppression using particle ID. Right:
 Signal and background distributions in the $M_{\rm trk}-M^2_{\pi\pi}$-plane.}

\label{fig:2}
\end{figure}

Discrimination of $\pi^+\pi^-\gamma$
from $e^+ e^-\to e^+ e^-\gamma$ events
is done via particle identification
based on the time of flight, on the shape and the energy
of the clusters associated to the tracks. In particular, electrons deposit
most of their energy in the first planes of the calorimeter while
minimum ionizing muons and pions release uniformly the
same energy in each plane. Events with at least one
of the two tracks not identified as electron are selected. This criterion
results in a rejection power of 97\% for $e^+ e^-\gamma$ events, while retaining a selection efficiency of $\sim 100\%$ for $\pi^+\pi^-\gamma$ events
(see \fig{fig:2}, left).

Contaminations from the processes $e^+e^-\to\mu^+\mu^-\gamma$ and
$\phi\to\pi^+\pi^-\pi^0$ are rejected by cuts on the track mass variable, $m_{trk}$,
defined by the four-momentum conservation, assuming a final state consisting
of two particles with the same mass and one photon, and on the missing mass, $m_{miss}=
\sqrt{E^2_{X}-|\vec{P}_{X}|^2}$, defined assuming the process to be $e^+ e^-\to\pi^+\pi^-X$.

\vspace{-0.2cm}

KLOE first published in 2005 results obtained with 141~pb$^{-1}$ collected in 2001
\cite{sighad}, which we will refer to as KLOE05 in the following. The statistical errors ranged from \ab2\% at
the lower limit in $s_\pi$ to \ab0.5\% at the $\rho$ peak.
The experimental systematic uncertainties were mostly flat in $s_\pi$
and amounted to 0.9\%.
The luminosity was estimated using large-angle Bhabha scattering and
contributed an additional, dominantly theoretical uncertainty of 0.6\%. The cross section $\sigma_{\pic}$ was obtained via \eq{eq:radiator}. The contribution
to the systematic uncertainty on $\sigma_{\pic}$ from FSR was
0.3\%, and the theory accuracy of $H$ was 0.5\%.
Finally, for the evaluation of $a_\mu^{\rm had}$, it was necessary to
remove the effects of vacuum polarization in the photon propagator
for the process $e^+e^-\to\pic$ itself by correcting for the running
of $\alpha_{\rm em}$. This contributes 0.2\% to the systematic
uncertainty. Thus from KLOE05 analysis we obtained for the contribution to $a_\mu^{\pi\pi}$ in the energy interval
$0.35 < s < 0.95$~GeV$^2$:
\begin{equation}
a_\mu^{\pi\pi}(0.35 < s < 0.95~{\rm GeV}^2) =
\pt{(388.7 \pm 0.8_{\rm stat} \pm 3.5_{\rm syst} \pm 3.5_{\rm th})},{-10},.
\label{eq:amu2001}
\end{equation}
Inclusion of the KLOE05 data together
with the CMD-2 data in the evaluation of $a_\mu^{\rm had}$ increased the
discrepancy between the calculated and measured values
of $a_\mu$ to \pt{(25.2\pm9.2)},{-10}, (\ab$2.7\sigma$) \cite{Hoe04}.
Furthermore, the precise KLOE spectrum exhibited a pronounced deviation
from those derived from $\tau$ lepton's spectral functions that led to the later being excluded in $a_\mu$ compilations \cite{Hoe04}.

\begin{figure}[htb]
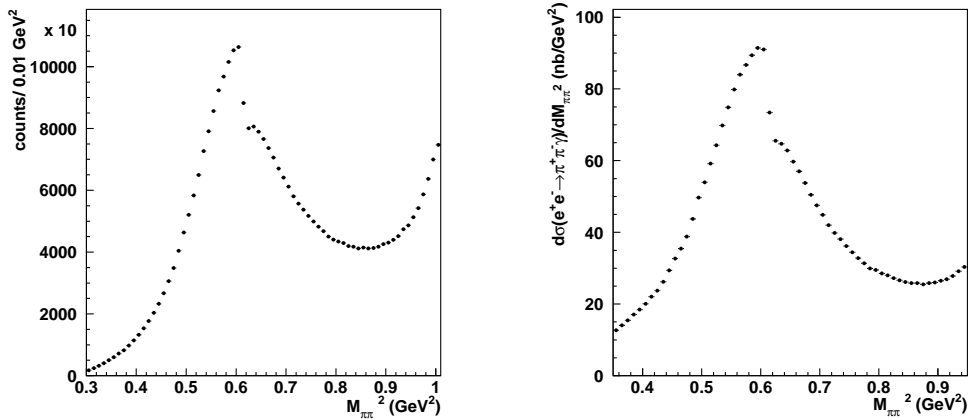

\centering \figb countspectrum;6.5;\kern.5cm\figb diffxsecppg;6.5;
\caption{ Left: KLOE08 $\pic\gam$ events per 0.01 GeV$^2$. Right: KLOE08 $\pic\gam$ differential cross section}
\label{sigppg}
\end{figure}

\begin{figure}[htb]
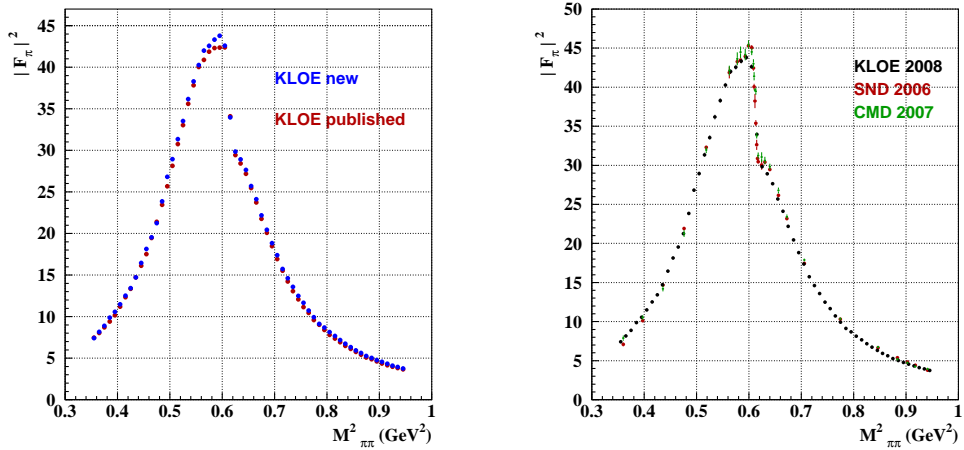

\centering\figb kloe02vs01;6.5;\kern.5cm\figb kloecomp;6.5;
\caption{Left: KLOE05 updated form factor in comparison with KLOE08. Right:
  comparison between KLOE 2008, SND 2006, and CMD 2007 form factors.}
\label{comp01_02}
\end{figure}

KLOE in 2008 completed the analysis of 241 pb$^{-1}$ data taken in 2002,
which we will refer to as KLOE08 in the following.
These data benefit from cleaner and more stable running conditions
of \DAF\, resulting in less machine background.
In particular the following changes are applied with respect to
the data taken in 2001:
i) an additional trigger level was implemented at the end of 2001
to eliminate a 30\% loss due
to pions penetrating up to the outer calorimeter plane and being misidentified
as cosmic rays events;
ii) the offline background filter, which
contributed the largest experimental systematic uncertainty (0.6\%) to the published
work~\cite{sighad}, has been improved and includes now a downscale algorithm providing an unbiased control sample. This greatly facilitates
the evaluation of the filter efficiency which increased from 95\% to 98.5\%,
with negligible systematic uncertainty;
iii) in addition, the knowledge of the detector response and
of the KLOE simulation program has been improved.
Thus the relative systematic errors on
the extraction of $a_\mu^{\pi\pi}$ in the mass range
[0.35,0.95] GeV$^2$ decreased from 1.3\% for the 2001
to 0.9\% for 2002 data.
Finally, we want to stress that the dominant term contributing to the
systematic error in KLOE08 analysis is given by the 0.5\% uncertainty quoted by the authors of \texttt{Phokhara}~\cite{Czyz:2005as}.
The KLOE08 results are shown in \fig{sigppg}. On the left the number
of $\pic \gam$ events per 0.01 GeV$^2$ are shown. The $\pic \gam$ differential cross section is shown in the right.
Finally, while performing the 2008 analysis we revisited the published 2001
data analysis, named KLOE05 updated, and we found a bias in the evaluation
of the trigger correction that affects mostly the low $M_{\pi \pi}$ region.
Correcting for this effect and normalizing to the new Bhabha cross section
we updated the published spectrum to compare with KLOE08.
\Fig{comp01_02} , left shows the pion form factor evaluated from these two sets of data.
Comparison among $a_\mu^{\pi\pi}([0.35,0.95] {\rm GeV}^2)$ values in units of
$10^{10}$ evaluated with the KLOE small angle $\gamma$ selection analyses:

KLOE05: $388.7~\pm~0.8_{\rm stat}~\pm~3.5_{\rm sys}~\pm~3.5_{\rm th}$;

KLOE05 Updated: $384.4~\pm~0.8_{stat}~\pm~3.5_{\rm sys}~\pm~3.5_{\rm th}$;

KLOE08:  $387.2~\pm~0.5_{stat}~\pm~2.4_{\rm sys}~\pm~2.3_{\rm th}$.

We see that in terms of $\sigma_{\pi\pi}$ the net shift of the updated
value is about one standard deviation below
the KLOE05 one and it is also in excellent agreement with the KLOE08 analysis.

Finally, we make a comparison of the most recent
$a_\mu^{\pi\pi}$ evaluations released by the CMD-2~\cite{Akhmetshin:2006bx} and
SND~\cite{Achasov:2006vp} experiments with the KLOE08 \cite{sighad07},
 in the mass range $M_{\pi\pi} \in [630,958]$ MeV.

CMD-2: $361.5~\pm~1.7_{stat}~\pm~2.9_{sys}$;

SND: $361.0~\pm~2.0_{stat}~\pm~4.7_{sys}$;

KLOE08: $356.7~\pm~0.4_{stat}~\pm~3.2_{sys}$;

The CMD-2 and SND dispersion integrals are performed
with the trapezoid rule, the KLOE value is done extrapolating
our bins to match the [630,958] MeV mass range, and summing directly
the bin contents of the $\dd\sigma_{\pi\pi\gamma}$ differential
spectrum, weighed for the kernel function.

The CMD-2 and SND values agree with the KLOE08 result within one standard
deviation. Thus, inclusion of the latest KLOE results in the world average
definitely established without a doubt the discrepancy between the measured
and calculated values of $a_\mu$, leading to hopes that there is a glimpse of which direction we should search for new physics.

\vspace{-0.2cm}

\subsubsection{Large angle $\gamma$ selection and data analysis}
\label{sec:largeang}
The contribution to $a_\mu^{\rm had}$ from $e^+e^-\to\pic$ in the interval
in $s$ between threshold and 0.35~GeV$^2$ is approximately \pt{1000},{-11},.
To explore the low-mass part of the spectrum, KLOE selects events requiring
a photon in the calorimeter with energy  greater than 50 MeV and
$50^\circ<\theta_\gamma<130^\circ$. With the photon explicitly detected,
about 40\% of the background from $\f\to\pic\po$ events is rejected by
kinematic closure. A further variable, the 3-dimensional angle $\Omega$ between the missing momentum and the photon, which is peaked at 0 degrees for signal events, removes most of the remaining $\f\to\pic\po$ background.

There are however two sources of irreducible background
: FSR and $\f\to f_0\gam\to\pic\gam$. FSR here is a very complicated problem: for $\theta_\gamma > 40\deg$
ISR and FSR events contribute nearly equally to the $\pic\gam$ spectrum
in the tails of the $\rho$, and the accuracy of the generator used to
obtain FSR corrections is critical.

As discussed by Binner et al. in Ref.~\cite{BKM99}, the ISR-FSR
interference results in a measurable charge asymmetry, as well as a distortion in the mass spectrum. The measured charge asymmetry can be used as a
useful gauge of the accuracy of the simulation.
KLOE has performed comparisons of this type in the
analysis of $\f\to\pic\gam$ as discussed in the previous section on scalar mesons \cite{sighad07}.

\begin{figure}[htb]
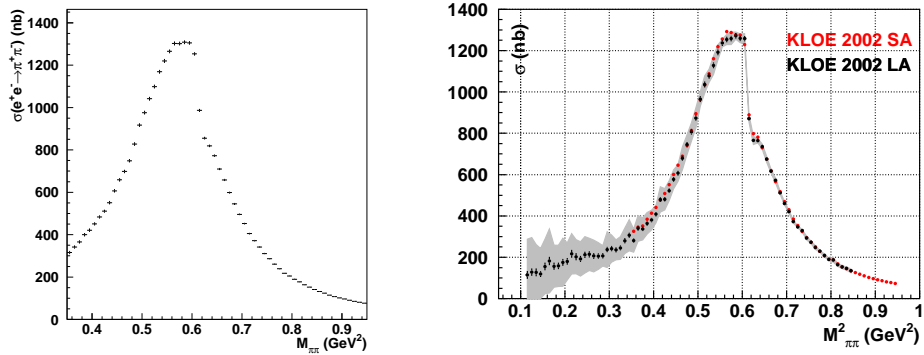

\centering\figb totalxsecpp;5.5;\kern.5cm\figb sigppsmala;7;
\caption{left: $\pi\pi$ cross section from small angle $\gamma$ selection, right: comparison between small and large $\gamma$ angle selections. }
\label{compsa_la}       
\end{figure}

We have made a measurement of $\sigma_{\pi\pi}$
with photon emitted at large angle using the data collected in 2002. Since
it is an independent event selection scheme, by comparing this result with
the small angle $\gamma$ analysis, we can also test the present models of FSR contributions. \Fig{compsa_la} shows
both spectra; note that the $\pi\pi$ threshold is reachable with the large
angle selection.
To make a quantitative comparison, at present we are limited by the unknown
interference between the FSR process and the resonant decays
$\phi\to\pi^+\pi^-\gamma$ as mentioned before.
Therefore we limit the comparison to the $m_{\pi\pi}^2$ range
where resonance contributions are negligible, that is, $a_{\mu}$ to
[0.5,0.85] GeV$^2$ where from the "small angle selection" we obtain $255.4~\pm~0.4_{stat}~\pm~2.5_{sys}$ and for the "large angle selection",   $252.5~\pm~0.6_{stat}~\pm~5.1_{sys}$, in excellent agreement.

The main source of systematic uncertainty in the large $\gamma$ angle selection
is the $f_0(980)$ background subtraction.
\subsubsection{Pion form factor from $R$}
\label{sec:ppff}
We can also obtain the pion form factor directly from measuring the ratio,
R, bin-by-bin of the observed pion and muon (radiative) cross
sections. In the absence of photons from final state radiation, one
finds~\cite{czyz:2004}
 \begin{equation}
|F_\pi(s')|^2 = \frac{4(1+2m_\mu^2/s')\beta_\mu}{\beta_\pi^3} \x
\frac{\dif\sigma({\pi\pi}+\gamma_{\mathrm{ISR}}(\gamma))}{\dif\sigma({\mu\mu}+\gamma_{\mathrm{ISR}}(\gamma))},
\label{eq:ratio1}
\end{equation}
where $m_\mu$ is the muon mass, and $\beta_\mu$, $\beta_{\pi}$ are the
muon and pion velocities in the center-of-mass frame.

The unavoidable presence of photons from final
state emission of both the pions and the muons in the data leads to a deviation
from the ``ideal'' \eq{eq:ratio1} which requires
treatment of final state radiation (FSR) with care. The advantage of this method is that
one does not need an absolute luminosity measurement, thus providing a totally independent
measurement of the pion form factor from those obtained using small or large angle $\gamma$ selections.

\begin{figure}[htb]
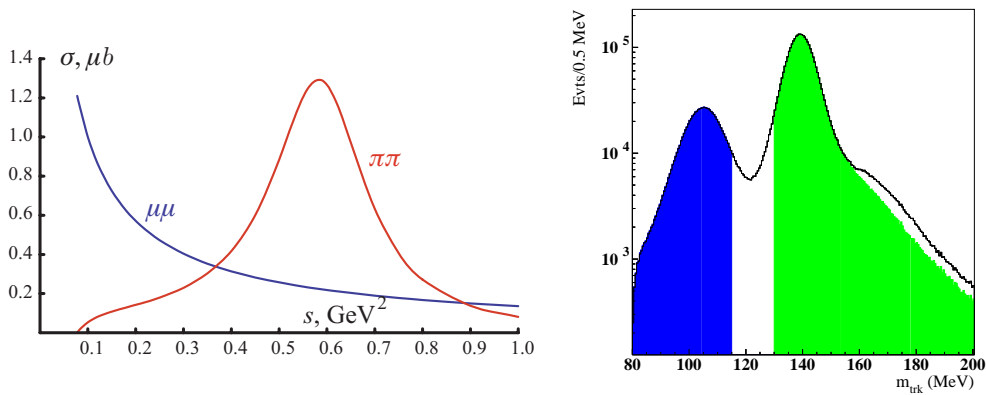

\centering\figb sigppmm-2;7;\kern.5cm\figb defmtrk;5.5;
\caption{Left: $\pi\pi$ versus $\mu\mu$ cross section as a function of energy. Right: . Discrimination of the $\mu^+\mu^-\gamma$ (filled left histogram) from the
$\pi^+\pi^-\gamma$ (filled right histogram) events.}
\label{compsamumu}       
\end{figure}

The experimental challenge of measuring the pion form factor using the ratio is the overwhelming $\mu\mu$ cross section over $\pi\pi$ cross section at low momentum, see \fig{compsamumu}, left. Furthermore, there is an overlap region between the $\pi\pi\gamma$ events and $\mu\mu\gamma$ events which makes precision event counting/separation a delicate operation. \fig{compsamumu},  right, shows the separation between $\pi^+\pi^-\gamma$ and $\mu^+\mu^-\gamma$ events, achieved
selecting different $m_{trk}$ regions around the missing mass peaks. The residual contamination of $\pi^+\pi^-\pi^0$ events are seen at high $m_{trk}$ values.

We are developing additional tools for $\pi\mu$ separation and the results from extracting $\left|F_\pi(s)\right|^2$ spectra from using the ratio is extremely promising ~\cite{ppg08}. We are confident of obtaining $a_\mu^{\pi\pi}$ from yet another independent method by the end of 2008.

\section{Conclusions and outlook}
\label{sec:concl}

The contributions by KLOE to particle physics during the years 2002-2008 are of great importance. The study of the decays of the kaons have demonstrated unitarity of the CKM matrix with accuracy better than one per mil and universality of the couplings between leptons and quarks. New mass and lifetime values have been obtained. Studies of the $\eta$ meson have reached new levels of precision. KLOE has made the most accurate measurement of the hadronic cross section pertaining to the calculation of the muon anomaly, and has made the most detailed studies on the nature of light scalar mesons.

A summary of all the masses, lifetimes, widths and branching ratios
measured by KLOE to date is shown in table~\ref{TAB:res}.
Results of tests on conservation of discrete symmetries, as
well as on the unitarity of the CKM matrix are shown in table~\ref{TAB:cons}.

\begin{table}[htb]
{\parbox{\textwidth}
{\centering
\begin{tabular}{lll}
        \hline
  {\bf Masses (MeV)} &  \multicolumn{2}{c}{}\\
                      &  $K^0$   & 497.583 $\pm$ 0.021 \\
                      & $\eta$    &  547.874 $\pm$ 0.032 \\
  \hline
 {\bf Lifetimes (ns)} & \multicolumn{2}{c}{}\\
                      & $\K_L$   & 50.84 $\pm$ 0.23    \\
                      & $K^\pm$    &  12.347 $\pm$ 0.030 \\
   \hline
  {\bf Widths (keV)}  & \multicolumn{2}{c}{}       \\
                      & $\phi \to e^{+}e^{-}$ & 1.32 $\pm$ 0.06 \\
   \hline
   {\bf Branching Ratios } & \multicolumn{2}{c}{} \\
    $\phi$ decays          & $\phi \to \eta'\gamma$ & (6.20 $\pm$ 0.27)$\times$10$^{-5}$ \\
                           & $\phi \to \pi^{0}\pi^{0}\gamma$ & (1.07 $\pm$ 0.07) $\times$10$^{-4}$ \\
                           & $\phi \to \eta\pi^{0}\gamma$ & (7.01 $\pm$ 0.24)$\times$10$^{-5}$ \\
                           & $\phi \to \omega\pi^{0}$ & (4.4 $\pm$ 0.6) $\times$10$^{-5}$ \\
                           & $\phi \to K^{0}\overline{K^{0}}\gamma$ & $<$ 1.8 $\times$10$^{-8}$ at 90$\%$ C.L. \\
   \kl\ decays             & {\em $K_L \to \pi e \nu$}   & 0.4008 $\pm$ 0.0015     \\
                           & {\em $K_L \to \pi\mu\nu$}   & 0.2699 $\pm$ 0.0014    \\
                           & {\em $K_L \to 3\pi^0$}   & 0.1996 $\pm$ 0.0020     \\
                           & {\em $K_L \to \pi^{+}\pi^{-}\pi^{0}$}   & 0.1261 $\pm$ 0.0011     \\
                           & {\em $K_L \to \pi^{+}\pi^{-}$} & (1.963 $\pm$ 0.21)$\times$10$^{-3}$ \\
                           & {\em $K_L \to \gamma\gamma$}   & (5.569 $\pm$0.077)$\times$10$^{-4}$ \\
   \ks\ decays             & {\em $K_S \to \pi^{+}\pi^{-}$}   & 0.60196 $\pm$ 0.00051     \\
                           & {\em $K_S \to \pi^{0}\pi^{0}$}   & 0.30687 $\pm$ 0.00051     \\
                           & {\em $K_S \to \pi e \nu$}   & (7.05 $\pm$ 0.09)$\times$10$^{-4}$  \\
                           & {\em $K_S \to \gamma\gamma$}   & (2.26 $\pm$ 0.13)$\times$10$^{-6}$ \\
                           & {\em $K_S \to 3\pi^{0}$} & $<$1.2$\times$10$^{-7}$ at 90$\%$ C.L. \\
                           & {\em $K_S \to e^{+}e^{-}(\gamma)$} & $<$9.3$\times$10$^{-9}$ at 90$\%$ C.L. \\
  \kpm\ decays             & {\em $K^{+} \to \mu^{+}\nu(\gamma)$}   & 0.6366 $\pm$ 0.0017     \\
                           & {\em $K^{+} \to \pi^{+}\pi^{0}(\gamma)$}   & 0.2067 $\pm$ 0.0012     \\
                           & {\em $K^{+} \to \pi^{0}e^{+}\nu(\gamma)$}   & 0.04972 $\pm$ 0.00053     \\
                           & {\em $K^{+} \to \pi^{0}\mu^{+}\nu(\gamma)$} & 0.03237 $\pm$ 0.00039     \\
                           & {\em $K^{+} \to \pi^{+}\pi^{0}\pi^{0}$}   & 0.01763 $\pm$ 0.00034  \\
  $\eta$ decays            & {\em $\eta \to \pi^{+}\pi^{-}e^{+}e^{-}$}  & (2.4 $\pm$ 0.4) $\times$10$^{-4}$\\
                           & {\em $\eta \to \pi^{0}\gamma\gamma$}  & (8.4 $\pm$ 3.0)$\times$10$^{-5}$ \\
 $\omega$ decays           & {\em $\omega \to \pi^{+}\pi^{-}\pi^{0}$} & 0.9024 $\pm$ 0.0018 \\
                           & {\em $\omega \to \pi^{0}\gamma$} & 0.0809 $\pm$ 0.0014 \\
\hline
 \end{tabular}}}
\caption{Summary of KLOE results on masses, lifetimes, widths and branching ratios measurements.}
\label{TAB:res}
 \end{table}

 \begin{table}[htb]
\begin{footnotesize}
{\parbox{\textwidth}
{\centering
\begin{tabular}{lll}
        \hline
 {\bf C symmetry} & \multicolumn{2}{c}{}       \\
                           & {\em $\eta \to 3\gamma$}  & $<$  1.6 $\times$10$^{-5}$ \\
                           & {\em A$_{LR}(\eta \to \pi^{+}\pi^{-}\pi^{0})$}   & (0.09$^{+0.13}_{-0.17}$) $\times$10$^{-2}$ \\
                           & {\em A$_{Q}(\eta \to \pi^{+}\pi^{-}\pi^{0})$}   & (-0.05 $\pm$ 0.11) $\times$10$^{-2}$   \\
                           & {\em A$_{S}(\eta \to \pi^{+}\pi^{-}\pi^{0})$}   & (0.08$^{+0.13}_{-0.16}$) $\times$10$^{-2}$ \\
  \hline
 {\bf CP symmetry} & \multicolumn{2}{c}{}        \\
                   & {\em $|\epsilon|$ (from BR($\kl\to \pic$)}  & (2.216 $\pm$ 0.013) )$\times$10$^{-3}$    \\
                   & {\em $\Re(\eps)$ (from Bell-Steinberger relation)}  &  (159.6 $\pm$ 1.3)$\times$10$^{-5}$ \\
                   & {\em $\eta \to \pi^{+}\pi^{-}$}  & $<$  1.3 $\times$10$^{-5}$ \\
  \hline
  {\bf CPT symmetry} & \multicolumn{2}{c}{}         \\
                     & {\em $\tau(K^{+})/\tau(K^{-})$}  & (1.004 $\pm$ 0.004) \\
                     & {\em $M(\rho^{+})-M(\rho^{-})$}  & (1.5 $\pm$ 1.1) MeV \\
                     & {\em $\Gamma(\rho^{+})-\Gamma(\rho^{-})$}  & (1.8 $\pm$ 2.1) MeV \\
                     & {\em $\Im(\delta)$ (from Bell-Steinberger relation)}  &  (0.4 $\pm$ 2.1)$\times$10$^{-5}$ \\
                     & {\em $\mathrm{Re}\,\delta+\mathrm{Re}\,x_{-}=(A_{S}-A_{L})/4$} & (-0.5$\pm$2.5)$\times$ 10$^{-3}$\\
    with CP symmetry & {\em $\mathrm{Re}\,\epsilon-\mathrm{Re}\,y=(A_{S}+A_{L})/4$} & (1.2$\pm$2.5)$\times$ 10$^{-3}$\\
\hline
{\bf Lepton universality} & \multicolumn{2}{c}{}         \\
                          & $r_{\mu e}$ (from \fVus\ for $K^{\pm0}_{e3}$ and $K^{\pm0}_{\mu3 }$)&  1.000 $\pm$ 0.008 \\
                          & \rk\ & $(2.55\pm0.05\pm0.05)\times10^{-5}$ \\
\hline
{\bf CKM matrix unitarity} & \multicolumn{2}{c}{}         \\
                           & \fVus\            &  0.2157$\pm$0.0006\\
                           & \fVusVud\         & 0.2766$\pm$ 0.0009\\
                           & $|\Vus|^2+|\Vud|^2-1$ (\Vud\ from~\cite{vud07ht}) &\hspace{-0.5cm}$-0.0004\pm0.0007(\ab0.6\sigma)$ \\
  \hline
 \end{tabular}}}
\end{footnotesize}
\caption{Summary of KLOE results on symmetry tests.}
\label{TAB:cons}
 \end{table}

It is interesting to note that over half of the results noted above are based on analysis of one fifth of the total KLOE data sample. Several of the measurements have reached the state where additional statistics will not improve accuracy due to theoretical or intermediate state uncertainties as described in the specific sections. There are measurements that could profit with respect to statistical and systematic errors continuing the studies of the entire body of data. Furthermore, several new measurements are only possible using the whole data set whose analysis is expected to be completed in the coming couple of years. These are discussed in the following three subsections. In the fourth subsection we describe a future project KLOE-2 that entails minor upgrade of KLOE in order to complete the original physics program proposed in 1992 but had been stymied by \DAF's limited luminosity. Thus it is clear that we have our hands full for the foreseable future!

\subsection{Kaon sector}
\label{sec:kaonf}
\begin{enumerate}
\item Studies on rare kaon decays can benefit of the use of the entire KLOE statistics. The accuracy in the determination of the branching ratio of
$\ks\to\pi^+e^-\bar\nu(\gam)$ events and of the related charge asymmetry will scale with statistics with respect to those shown in sect.~\ref{sec:ksl3},  thus reaching $\sim$0.5$\%$ (fractional) and 5$\times$10$^{-3}$ respectively.
\item We have observed the $\ks\to\pi^+\mu^-\bar\nu(\gam)$ decay channel
and will aim at reaching a precision better than 2$\%$ on its BR.

\item We plan to improve our limit on $\ks\to3\po$ decays
(see sect.~\ref{sec:ks3p0}). Preliminary studies have shown that, by slightly modifying the analysis cuts, one can improve on background rejection, while leaving the signal efficiency almost unaffected. This means that we can hope to set the limit on the  corresponding BR at the level of $\sim$2$\times$10$^{-8}$.

\item Moreover, we can search for $\ks \to\pi^{+}\pi^{-}\pi^{0}$ events,
which proceed mainly via a CP-conserving, $\Delta$I = 3/2 transition. A preliminary, but rather detailed, analysis shows that KLOE should be able to measure the corresponding branching ratio with a $\sim$60$\%$ precision, comparable to the precision with which this BR has been measured by three experiments.

\item The CP violating $K_L \to 2\pi^0$ events can be measured with a
\ab0.5\% accuracy, thus setting at the few per mil level the precision on~\rep\ which KLOE can reach.

\item The study of the main $K_L$ decay channels, will allow us to
improve our accuracy in the measurement of the $K_L$ lifetime
by a factor of $\sim$2. We can also improve the precision in the
measurement of the scalar form factor for semileptonic decays by the same amount.

\item The form factor slopes for \kpm\ can be measured as well, using
about 5.9$\times 10^{6}$ $K^{\pm}_{e3}$ and about 2.5$\times 10^{6}$ $K^{\pm}_{\mu3}$ decays from the entire data sample. It is important to note that in the scalar form factor measurement from \kpm\ there is no $\pi/\mu$
ambiguity.

\item The accuracy on the \kpm\ lifetime can be improved by a factor \ab\ 1.5. As discussed in sect.~\ref{kloevus}, these are key ingredients for the determination of the CKM matrix element $V_{us}$.

\item Semi-rare charged kaon decay channels can also be studied. In particular, we are planning to measure the branching ratio of the decay into three charged pions. With the present data sample a few per mil statistical accuracy can be reached. Using the same data set the measurement of the branching ratio of the $K^{\pm}\to\pi^0\pi^0 e^\pm \nu_{e}$ decay can reach a 10\% statistical accuracy. This will allow us to complete the measurement of the main \kpm\ branching ratios.
\end{enumerate}

\subsection{Light meson spectroscopy}
\label{sec:lightmsf}

Many analyses on light meson spectroscopy could be refined in the future.
Although the level of accuracy for the measurements on the scalar sector
has already reached its limit, a combined analysis of different final states, such as $\phi\to\pi^+\pi^-\gamma$/$\phi\to\pi^0\pi^0\gamma$, will give more insight into the structure of light scalar mesons.

On the other hand, the pseudoscalar mesons could benefit from the larger statistics of the whole KLOE data sample.

\begin{enumerate}
\item In particular, we can complete the study of the main $\eta$ decay channels by measuring the BR's of $\eta \to \pi^+\pi^-\gamma$,
$\eta \to e^+ e^- \gamma$ with a statistical precision of few per mil.
\item Moreover, the knowledge on numerous rare $\eta$ decay channels can be improved. Apart from lowering the limits on C (P,CP) violating decays
$\eta\to 3\gamma$ ($\eta \to \pi^+\pi^-,\pi^0\pi^0$), we could reach a
statistical precision of 3.5\% on the BR($\eta\to \pi^+\pi^-\break e^+e^-$) and
measure for the first time the asymmetry between the $\pi\pi$ and $ee$ decay planes. The knowledge of the systematics for the latter decay could reach a few \% accuracy by relying on large data/MC control samples.
An observation at the level of 100 events is expected for the double
Dalitz decay $\eta \to e^+ e^- e^+ e^-$.
\item Finally, the BR($\eta \to \pi^0 \gamma \gamma$) will reach a 1.5\%
statistical accuracy. Improvements of the analysis based on a new
clustering scheme and on a photon veto at small angle could reduce
the related systematics and background contamination, allowing also the
measurement of the $\gamma\gamma$ invariant mass spectra.

\item For the $\eta'$, we have in hand about 450,000 of these mesons,
so precise measurements could now be carried out. We plan to do
this in three ways:

i. looking at the $\eta'$ dynamics by studying the Dalitz plot for
   the $\eta' \to \eta \pi^+ \pi^-$, $\eta' \to \pi^+ \pi^- \pi^0$
   processes;

ii. measuring ratio of branching fractions for some of the main decay
   channels, such as $\eta' \to \pi^+ \pi^- \eta$,
   $\eta' \to \pi^0 \pi^0 \eta$, $\eta' \to \omega \gamma$, thus
   improving their BR accuracy;

iii. searching for never heretofore observed decay channels, whose upper limits are   at the \% level ($\eta' \to \pi^+ \pi^- \pi^0$,
   $\eta' \to \pi^+ \pi^- \pi^+ \pi^-$)
\end{enumerate}
\subsection{Pion form factor at about 1 GeV}
\label{sec:ppff1GeV}
In early 2006, KLOE also collected \ab200 pb\up{-1} of data at $\sqrt s =
1000$~MeV, at which $\sigma(\epm\to\f\to\pic\po)$ drops to 5\% of its peak
value. These data by themselves should determine the contribution to $a_\mu^{\pi\pi}$ normalizing to absolute luminosity, with a statistical error of \ab\pt35,-11,.  
This will give us another check on our measurements using the previously mentioned three methods: small angle selection, large angle selection and $\pi\pi\gamma$ to $\mu\mu\gamma$ ratio.

\subsection{KLOE-2}
\label{sec:kloe2}
In late 2007 an important accelerator experiment has begun at \DAF, with the goal of increasing the machine's luminosity by a factor of at least three, from the implementation of an innovative interaction scheme~\cite{ref:panta}.

The exciting perspective of collecting much more data  within a few years, inspired many KLOE members, together with several new collaborators, to form a new enterprise with the goal of continuing the KLOE physics program with an upgraded detector~\cite{ref:k2}.

This project, dubbed with a certain lack of fantasy KLOE-2, foresees the insertion of new subdetectors in the inner part of KLOE, between the beryllium beam pipe and the internal wall of the drift chamber.
An inner tracking chamber, based on the GEM technology is being developed at LNF. Crystal calorimeters will help catching photons emitted at very low polar angles. These detectors will increase acceptance for many
physics channels, especially for $K_S$, $\eta$ and $\eta$' decays.
The implementation of a detector to tag $\gamma\gamma$ interactions is also under study.

KLOE-2 is an ambitious project, which requires a few years of preparation and a longer period of data taking: the goal is to acquire at least 20-30 fb$^{-1}$ by year 2012.

\section*{Appendix A. The KLOE Collaboration: inclusive to 2008}

F.~Adinolfi~$^{j,k}$,
F.~Ambrosino~$^{e,f}$,
A.~Antonelli~$^a$,
M.~Antonelli~$^a$,
F.~Archilli~$^{j,k}$,
C.~Bacci~$^{l,m}$,
M.~Barva~$^{l,m}$,
P.~Beltrame~$^b$,
G.~Bencivenni~$^a$,
S.~Bertolucci~$^a$
% ...
I-00044 Frascati, Italy\\
\llap{$^b$}Institut f\"ur Experimentelle Kernphysik, Universit\"at Karlsruhe, D-76128 Karlsruhe,Germany\\
\llap{$^c$}Dipartimento di Fisica dell'Universit\`a del Salento, Via Arnesano, I-73100 Lecce, Italy\\
\llap{$^d$}INFN Sezione di Lecce, Via Arnesano, I-73100 Lecce, Italy\\
\llap{$^e$}Dipartimento di Scienze Fisiche dell'Universit\`a  di Napoli ``Federico II'', Via Cintia, I-80126 Napoli, Italy\\
\llap{$^f$}INFN Sezione di Napoli, Via Cintia, I-80126 Napoli, Italy\\
\llap{$^g$}Dipartimento di Energetica dell'Universit\`a di Roma ``La Sapienza'', P. Aldo Moro 2, I-00185 Roma, Italy\\
\llap{$^h$}Dipartimento di Fisica dell'Universit\`a di Roma ``La Sapienza'', P. Aldo Moro 2, I-00185 Roma, Italy\\
\llap{$^i$}INFN Sezione di Roma, P. Aldo Moro 2, I-00185 Roma, Italy\\
\llap{$^j$}Dipartimento di Fisica dell'Universit\`a di Roma ``Tor Vergata'',
Via della Ricerca Scientifica 1, I-00133 Roma, Italy\\
\llap{$^k$}INFN Sezione di Roma Tor Vergata, Via della Ricerca Scientifica 1, I-00133 Roma, Italy\\
\llap{$^l$}Dipartimento di Fisica dell'Universit\`a di Roma ``Roma Tre'', Via della Vasca Navale 84, I-00146 Roma, Italy\\
\llap{$^m$}INFN Sezione di Roma Tre, Via della Vasca Navale 84, I-00146 Roma, Italy\\
\llap{$^n$}Physics Department, State University of New York at Stony Brook, Stony Brook, NY 11794-3840  USA\\
\llap{$^o$}Institute of High Energy Physics of Academia Sinica,  P.O. Box 918 Beijing 100049, P.R. China\\%
\llap{$^p$}Physics Department, University of Virginia, Virginia, USA\\
\llap{$^q$}Institute for Theoretical and Experimental Physics, B. Cheremushkinskaya ul. 25 RU-117218 Moscow, Russia\\%
\llap{$^r$}Dipartimento di Fisica dell'Universit\`a e Sezione INFN di Trieste, Trieste, Italy\\
\llap{$^s$}Dipartimento di Fisica dell'Universit\`a e Sezione INFN di Pisa, Pisa, Italy\\
\llap{$^t$}Dipartimento di Fisica dell'Universit\`a e Sezione INFN di Bari, Bari, Italy\\
\llap{$^u$}High Energy Physics Institute, Tbilisi State University, Tbilisi, Georgia\\
\section*{Acknowledgements}

We thank the \DAF\  team for their efforts in maintaining low background running conditions and their collaboration during all data-taking. We want to thank our technical staff: G.F.Fortugno for his dedicated work to ensure an efficient operation of the KLOE Computing Center; M.Anelli for his continuous support to the gas system and the safety of the detector; A.Balla, M.Gatta, G.Corradi and G.Papalino for the maintenance of the electronics; M.Santoni, G.Paoluzzi and R.Rosellini for the general support to the detector; C.Piscitelli for his help during major maintenance periods. This work was supported in part by DOE grant DE-FG-02-97ER41027; by EURODAPHNE, contract FMRX-CT98-0169; by the German Federal Ministry of Education and Research (BMBF) contract 06-KA-957; by Graduiertenkolleg `H.E. Phys. and Part. Astrophys.' of Deutsche Forschungsgemeinschaft, Contract No. GK 742; by INTAS, contracts 96-624, 99-37; and by the EU Integrated Infrastructure Initiative HadronPhysics Project under contract number RII3-CT-2004-506078.

\end{document}
\endinput